\documentclass[12pt,onecolumn]{article}
\usepackage{amsmath}
\usepackage{amssymb}
\usepackage{graphicx}
\usepackage{caption2}
\usepackage{amsfonts}
\usepackage{cite}

\newcommand{\be}{\begin{equation}}
\newcommand{\ee}{\end{equation}}
\newcommand{\bea}{\begin{eqnarray}}
\newcommand{\eea}{\end{eqnarray}}
\newcommand{\bef}{\begin{figure}}
\newcommand{\ef}{\end{figure}}
\newcommand{\bt}{\begin{tabular}}
\newcommand{\et}{\end{tabular}}
\newcommand{\bno}{\begin{enumerate}}
\newcommand{\eno}{\end{enumerate}}

\def\3{\ss}

\catcode`\"=\active
\def"{\accent'177}

\begin{document}

\begin{center}
{\bf\Large Do peaked solitary water waves indeed exist? }

\vspace{0.5cm}

Shijun  Liao $^{1,2,3}$  \\
 \vspace{0.5cm}

$^1$ State Key Laboratory of Ocean Engineering, Shanghai 200240, China 

$^2$ School of Naval Architecture, Ocean and Civil Engineering \\
Shanghai Jiaotong University, Shanghai 200240, China 

$^3$  Dept. of Mathematics,  Shanghai Jiaotong University, Shanghai 200240, China


\vspace{0.5cm}
(Email address:  sjliao@sjtu.edu.cn)

\end{center}

\begin{abstract}

{\em  Many   models of shallow water waves, such as the famous Camassa-Holm equation,  admit peaked solitary waves.   However, it is an open question whether or not the widely accepted   peaked solitary waves can be derived from the fully nonlinear wave equations.   In this paper,  a  unified  wave model  (UWM)  based on the  symmetry and the fully  nonlinear  wave  equations   is put forward for progressive waves with permanent form in finite water depth.   Different from  traditional wave models,  the flows  described by the  UWM  are  not  necessarily  irrotational  at crest,  so that it is more general.    The  unified  wave model admits not only the traditional  progressive waves with smooth crest, but also a new kind of  solitary waves with peaked crest that include the famous peaked solitary waves given by the Camassa-Holm equation.  Besides,  it is proved that  Kelvin's theorem still holds everywhere for the newly found peaked solitary waves.   Thus,  the  UWM unifies, for the first time,  both of the traditional  smooth  waves and the  peaked  solitary waves.  In other words,  the peaked solitary waves are consistent with the traditional smooth ones.  So,   in the frame of inviscid fluid,   the peaked solitary waves are as acceptable and reasonable as the traditional smooth ones.   It is found that the peaked solitary waves  have some unusual and unique characteristics.    First of all, they   have  a  peaked  crest  with  a  discontinuous  vertical velocity at crest.     Especially, unlike the traditional smooth waves  that are  dispersive with wave height,  the  phase speed of the peaked solitary waves has nothing to do with wave height, but depends (for a fixed wave height) on its decay length, i.e. the actual wavelength:  in fact, the peaked solitary waves are dispersive with  the actual wavelength when wave height is fixed.    In addition,  unlike traditional smooth waves whose kinetic energy decays exponentially from free surface to bottom,    the kinetic energy of the peaked solitary waves  either  increases  or  almost keeps the same.  All of these  unusual properties show the novelty of the peaked solitary waves,  although it is  still  an open question whether or not  they are reasonable in physics if the viscosity of fluid and surface  tension  are considered. }

\end{abstract}

\hspace{-0.75cm}{\bf Key words}  unified wave model (UWM),   progressive wave,   solitary  peaked wave, homotopy analysis method (HAM)

\section{Introduction}

Since the solitary surface wave was discovered by John Scott Russell \cite{Russell1845} in 1834,   various types of solitary waves have been found.   The mainstream models of shallow water waves, such as the Boussinesq equation  \cite{Boussinesq1872}, the KdV equation  \cite{KdV1895}, the  BBM equation  \cite{Benjamin1972} and so on, admits   dispersive  {\em smooth}  periodic and  solitary  waves with permanent form: the wave elevation is {\em infinitely} differentiable {\em everywhere}.   Especially,  the phase  speed  of  these smooth waves is  closely  related to wave height:  a  smooth progressive wave with higher amplitude always propagates faster.      Today, such  kind  of  amplitude-dispersive  periodic and solitary waves with smooth crest  are the mainstream  of  water waves.

However, the discontinuity of water wave elevation appears accidentally in theory.     It is well-known that the famous  limiting  gravity  wave  has  a  corner crest  with  120  degree,  as  pointed out by  Stokes \cite{Stokes1894} in 1894.   It is a pity that  the  importance  of  such  kind  of  discontinuity  seems to  be  neglected, mainly because    Stokes'  limiting  gravity wave  \cite{Stokes1894}  has been never observed in practice.   In 1993,  almost one hundred  years  later,  Camassa \& Holm \cite{Camassa1993PRL}   proposed  a  model for shallow water waves with small amplitude, called   today   the  Camassa-Holm (CH) equation
\begin{equation}
u_t + 2 \omega u_X - u_{XXt} + 3 u u_X = 2 u_X u_{XX} + u u_{XXX},    \label{geq:CH}
\end{equation}
where $u(X,t)$ denotes the wave elevation, $X$ and $t$ are the spatial and temporal variables,  $\omega$ is a constant related to the critical  phase speed of shallow water waves,   respectively.   As pointed out by Camassa \& Holm \cite{Camassa1993PRL},  the CH equation (\ref{geq:CH})  has the  solitary wave  when $0 \leq \omega <1/2$.
Especially,   in the limit $\omega \to 0$,   the CH equation (\ref{geq:CH}) admits the peaked  solitary wave   in the closed-form
\begin{equation} 
 u(X,t)  =  c \; \exp(-|X - c \;  t|),   \label{res:u:CH}
 \end{equation}
whose first derivative is  {\em discontinuous}  at  the  crest $X=c\; t$,  where $c$ denotes the phase speed.   Unlike the KdV equation and Boussinesq equation,  the  CH equation  (\ref{geq:CH})   can model  phenomena of soliton interaction and wave breaking, as mentioned  by  Constantin \cite{Constantin2000}.   Mathematically,  the CH equation (\ref{geq:CH}) is integrable and bi-Hamiltonian,  thus possesses an infinite number of conservation laws in involution, as pointed out by  Camassa \& Holm \cite{Camassa1993PRL}.   In addition,  it is associated with the geodesic flow on the infinite dimensional Hilbert manifold of diffeomorphisms of line, as mentioned by  Constantin \cite{Constantin2000}.   Thus, the CH equation  (\ref{geq:CH})   has many intriguing physical  and  mathematical properties.   A few researchers even believed that  the CH equation (\ref{geq:CH}) ``has the potential to become the new master equation for shallow water wave theory'' \cite{Fushssteiner1996}.      

Note that  the discontinuity also widely appears in practical flows,  such as dam break in hydrodynamics and shock wave in aerodynamics.   In fact, such kind of discontinuous problems  belong to the so-called Riemann problem \cite{Wu-2008IJNMF, Rosatti-2010JCP}, a classic research field.   So,  the discontinuity of  wave elevation  seems to  be  reasonable  not only in mathematics but also in physics.

Thousands of articles for peaked solitary waves  have been published.   
Currently,  the closed-form solutions of peaked solitary waves of the Boussinesq equation, the KdV equation, the BBM equation,  the modified KdV equation and KP equation are derived  by Liao \cite{Liao2012-SciChinaG, Liao2013-SciChinaG} using the symmetry and the corresponding wave models mentioned above.    
Besides,  Kraenkel and Zenchuk  \cite {Kraenkel1999}  gave  the explicit  cusped solitary waves  of the CH equation (\ref{geq:CH}) when $\omega\neq 0$, called cuspon.    The so-called  cuspon is a kind of solitary wave with the 1st derivative  going to {\em infinity} at crest.  Note that the peakon has a {\em finite} 1st derivative, but the cuspon has an {\em infinite} 1st derivative at crest.  Thus, peakons and cuspons are completely different two kinds of discontinuous solutions of the CH equation (\ref{geq:CH}).    Therefore, nearly {\em all} mainstream models of the shallow water waves admit the peaked solitary waves, as pointed out  by Liao \cite{Liao2012-SciChinaG, Liao2013-SciChinaG}.  

However, there are many open questions about peaked  solitary waves.  Where does this kind of discontinuity come from?   Are these peaked solitary waves as acceptable and reasonable as the traditional smooth ones?   Are  there  any  other  peaked waves  in  {\em finite}  water  depth?
It should be emphasized  that all of the above-mentioned mainstream models of shallow water waves are approximations of the fully nonlinear wave equations under assumptions of the existence of some small physical parameters (such as small wave amplitude,  shallow water and so on).    Can we gain the peaked solitary waves from the fully  nonlinear water wave equations?   To the best of the author's knowledge,  this is still an open question up to now.    This situation  is  rather  strange.   Logically,  if the peaked solitary waves given by the simplified wave models are acceptable and reasonable,  the  exact fully nonlinear wave equations should also admit it.  Otherwise,  we must  check carefully  the  acceptableness  and  reasonableness of  the   peaked/cusped solitary waves  reported/studied  in  thousands of published articles.     

 In this article, a positive answer to this open question is given.   Proposing  a  unified  wave model  (UWM) based on the symmetry and the fully nonlinear wave equations,   we  gain  a new type of  solitary surface waves in {\em finite} water depth,   which have a peaked crest and  possess  many unusual characteristics quite different from the traditional smooth ones.     The  UWM  admits   two different kinds of waves:  one is {\em infinitely} differentiable everywhere, with a phase speed closely related to the wave height,  the other has a  peaked crest and especially the {\em discontinuous} vertical velocity at crest, with a phase speed having nothing to do with wave height!   The former  are exactly the same as the traditional periodic and solitary waves with smooth crest found  in textbooks.  However,  the latter has many completely new properties and has never been reported, to the best of our knowledge.   In theory, the  UWM  {\em unifies} both of the traditional smooth waves and the peaked solitary waves, for the first time.  Therefore,   in essence, the peaked solitary waves are  {\em consistent} with the traditional smooth ones, and thus are as acceptable and reasonable as them.  
 
This article is outlined as follows.  
In \S~\ref{formula},  a  unified  wave model  (UWM) based on the symmetry and the fully nonlinear wave equations  for progressive gravity waves with permanent form in finite water depth is described, which admits  not only the   traditional  periodic and solitary  progressive waves with smooth crest  (see \S~3),  but also a new type of peaked solitary waves (see \S~4).  
  In \S~\ref{LinearTheory}, a new type of peaked solitary surface waves are first obtained  by means of  the linearized  UWM.   In \S~\ref{NonlinearTheory},   such kind of  peaked solitary surface waves are confirmed by the exact  UWM.    This kind of  peaked  solitary surface waves have some unusual characteristics quite different from traditional smooth ones, as described in \S~\ref{Characteristic}.   Besides, it  is  proved  in \S~4.4 that  the  Kelvin's theorem  still  holds  everywhere for these peaked solitary waves.    The concluding  remarks  and  discussions  about  the newly found  peaked solitary waves are given in \S~\ref{Discussion}.   The  corresponding nonlinear partial differential equations are solved in \S~4.2 by means of  an analytic approximation method for highly nonlinear problems, namely the homotopy analysis method (HAM)  \cite{ Liao1997NLM, Liao1999JFM,   LiaoBook2003,  Liao2003JFM, Liao2006SAM,  LiaoBook2012, KVBook2012},  which has nothing to do with small/large physical parameters at all and besides can guarantee the convergence of solution series.  

\section{A  unified  wave model based on symmetry \label{formula}}  %

First of all,  we describe a  unified  wave model  (UWM) for progressive waves with permanent form in water of {\em finite} depth.   Note that, mathematically,  we must be extremely careful so that  the  discontinuous  solutions are not lost. 

Consider a  progressive surface gravity wave propagating on a horizontal  bottom  with a constant phase speed $c$ and a  permanent  form in a {\em finite} water depth $D$.   For simplicity,  let us  consider  the  problem  in the frame moving with the  phase speed $c$.   
Let $x, z$ denote the horizontal and  vertical dimensionless co-ordinates (using $D$ as the characteristic length),   with  $x=0$ corresponding to the wave crest,  $z=-1$ to the bottom, and  the $z$  axis upward, respectively.     Assume that the wave elevation has a symmetry about the crest at  $x=0$.  
   Due to this symmetry, we   need consider the interval $x \geq 0$ only.    Assume that the fluid  in the domain  $x >0$ is inviscid and incompressible,  the flow is irrotational, and surface tension is neglected.   
It should be emphasized  here that the flow at $x = 0$ is {\em not} necessarily irrotational.    In other words,  there might exist the vorticity at $x = 0$.    Thus, this model  based on the symmetry is more general than others, as shown below.      

 Let   $g$  denote the acceleration due to gravity,  
 $\phi$ the velocity potential (in the interval  $0<x <+\infty$),   $\zeta$  the free surface,   respectively.    All of these variables are dimensionless by means of $D$  and  $\sqrt{g D}$ as the characteristic  scales of length  and  velocity.       Due to the symmetry $\zeta(-x)=\zeta(x)$,   we  only  need  consider  the flow  in the interval  $x\in(0,+\infty)$.   Since  the fluid  in the domain $x >0$ is inviscid and incompressible,  besides the flow is irrotational and the surface tension is neglected,   it is easy to have the governing equation 
 \begin{equation}
\nabla^2\phi(x,z)=0,  \hspace{1.0cm} z \leq \zeta(x),   0   <  x <+\infty, \label{geq:phi}
\end{equation}
subject to the  boundary conditions on the unknown free surface $z=\zeta(x)$:
\begin{equation}
\alpha^2 \frac{\partial^2 \phi}{\partial x^2} + \frac{\partial \phi}{\partial z} -\alpha \frac{\partial}{\partial x} \left(  \nabla \phi  \cdot \nabla \phi \right) +\nabla \phi \cdot \nabla \left( \frac{1}{2} \nabla \phi  \cdot \nabla \phi \right) =0,  \hspace{1.0cm}  0 < x <+\infty,  \label{bc:phi}
\end{equation}
\begin{equation}
 \zeta  - \alpha \frac{\partial \phi}{\partial x} + \frac{1}{2} \nabla \phi  \cdot \nabla \phi  = 0,  \hspace{1.0cm}  0 < x <+\infty,
\label{bc:zeta}
\end{equation}
and the bottom condition
\begin{equation}
\frac{\partial \phi}{\partial z} = 0, \hspace{1.0cm}  z = -1,   \; \;  0 <  x <+\infty,  \label{bc:bottom}
\end{equation}
 where  $\nabla^2$ is a Laplace operator, and 
  \begin{equation}
\alpha = \frac{c}{\sqrt{g D}}
\end{equation}
is the dimensionless wave-speed, respectively.    Note that all of the above-mentioned governing equation and boundary conditions are defined in the domain $0 < x < +\infty$.    In fact, they are just the fully nonlinear wave equations,  except the restricted interval $0 <  x  < +\infty$.     Besides,  we have either the periodic condition
\begin{equation}
\phi(x,z) = \phi(x + \lambda, z), \hspace{1.0cm}  -1\leq z \leq \zeta(x)  \label{bc:periodic}
\end{equation}
for the periodic progressive waves, where $\lambda = 2\pi/k$ is the wavelength,  or the decay condition
\begin{equation}
\phi(\pm \infty, z) = 0, \hspace{1.0cm}  -1\leq z \leq \zeta(x)   \label{bc:decay}
\end{equation}
for the solitary waves.   On the vertical boundary $x = 0$,  we have the additional condition
\begin{equation}
u(0,z) = \lim_{x\to 0}\frac{\partial \phi}{\partial x}  =  U(z),   \hspace{1.0cm}  z\leq \zeta(x),    \label{bc:left}
\end{equation}
where $U(z)$  is such an {\em unknown} horizontal velocity at $x=0$  that  the corresponding velocity potential $\phi(x,z)$ and the  progressive wave elevation  $\zeta(x)$  with {\em permanent} form  exist.
Furthermore,   let $H_w$ denote the dimensionless wave-elevation at $x=0$,  corresponding to the wave crest.  Then,  for  given $H_w$, one has an addition condition
 \begin{equation}
\lim_{x\to 0} \zeta(x)  =  H_w.    \label{bc:Hw}
 \end{equation}
In addition, the wave elevation must be bounded, i.e.
 \begin{equation}
 |\zeta(x)| < C,   \hspace{1.0cm}  0 <  x <  +\infty,   \label{bc:bounded}
 \end{equation}
 for a large enough constant $C$.    
 
 Let $u(x,y)$ and $v(x,y)$ denote the horizontal and vertical velocity of fluid, respectively.    Due to the symmetry, in the domain $x\in(-\infty,+\infty)$, we have the symmetry 
 \begin{equation}
 \zeta(x) = \zeta(-x), \;\;\;  u(x,z)  =  u(-x,z), \;\;\;  v(x,z)  = -v(-x,z),    \label{symmetry}
 \end{equation}
 which gives at the crest ($x=0$) that 
 \begin{equation}
 v(0,z) = -v(0,z), \;\;\; \mbox{i.e.} \;\; v(0,z) = 0.  \label{v:x=0}
 \end{equation}
Since the flow is irrotational in the interval $x\in(0,+\infty)$, 
the  corresponding  velocities $u(x,z)$ and $v(x,z)$  are  given by
 \[    u(x,z) = \frac{\partial \phi}{\partial x}, \;\;\;  v(x,z) = \frac{\partial \phi}{\partial z},  \hspace{1.0cm}    0 <  x <+\infty.       \]
 Note that, at crest ($x=0$), we have the boundary condition $v=0$ due to the symmetry, and the boundary condition  $u = U(z)$ defined by the limit (\ref{bc:left})  as $x\to 0$.   It should be emphasized that, in order not to lose the solutions that are discontinuous  at crest ($x=0$),  the boundary condition (\ref{bc:left}) is defined by a limit.   In this way,   the problem  is  described   more   generally  than the traditional theories.

 Note that, according to the symmetry (\ref{symmetry}) about crest (at $x=0$),  the wave elevation $\zeta(x)$ and the horizontal velocity $u$  are continuous at the vertical boundary $x=0$.   In fact,  due to the symmetry (\ref{symmetry}), the boundary condition (\ref{bc:left}) is equivalent to the continuous condition of the horizontal velocity $u$ at  $x=0$.  It is a common knowledge that the Laplace equation (\ref{geq:phi}) needs   only  {\em one} boundary condition at each boundary.   Therefore,  at  $x=0$,    the boundary condition (\ref{bc:left})  is  {\em sufficient} for the Laplace equation (\ref{geq:phi}) so that any other conditions for  the smoothness of the horizontal velocity $u$  (as $x\to 0$)   are {\em unnecessary}:    the higher-order derivatives of the horizontal velocity
 \[   \lim_{x\to 0}\frac{\partial^2 \phi}{\partial x^2}, \;\;  \lim_{x\to 0}\frac{\partial^3 \phi}{\partial x^3}, \;\;  \lim_{x\to 0}\frac{\partial^4 \phi}{\partial x^4}, \cdots, \]
are {\em unnecessary} to be continuous, since  the  boundary condition (\ref{bc:left})   is  mathematically  {\em enough}  for the Laplace equation (\ref{geq:phi}).   Thus,  any other boundary conditions (such as  that $\phi$ and $\zeta$  should  be  {\em infinitely} differentiable at $x=0$) may lead to the {\em loss}  of the  discontinuous  solutions and thus {\em must} be avoided.     In other words,   in the frame of the  UWM,  both of  $\phi$ and $\zeta$ are {\em unnecessary} to be  {\em infinitely} differentiable at crest ($x=0$).  This is mainly because the velocity potential $\phi$  satisfies Laplace equation  only in the domain $x>0$ and $x<0$ so that  $\phi$  is smooth  only in $x>0$ or $x<0$, but {\em not} necessarily analytic at $x=0$.  

The two nonlinear boundary conditions (\ref{bc:phi}) and (\ref{bc:zeta})  must be satisfied on the unknown free surface $z=\zeta(x)$.   This leads to the mathematical  difficulty to solve the nonlinear partial differential equations (PDEs).  In case of small wave-amplitude,  the linear boundary condition
\begin{equation}
\alpha^2 \frac{\partial^2 \phi}{\partial x^2} + \frac{\partial \phi}{\partial z}  =0, \hspace{1.0cm} \mbox{on $z = 0$},  \; 0<x<+\infty,  \label{bc:phi:linear}
\end{equation}
is a good approximation of  (\ref{bc:phi}), and
\begin{equation}
\zeta(x) = \left. \alpha \frac{\partial \phi}{\partial x}  \right|_{z=0}, \;\;\; 0 < x <+\infty,  \label{bc:zeta:linear}
\end{equation}
is a good approximation of (\ref{bc:zeta}), respectively. The above two linearized free-surface boundary conditions,  combined with the Laplace equation (\ref{geq:phi}),  the bottom condition (\ref{bc:bottom}) and either the periodic condition (\ref{bc:periodic}) or the decay condition (\ref{bc:decay}),   provide us the so-called linear  UWM.

\section{Smooth gravity waves given by the  UWM}

Based on the traditional linear  or fully nonlinear wave equations,  thousands of  articles  have been published for the smooth  periodic and solitary progressive  waves.  Mostly,    these traditional smooth progressive waves  are based on  the base functions
\begin{equation}
\cosh[n k (z+1)] \; \sin( n k x)  , \;\;\; n \geq 1, \label{base:traditional}
\end{equation}
for the velocity potential $\phi$,  which {\em automatically}  satisfy  the  Laplace  equation (\ref{geq:phi}), the bottom condition (\ref{bc:bottom}) and the periodic condition (\ref{bc:periodic}),  where  $k$ denotes the wave number and $n\geq 1$ is an integer.      For smooth periodic progressive waves with small wave-amplitude,   substituting the velocity potential
 \begin{equation}
 \phi(x,z) = \alpha  A_0   \cosh[k(z+1)]  \;  \sin(k x)  \label{def:phi:A}
 \end{equation}
 into  the  linear boundary condition  (\ref{bc:phi:linear}),  one has the dimensionless  phase speed
  \begin{equation}
 \alpha = \sqrt{\frac{\tanh(k)}{k}} \leq  1,  \label{property:PhaseSpeed:A}
 \end{equation}
say,  the phase speed of a smooth periodic progressive wave  in a finite water depth $D$ is always less then $\sqrt{g D}$.  Besides, substituting (\ref{def:phi:A}) into (\ref{bc:zeta:linear}) gives the wave elevation with small amplitude
\begin{equation}
\zeta = \frac{H_w}{2} \; \cos(k x),  \label{elevation:linear:A}
\end{equation}
where  $H_w =  2 A_0 \sinh (k)$.  The corresponding horizontal velocity  reads
 \begin{equation}
 u(x,z) 
 = \frac{H_w}{2\alpha \cosh(k)}  \cosh[k(z+1)]\cos(k x), \;\;
  \end{equation}
which gives
  \begin{equation}
  \frac{u}{U_0} = \frac{\cosh [k(z+1)] \; \cos (k x)}{\cosh(k)}, \;\;\;  0 < x <+\infty, 
  \end{equation}
  where $U_0 = H_w    /(2\alpha)$.    
   At  $x=0$,  we have the corresponding horizontal velocity
  \[   U(z) = \lim_{x\to 0} u(x,z) 
  = \frac{U_0}{ \cosh(k)}  \cosh[k(z+1)].  \]
Note that the above velocity potential  (\ref{def:phi:A}) and wave elevation (\ref{elevation:linear:A})   {\em automatically} satisfy the symmetry (\ref{symmetry}).     Besides, at $x=0$,  the horizontal velocity $u = U(z)$ is gained by the limit (\ref{bc:left}),  and the boundary condition  $v = 0$  is  {\em automatically}  satisfied.   In other words,  first solving  the linearized PDEs (\ref{geq:phi}) -- (\ref{bc:bounded}) in the interval $0  <  x < +\infty$ and then expanding the results to the interval $-\infty < x < 0$ by means of the symmetry (\ref{symmetry}),   we  can gain exactly the {\em same} results  by  the linear  UWM as the traditional ones, i.e. the Airy wave.   In this case,  the velocity potential $\phi$ and the wave elevation are smooth  {\em everywhere}.   All of these indicate that   
the linear  UWM is consistent with the traditional  theories for linear smooth periodic progressive waves.     

Note that, for given $x$,  the horizontal velocity $u$ and the kinetic energy of the smooth periodic progressive Airy's waves decrease  {\em exponentially} from free surface ($z=0$) to the bottom ($z=-1$).  Especially,  based on the base functions (\ref{base:traditional}), the elevation of the Airy's wave and the corresponding velocities are {\em infinitely} differentiable {\em everywhere},  although  the  UWM does not contain any smoothness conditions at all.

For the traditional periodic progressive surface waves with  large amplitude, the fully nonlinear wave equations must be considered.  As pointed out by  Cokelet \cite{Cokelet1977}, the phase speed $c$ of the  smooth  periodic progressive    waves depends not only on the water depth $D$ and the wave number $k$ but also on the wave height $H_w$:   the larger the wave amplitude, the faster the periodic wave propagates.  In other words,  the  smooth   periodic  progressive  waves  are  dispersive with wave height.     Like the linear Airy waves,  the nonlinear periodic progressive surface waves have a smooth crest with the exponentially decaying velocity $u(x,z)$ and kinetic  energy  from free surface to bottom.   Besides,  they are also infinitely differentiable {\em everywhere}, although the boundary conditions of  such kind of  smoothness do {\em not} exist at all for all traditional wave theories.   

It is very interesting that, using the  UWM, i.e. first  solving  the fully nonlinear wave equations  (\ref{geq:phi}) -- (\ref{bc:bounded}) in the interval $0 < x < +\infty$ and then expanding the results to the interval $-\infty <  x < 0$ by means of the symmetry (\ref{symmetry}),   we can gain exactly  the  {\em same}  results as  the  traditional  nonlinear  periodic  progressive  waves.   Besides,  the corresponding boundary condition  $v=0$  is satisfied  {\em automatically}  at crest ($x=0$).
 To shorten the length of this article,  we neglect the detailed mathematical derivation for them.   Here,  we only  emphasize that  the  UWM   indeed  admits  the  traditional  nonlinear  periodic progressive  waves  with smooth crest.      

  It should be emphasized that, in the frame of the traditional linear wave theories,  solitary waves have {\em never}  be reported, to the best  of the author's knowledge.  For details, please refer to Mei {\em et al.} \cite{MeiBook2005}.   Solitary waves for nonlinear and amplitude-dispersive long waves had been found  by Boussinesq \cite{Boussinesq1872} and  Rayleigh \cite{Rayleigh1876}.  For amplitude-dispersive long waves of permanent form,   the so-called KdV equation  \cite{KdV1895}  gives the periodic cnoidal  wave for a finite wavelength $\lambda$, which  tends  to   the solitary wave
\begin{equation}
\zeta(x) = H_w  \;  \mbox{sech}^2 \left[ \frac{\sqrt{3H_w}}{4}  \; \left(x - c \; t \right) \right]  \label{zeta:KdV}
\end{equation}
with the phase speed
\begin{equation}
c = \sqrt{ 1+ H_w}  \label{c:traditional:soliton}
\end{equation}
as $\lambda\to +\infty$.      
Both of  the above-mentioned solitary wave and the nonlinear periodic progressive waves are special cases  of  the so-called cnoidal  waves.   By means of perturbation methods and using the fully nonlinear wave equations,  Fenton \cite{Fenton1972JFM,Fenton1979JFM}  gave  a high-order cnoidal wave theory and a  ninth-order solution for the solitary wave in the form
\begin{equation}
\zeta(x) = \sum_{i=1}^{+\infty} \sum_{j=1}^{i} a_{i,j} \;  \epsilon^i \left[\mbox{sech}^2(\beta x) \right]^j,
\end{equation}
where $a_{i,j}, \epsilon, \beta$ are constants.  It should be emphasized that all of these traditional  cnoidal  and solitary waves have a  smooth crest:   $\zeta(x)$  is  {\em infinitely} differentiable  {\em everywhere}.  Besides,  the  horizontal velocity $u(x,z)$ and the kinetic energy always decay {\em exponentially} from free surface to bottom.  Furthermore,  the phase speed is dependent upon wave height, say,  the traditional  cnoidal  and solitary waves are dispersive with wave height.  Finally, to the best of author's knowledge,  all traditional solitary  surface  waves   have a  crest higher than the still water:  solitary waves in the form of depression have been reported only for interfacial waves, but never  for the  surface  waves.   

 It is very interesting  that, using the  UWM,   i.e.  first  solving  the PDEs (\ref{geq:phi}) --  (\ref{bc:bounded})  (or the KdV equation) in the domain $0 <  x < +\infty$ in a traditional way and then expanding these results to the domain $-\infty < x < 0$ by means of the symmetry (\ref{symmetry}), we can gain exactly the  {\em same} results as  the  traditional  cnoidal  and solitary waves with smooth crest!   Especially,  the boundary condition $v=0$  at crest ($x=0$)  and the  required smoothness in the whole domain are {\em automatically}  satisfied.    This is mainly  because the corresponding base functions such as  (\ref{base:traditional}) are smooth {\em everywhere},  and besides {\em automatically} satisfy the symmetry condition (\ref{symmetry}) and the boundary condition $v=0$ at crest.    To shorten the length of this article, the detailed mathematical derivations are neglected here.   All of these indicate that the  UWM  based on  the  symmetry  (\ref{symmetry}) and the fully nonlinear wave equations (\ref{geq:phi}) --  (\ref{bc:bounded})  is  indeed  {\em consistent} with {\em all} traditional  theories  for  the smooth progressive waves with permanent form.

Indeed,  the  traditional periodic and solitary progressive waves are smooth {\em everywhere}.   However,  this kind of smoothness is  {\em not}  absolutely necessary,  since all of these traditional smooth waves can be obtained in the frame of the traditional wave theories  {\em and}  the  UWM  {\em without} any smoothness conditions  at all!   In essence,  such kind of perfect smoothness of the wave elevation and velocities  come  {\em automatically}  from the base functions (\ref{base:traditional}),  which are infinitely differentiable  everywhere.

In summary,  the  unified  wave model  (UWM)   admits  {\em all}  traditional  smooth  periodic and solitary progressive waves with  permanent  form!     Thus, the  UWM  based on the symmetry and the fully nonlinear wave equations defined  in  $0<x<+\infty$ is {\em consistent} with the traditional  wave  theories  defined in the whole domain $-\infty < x < +\infty$.   This  shows  the  generality  of  the  UWM.  

\section{Peaked solitary waves given by the  UWM} 

In this section, we  illustrate  that  the   unified  wave model  (UWM) based on symmetry (\ref{symmetry}) and the fully nonlinear wave equations also admits the peaked solitary waves,  which include the famous peaked solitary wave (\ref{res:u:CH}) of the CH equation (\ref{geq:CH})  but  have many unusual characteristics quite different from the traditional smooth ones.  Thus,  the UWM {\em unifies} the traditional smooth waves and the peaked solitary ones,  for the first time, to the best of the author's knowledge.        

Mathematically, like the base functions (\ref{base:traditional}) that are widely used for the  smooth  periodic and solitary progressive waves, the following base functions
\begin{equation}
  \cos[n k (z+1)] \; \exp(- n k x), \;\; n\geq 1, \; k >0,\; 0 \leq  x < +\infty,   \label{base:new}
\end{equation}
also {\em automatically} satisfy the Laplace equation (\ref{geq:phi}) in the domain $0<x<+\infty$, the bottom condition (\ref{bc:bottom}) and the bounded condition (\ref{bc:bounded}), too.   However,  different from the smooth base functions (\ref{base:traditional}),  the above base function decays exponentially in the $x$ direction and satisfies the decay boundary condition (\ref{bc:decay}) at infinity.  Thus,  it is  more convenient to  strictly  express solitary waves that have truly only one crest.   
Note that,  unlike the  base function (\ref{base:traditional}), its derivatives  with respect to $x$  are {\em not} differentiable at crest ($x=0$).
Even so,  the base function (\ref{base:new}) with discontinuity of  the 1st-derivative at crest was  widely used as  the so-called evanescent (or non-propagating) mode \cite{McKee1988}  in the problems of {\em linear}  water  wave  diffraction-refraction by discontinuous bed undulations \cite{Massek-CE1983, Kirby-JFM1983, Kirby-JFM1986, Porter-JFM1995}, or {\em linear} waves propagating over a bed consisting of substantial variations in water depth \cite{Mattioli-1990, Mattioli-1991, Massel-CE1993}, and so on.   The solutions of these problems contain not only the propagating waves expressed by the base function (\ref{base:traditional}), but also the non-propagating (or evanescent) waves expressed by  the base function  (\ref{base:new}), which represent localized  effects and depend on the local bottom geometry \cite{Massel-CE1993}.   However, to the best of the author's knowledge,    the  evanescent  base  functions (\ref{base:new})   have  never   been applied to express  progressive  solitary  waves with  permanent form  propagating in a constant,  finite water depth.

Can we  find  any peaked solitary waves in the frame of the  UWM  based on the symmetry and  the fully nonlinear wave equations  (\ref{geq:phi}) --  (\ref{bc:bounded})  by means of the  evanescent  base function (\ref{base:new})?   The  answer  is  positive:   the   UWM  indeed  admits  a  kind  of solitary surface waves with peaked crest, which not only include the famous solitary wave (\ref{res:u:CH}) of the CH equation (\ref{geq:CH}) but also have many  unusual  characteristics  quite  different from the  traditional smooth ones, as shown below.   Therefore, the  UWM  indeed unifies  the smooth waves and the peaked solitary waves as a whole, for the first time, to the best of the author's knowledge.    

\subsection{Peaked solitary  waves by the linear  UWM \label{LinearTheory}}

As mentioned in \S~2,  both of the base functions (\ref{base:traditional}) and  (\ref{base:new}) {\em automatically}  satisfy  the  Laplace  equation (\ref{geq:phi}) in the domain $0<x<+\infty$, the bottom condition (\ref{bc:bottom}) and the bounded condition (\ref{bc:bounded}).   In the frame of the linear  UWM,  the former satisfies the periodic boundary condition (\ref{bc:periodic})  and   gives  the traditional    Airy  waves,  which are infinitely differentiable everywhere,  as mentioned in \S~3.   The latter satisfies the decay condition (\ref{bc:decay}) and gives peaked solitary waves, as shown below.

In the interval  $0 < x < +\infty$,  we have the velocity potential in the form
 \begin{equation}
 \phi^+(x,z) = \alpha  A   \cos[k(z+1)]  \;  e^{- k x},   \;\;\;\;   0 <  x <+\infty,  \label{def:phi:linear:0}
 \end{equation}
where the superscript $+$ denotes a solution in the interval $x\in(0,+\infty)$,  $\alpha$ denotes the (dimensionless) phase speed,  $A$ is a constant related to the wave height, and  $k>0$ is  a  parameter  related to the phase speed $\alpha$, respectively.    Note that the above expression {\em automatically}  satisfies the Laplace equation (\ref{geq:phi}) in the domain $0<x<+\infty$, the bottom condition (\ref{bc:bottom}), the decay condition (\ref{bc:decay}) and the bounded condition (\ref{bc:bounded}).   Substituting (\ref{def:phi:linear:0}) into the linear boundary condition   (\ref{bc:phi:linear}) gives
\[   \alpha  k  A  \left(  \alpha^2 k \cos k -\sin k \right)\exp(-k x)  = 0, \;\;\;  0  <  x < +\infty,  \]
 which leads to the important relationship 
\begin{equation}
\alpha^2 = \frac{\tan k}{k},  \hspace{1.0cm} n\pi  <  k  <  n \pi + \frac{\pi}{2},     \label{property:PhaseSpeed:B}
\end{equation}
where $n\geq 0$ is an integer.   Similarly, defining  $\phi^-$  the potential function in the interval $x\in(-\infty, 0)$,  we gain exactly the same result.    Obviously, we have  $\alpha \geq 1$, i.e. $ c \geq \sqrt{g D}$.  This is different from the periodic Airy wave whose phase speed has the property  $c \leq \sqrt{g D}$.

\begin{figure}
\centering
\includegraphics[scale=0.5]{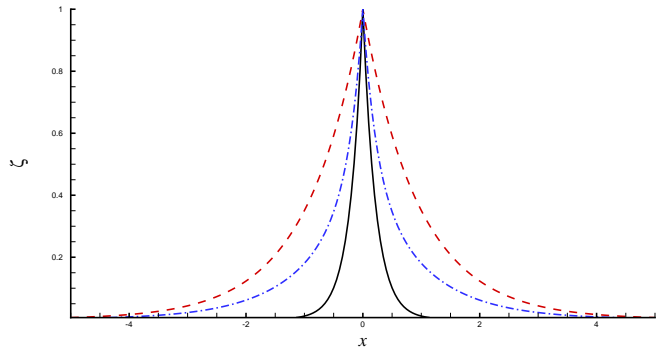}
\caption{$\zeta(x)/H_w$ in the case of  $k_0 = \pi/3, k_1 = 4.58117$ with $\alpha = 3^{3/4}/\sqrt{\pi}$.   Dashed line: $\exp({-k_0|x|})$; Solid line: $\exp(-k_1|x|)$;  Dash-dotted line:  $[\exp(-k_0 |x|) + \exp(-k_1|x|)]/2$.   }
\label{figure:zeta}
\end{figure}

 Given the dimensionless phase velocity $\alpha$, the  transcendental  equation (\ref{property:PhaseSpeed:B}) has an {\em infinite} number of solutions $k_n$, where
\begin{equation}
 n\pi   <  k_n  <  n \pi + \frac{\pi}{2}, \hspace{1.0cm} n \geq 0.
 \end{equation}
For the sake of simplicity, define the set
\begin{equation}
   {\bf K}_\alpha = \left\{ k_n:  \alpha^2 = \frac{\tan k_n}{k_n},  n\pi  <  k_n  <  n \pi + \frac{\pi}{2}, n=0,1,2,3,\cdots \right\}.   \label{def:set:K}
   \end{equation}
Write $k_\nu \in {\bf K}_\alpha$, where $\nu \geq 0$ is an integer.  Then, we have the solution
\begin{equation}
\phi^+(x,z) = \alpha  A_\nu \; \cos \left[ k_\nu (z+1)\right] \; \exp (-k_\nu x), \;\;\; 0 <  x < +\infty.
\end{equation}
According to the linearized boundary condition (\ref{bc:zeta:linear}), the corresponding elevation of the solitary wave reads
 \begin{eqnarray}
 \zeta^+(x) &=&   \left.  \alpha \frac{\partial \phi^+}{\partial x} \right|_{z=0} = -  \alpha^2 A_\nu  k_\nu  \cos k_\nu  \exp(- k_\nu x)\nonumber\\
 &=& - A_\nu  \sin k_\nu  \exp(- k_\nu x) , \;\;\; 0  <  x < +\infty,
 \end{eqnarray}
 where $\alpha^2 k_\nu \cos k_\nu =\sin k_\nu$ given by (\ref{property:PhaseSpeed:B}) is applied.  Due to the continuity of wave elevation,   we have   at $x=0$ the wave elevation 
 \begin{equation}
 \zeta(0) = \lim_{x\to 0} \zeta^+(x).  
 \end{equation}  
 Then,  according to (\ref{bc:Hw}),  we have the relation
 \begin{equation}
 H_w = - A_\nu  \sin k_\nu. \label{geq:A:0}
 \end{equation}
Thus, using the symmetry (\ref{symmetry}), we gain a  peaked solitary wave
  \begin{equation}
 \zeta(x)  = H_w  \;  e^{-k_\nu |x|}, \hspace{1.0cm}  -\infty < x < +\infty.      \label{zeta:linear:new}
 \end{equation}
This is a solitary wave that seriously has only one crest, but with a  {\em discontinuous}  1st-derivative $\zeta'(x)$ at crest!     For example, in the case of  $\alpha = 3^{3/4}/\sqrt{\pi}$, we have an {\em infinite} number of peaked solitary waves with  $k_0 = \pi/3, k_1 = 4.58117$, and so on: the corresponding elevations of the peaked solitary waves  are  as shown in Fig.~\ref{figure:zeta}, respectively.  Note that all of them have a peaked crest.  Besides,  the larger the value of $k_\nu$, the sharper the peaked solitary wave.  These are essentially  different from the traditional periodic and solitary progressive waves that are {\em infinitely} differentiable everywhere.    This clearly indicates the novelty of the peaked solitary waves given by the  UWM.  It should be emphasized that  (\ref{zeta:linear:new}) is exactly the same as the famous  peaked solitary wave (\ref{res:u:CH}) of the CH equation (\ref{geq:CH}) when $H_w = k_\nu = c$, with the definition  $x = X - c\; t$.  

\begin{figure}
\centering
\includegraphics[scale=0.5]{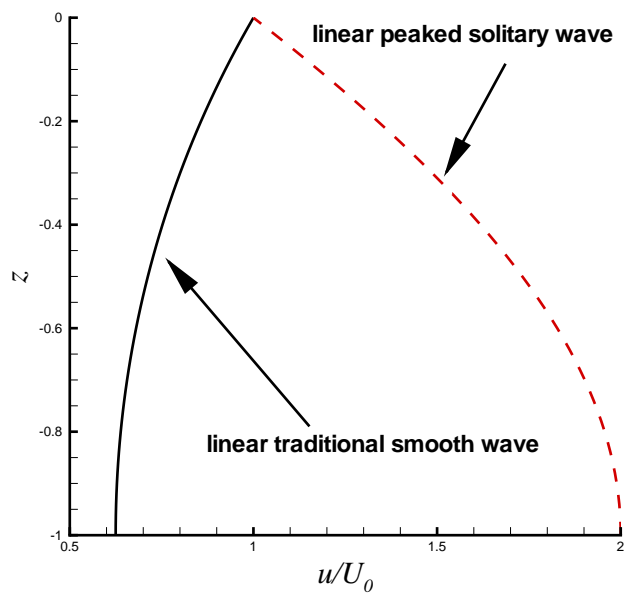}
\caption{  Velocity profile $u/U_0$ at $x=0$ in the case of  $k_0 = \pi/3$ with $U_0 = H_w/\alpha$.   Solid line: periodic Airy wave; Dashed line:  linear peaked solitary wave.   }
\label{figure:u}
\end{figure}

In the interval  $0 < x < +\infty$,  the corresponding horizontal velocity reads
 \begin{eqnarray}
 u^+(x,z) &=& \frac{\partial \phi^+}{\partial x} = \frac{\alpha k_\nu H_w   \cos[k_\nu (z+1)]  e^{-k_\nu x}}{\sin(k_\nu)}.  \label{u:peaked}
   \end{eqnarray}
 Using the symmetry (\ref{symmetry}),  we have
 \begin{eqnarray}
 u^-(x,z) &=&  u^+(-x,z) =  \frac{\alpha k_\nu H_w  \cos[k_\nu(z+1)]  e^{k_\nu x}}{\sin(k_\nu)} ,  \;\;\;  -\infty < x < 0. 
 \end{eqnarray}
At  $x=0$,  $u$  is  continuous  and  we  have  the corresponding  horizontal  velocity
 \begin{equation}
 U(z) = \lim_{x\to 0}  \frac{\partial \phi^+}{\partial x}  = \frac{\alpha k_\nu  H_w   \cos[k_\nu (z+1)] }{\sin(k_\nu)}.
 \end{equation}
 Thus, in the whole domain $-\infty < x < +\infty$,  we have a uniform expression
\begin{equation}
 \frac{u}{U_0} =   \frac{\cos[k_\nu(z+1)]  e^{-k_\nu |x|}}{\cos (k_\nu)}, \hspace{1.0cm}   x\in(-\infty, +\infty),    \label{u:linear:peaked}
 \end{equation}
where $U_0 = H_w   /\alpha$.   Thus, as long as the horizontal velocity at $x = 0$ is given by the above expression,  the  peaked solitary wave with the symmetry and the {\em permanent} surface form exists!   Note that,  for given $x$,  the horizontal velocity $u$ of the peaked solitary wave  {\em increases}  as $z$ varies from the free surface ($z=0$) to the bottom ($z=-1$): in other words, $u$ on the bottom is always larger than that on the free surface.    For example, when $k_0 = \pi/3$ (corresponding to the phase speed $c=3^{3/4}/\sqrt{\pi}$),  the horizontal velocity on the bottom  is always twice of that on the free surface, as shown in Fig. \ref{figure:u}.    This is quite different from the traditional smooth waves whose horizontal velocity $u$ on bottom is  always less than that on free surface.  This  indicates once again the novelty of the peaked solitary waves given by the  UWM.

Let $v^{+}(x,z)$ and $v^{-}(x,z)$ denote the vertical velocity in the interval $x> 0$ and $x <0$, respectively.   In the domain $0< x < +\infty$,  the corresponding vertical velocity reads
 \begin{eqnarray}
 v^+(x,z) &=& \frac{\partial \phi^+}{\partial z} = \frac{\alpha k_\nu H_w   \sin[k_\nu (z+1)]  e^{-k_\nu x}}{\sin k_\nu}.  \label{v:peaked:L}
   \end{eqnarray}
 Using the symmetry (\ref{symmetry}),  we have
 \begin{eqnarray}
 v^{-}(x,z) &=&  -v^+(-x,z) =  -\frac{\alpha k_\nu H_w  \sin[k_\nu (z+1)]  e^{k_\nu x}}{\sin k_\nu} ,  \;\;\;  -\infty < x < 0.  \label{v:peaked:R}
 \end{eqnarray}
As  $x\to 0$,  we have the limits 
\begin{equation}
\lim_{x\to 0}v^+= \frac{\alpha k_\nu H_w }{\sin k_\nu}   \sin[k_\nu (z+1)], \;\;\;  \lim_{x\to 0}v^- = -\frac{\alpha k_\nu H_w }{\sin k_\nu}   \sin[k_\nu (z+1)]
\end{equation}
so that
\begin{equation}
\lim_{x\to 0}v^+ = - \lim_{x\to 0} v^- \neq 0.
\end{equation}
However, according to the boundary condition (\ref{v:x=0}), the vertical velocity $v$ exactly equals to zero at $x = 0$, i.e. $v(0,z) = 0$.  Thus,  at the interface $x=0$,  the vertical velocity $v$ is {\em discontinuous}: $v$ changes sign as we cross the interface $x=0$.    Besides,  for given $z$,  the jump of $v$ at $x=0$, i.e.  \[ \lim_{x \to 0}(v^+ - v^-) = 2\lim_{x\to 0} v^+,\]     is  directly  proportional to $k_\nu$.   

Is such kind of discontinuity of the vertical velocity  acceptable in the frame of inviscid fluid?   
  It should be emphasized here that, in the frame of  the traditional  theories for smooth waves, there exist the similar  discontinuity and  jump  of velocity on an interface, which however are widely accepted by the scientific community.   For instance, on the  free  surface of  interfacial waves \cite{Grue-JFM1997, Choi1999,  Kataoka-JFM2008, Alam-JFM2012},  although the velocity normal to the interface is continuous,  the tangential velocity is discontinuous, as  pointed out by  Lamb \cite{Lamb1932} (\S 231, page 371): ``the tangential velocity changes sign as we cross the surface'',  but  ``in reality the discontinuity, if it could ever be originated, would be immediately abolished by viscosity''.    Similarly,  in the frame of the inviscid fluid,  the peaked solitary waves  should be acceptable, too.   We will discuss this problem later in details.  
  
\begin{figure}
\centering
\includegraphics[scale=0.35]{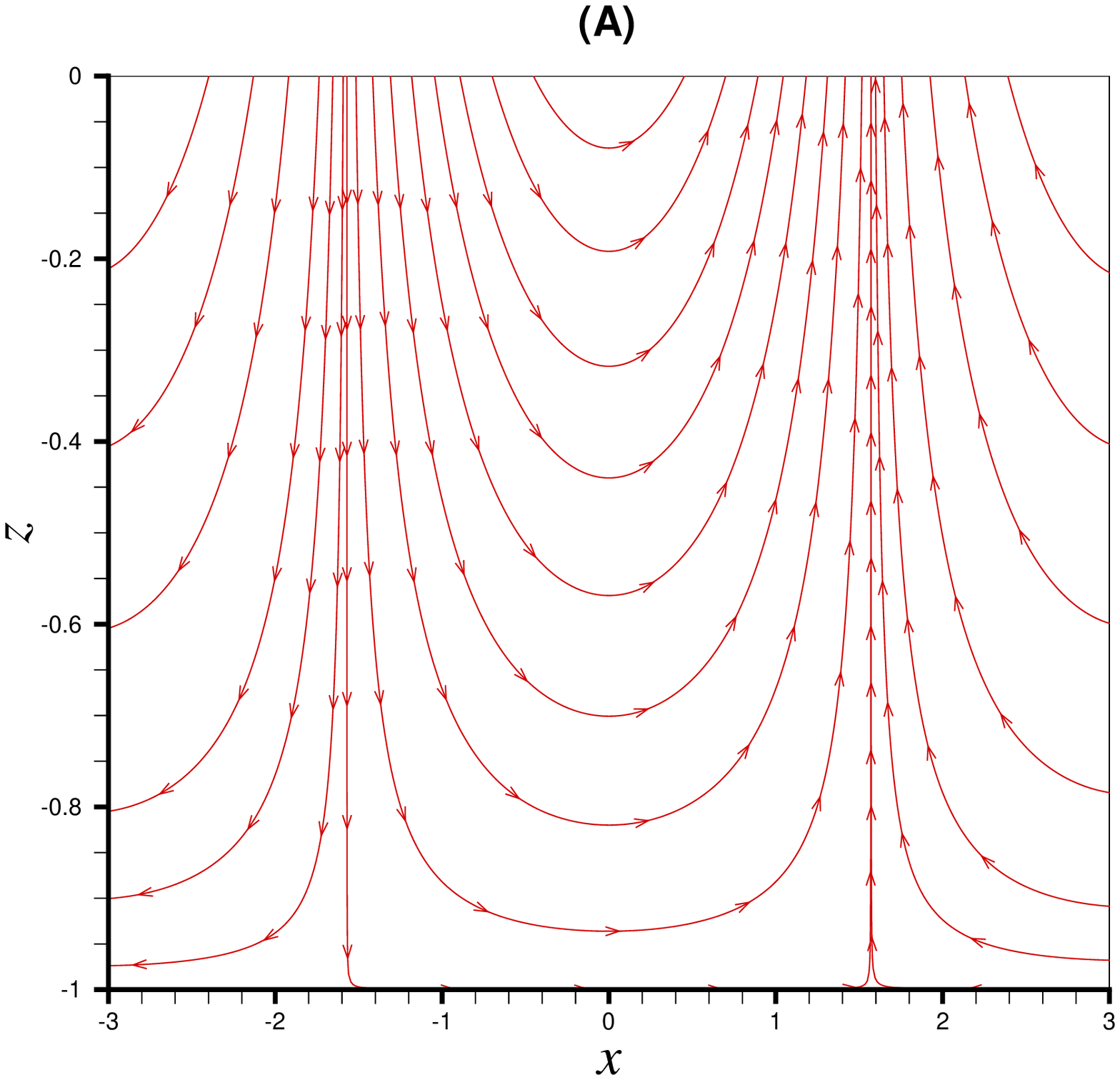} 
\includegraphics[scale=0.35]{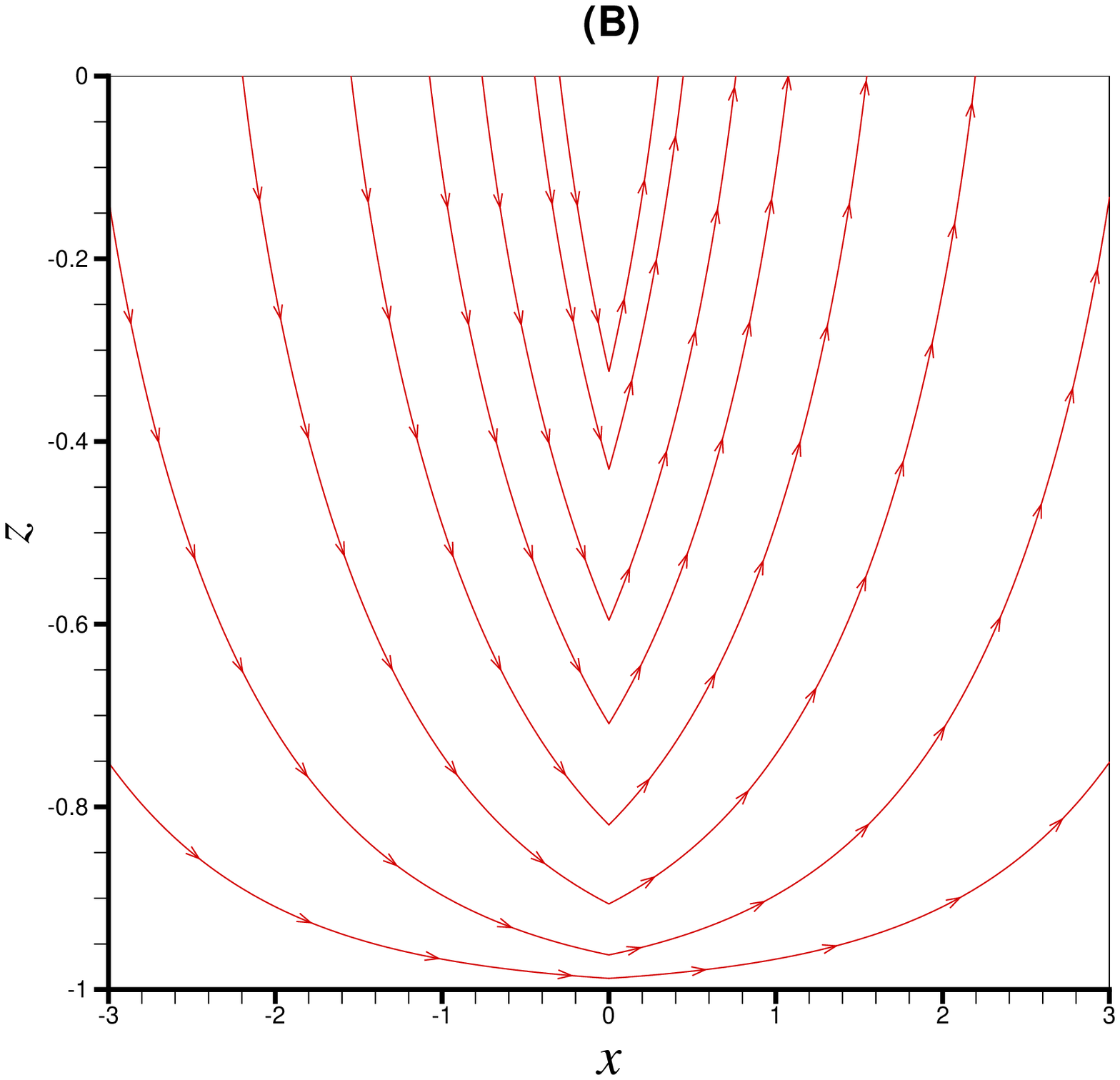} 
\caption{ Comparison of streamlines  given by the stream-function $\psi/(\alpha H_w)$ in case of $k = 1$.  (A):  Airy wave; (B): peaked solitary wave given by the linear UWM.    }
\label{figure:psi}
\end{figure}
  
  The stream function of the peaked solitary wave reads 
\begin{equation}
\psi(x,z) = \frac{\alpha H_w \sin[k_\nu(z+1)]e^{-k_\nu |x|}}{\sin(k_\nu)}, \;\;\;\;  |x|\in(0,+\infty).   \label{streamfunction:peaked}
\end{equation} 
For the linear Airy waves, we have the stream function 
\begin{equation}
\psi(x,z) = \frac{\alpha H_w \sinh[k(1+z)]\cos(k x)}{2 \sinh(k)},\;\;\;\; x\in(-\infty,+\infty).  
\end{equation}
The  streamlines for Airy wave are smooth {\em everywhere} and periodic in the horizontal direction, as shown in Fig.~\ref{figure:psi}.  However, the streamlines of the peaked solitary waves are neither smooth at crest ($x=0$)  nor  periodic in the horizontal direction.    Note that the velocity field of the traditional Airy waves is also smooth {\em everywhere}, but there exists the discontinuity of the velocity at crest $x=0$ for the peaked solitary waves.   This illustrates once again that the peaked solitary waves are indeed  quite  different from the traditional smooth ones.

Mathematically speaking, given a dimensionless phase velocity $\alpha$, there exist an {\em infinite} number of $k_\nu$ satisfying (\ref{property:PhaseSpeed:B}), and each of them corresponds to a solution of the linear   UWM.  This is quite different from the smooth periodic progressive Airy wave $\zeta(x) = A \cos (k x)$ that is {\em unique} for a given $\alpha$.  This is  because, given a dimensionless phase speed $\alpha$, the transcendental equation (\ref{property:PhaseSpeed:A}) has an {\em unique} solution, but the transcendental equation (\ref{property:PhaseSpeed:B}) has an {\em infinite} number of solutions!  Note that, unlike the smooth Airy wave theory,  the parameter $k_\nu$ of the peaked solitary wave  does not denote the wave number,  but the decay-rate of the wave elevation as $x\to +\infty$: the larger the value of $k_\nu$, more  quickly the wave elevation decays to zero.   When the wave height is fixed,  define the  decay length  
\begin{equation}
\lambda_a = 2|x_{B}|  \label{def:L:a}
\end{equation}
 of the peaked solitary waves,  where $x_B$ is determined by the  nonlinear  equation $\zeta(x_B) = 10^{-4}$.   Essentially,  for a fixed wave height,  we can regard the  decay  length $\lambda_a$  as  a  characteristic length of peaked solitary waves in the horizontal direction,   called the ``actual wavelength'',    although in theory wavelength of a solitary waves is infinite.  In this meaning,  for a fixed wave height,  the peaked solitary waves are dispersive with the actual wavelength $\lambda_a$.   Besides, according to (\ref{property:PhaseSpeed:B}),  the  peaked  solitary  wave can propagate very quickly even if  the wave height $H_w$ is small, since $\tan k_\nu/k_\nu \to +\infty$ as $k_\nu\to  \nu \pi +\pi/2$, where $k_\nu \in {\bf K}_\alpha$ and $\nu \geq 0$ is an integer.
 
The kinetic energy of the smooth progressive  Airy waves given by the linear   UWM  reads
 \begin{equation}
E_k = \frac{1}{2} \left( u^2 + v^2\right) = \left[\frac{\alpha H_w k}{4\sinh(k)} \right]^2 \; \left\{ \cosh[2k(1+z)]+\cos(2k x)\right\},
 \end{equation} 
 which is periodic in the horizontal direction and decreases exponentially as $z$ decreases from the free surface ($z= 0$) to the bottom ($z= -1$).   However,   according to (\ref{u:peaked}) and (\ref{v:peaked:L}) or (\ref{v:peaked:R}),  we have the kinetic energy of the peaked solitary waves given by the linear  UWM: 
 \begin{equation}
E_{k_\nu} =  \frac{1}{2} \left( u^2 + v^2\right) = \frac{(\alpha H_w k_\nu)^2}{2 \sin^2(k_\nu)} \exp(-2 k_\nu |x|),
 \end{equation}
 which is  {\em independent} of $z$ and decays exponentially as $|x|$ increases.   This is quite different from the traditional smooth Airy waves.    It  illustrates once again the novelty of the peaked solitary waves given by the  UWM.

Since the governing equation (\ref{geq:phi}), the two linearized free surface conditions (\ref{bc:phi:linear}) -- (\ref{bc:zeta:linear})  and all other conditions are linear, the solution of the peaked solitary wave can be expressed in a general form
\begin{equation}
\phi^+(x,z) = \alpha \sum_{n=0}^{+\infty} A_n \; \cos \left[ k_n (z+1)\right] \; \exp (-k_n x), \;\;\; 0 < x < +\infty.
\end{equation}
Using the linear boundary condition (\ref{bc:zeta:linear}), we have the corresponding elevation of the solitary wave
 \begin{eqnarray}
 \zeta^+(x) &=&   \left.  \alpha \frac{\partial \phi^+}{\partial x} \right|_{z=0} = -  \alpha^2 \sum_{n=0}^{+\infty} A_n  k_n  \cos k_n  \exp(- k_n x)\nonumber\\
 &=& -\sum_{n=0}^{+\infty} A_n  \sin k_n  \exp(- k_n x) \nonumber\\
 &=& \sum_{n=0}^{+\infty} b_n  \exp(- k_n x), \;\;\; 0  <  x < +\infty,  \label{elevation:linear}
 \end{eqnarray}
 where the relationship (\ref{property:PhaseSpeed:B}) is applied.  Using the symmetry (\ref{symmetry}), we have
\begin{eqnarray}
\zeta(x) &=&  \sum_{n=0}^{+\infty} b_n   \exp(- k_n |x|), \\
u(x,z)   &=& -\alpha \sum_{n=0}^{+\infty} A_n \; k_n \cos[k_n (z+1)] \exp(-k_n |x|), \\
v^+(x,z)  &=& -\alpha \sum_{n=0}^{+\infty} A_n \; k_n \sin[k_n (z+1)] \exp(-k_n x),\\
v^-(x,z)  &=&  \alpha \sum_{n=0}^{+\infty} A_n \; k_n \sin[k_n (z+1)] \exp( k_n x),
\end{eqnarray}
where the superscripts  $+$ and $-$ denote the interval $x \in (0,+\infty)$ and $x\in(-\infty,0)$, respectively.
Note that, using the restriction (\ref{bc:Hw}),  we have a linear algebraic equation
 \begin{equation}
 H_w = \sum_{n=0}^{+\infty} b_n, \label{geq:A:0}
 \end{equation}
 which has an {\em infinite} number of solutions of $b_n$ for a given value of $H_w$, where $n\geq 0$ is an integer.   In other words, from mathematical viewpoints, given a wave height $H_w$ and  a  phase speed $\alpha \geq 1$,  there exist an {\em infinite} number of peaked solitary waves. This is completely different from the smooth periodic progressive Airy wave that is {\em unique} for given phase speed and wave height.  This is because, given a phase speed $\alpha$, the transcendental equation (\ref{property:PhaseSpeed:A}) has an {\em unique} solution but  (\ref{property:PhaseSpeed:B}) has an {\em infinite} number of ones! Thus, different peaked solitary waves  might have exactly the same phase speed.  For example, in the case of  $k_0 = \pi/3, k_1 = 4.58117$ when $\alpha = 3^{3/4}/\sqrt{\pi}$, the peaked solitary wave
\[  \zeta(x) = \frac{1}{2} \left( e^{-k_0 |x|} + e^{-k_1 |x|} \right)   \]
is different from the peaked solitary waves  $\exp(-k_0 |x|)$ and $\exp(-k_1|x|)$ but has the same phase speed with them, as shown in Fig~\ref{figure:zeta}.
Note that, in the frame of the linear  UWM,  the crest of each peaked solitary wave may be at different position.  So, the peaked wave elevation can be generally expressed by
\begin{equation}
\zeta(x) = \sum_{n=0}^{+\infty} b_n \exp(-k_n |x-\xi_{0}|), \label{zeta:linear: unified }
\end{equation}
where $b_n$ and $\xi_n$ are real numbers.

Like traditional smooth Airy's wave,  since  the  elevation (\ref{zeta:linear:new}) of the peaked solitary wave  is  gained by the linear wave equations,  the value of $H_w$ can be negative, corresponding to a peaked solitary wave in the form of depression.  For example,  $\zeta(x)=-\exp(-|x|)/10$ is a  peaked solitary wave of depression.   Such kind of  peaked solitary waves have been  never  reported  for  surface  waves.   This once again indicates the novelty of the peaked  solitary waves given by the  UWM.
 
It is traditionally believed that solitary waves are always governed by nonlinear differential equations and exist in shallow water.  However,  we illustrate here  that solitary waves may exist even in the frame of the {\em linear} wave equations in water of {\em finite} depth.   It should be emphasized that the peaked solitary wave (\ref{zeta:linear:new}) includes   the famous peaked solitary wave (\ref{res:u:CH}) found by Casamma \& Holm \cite{Camassa1993PRL}.  This  might reveal  the origin  of the peaked solitary waves of the CH equation (\ref{geq:CH}), since the CH equation was derived from the fully nonlinear wave equations under some assumptions in shallow water.   Note that   the peaked solitary wave  (\ref{zeta:linear:new})  is valid not only in shallow water but also in water of finite  depth,  and thus  is  more general.   Logically speaking,  the peaked solitary wave  (\ref{zeta:linear:new})  is as acceptable as the peaked solitary waves (\ref{res:u:CH}) of the CH equation (\ref{geq:CH}), as shown later in more details.     

In summary,  the linear UWM  admits not only the traditional progressive waves with smooth crest, but also the peaked solitary waves that include the famous peaked solitary wave (\ref{res:u:CH}) given by the CH equation (\ref{geq:CH}) and besides  have many unusual characteristics.   Thus,  the UWM  unifies, for the first time,  the traditional smooth waves and the peaked solitary waves.  In  other  words,   the   linear  peaked   solitary   waves   are  consistent  with  the traditional smooth ones, and thus are as acceptable as the traditional ones  in the frame of inviscid  fluid.   It is found that  the linear peaked solitary waves have many unusual characteristics that are quite different from those of the traditional smooth ones.   First, its 1st-derivative of elevation is discontinuous at crest with a discontinuous vertical velocity.  Besides,   the phase speed of the peaked solitary waves has nothing to do with wave height: in fact,  it is dispersive (when wave height is fixed)  with the decay length, i.e. the so-called  ``actual wavelength''  $\lambda_a$.  In addition, unlike the traditional smooth waves whose kinetic energy is periodic in the horizontal direction and decreases exponentially as $z$ decreases from free surface to bottom,   the  kinetic energy of the linear peaked solitary waves decreases in the horizontal direction as $|x|$ increases, but  keeps the same  from free surface to bottom,  i.e.  it is {\em independent} of $z$ at all.   All of these reveal the novelty of the linear peaked solitary waves.  

In the following section, we  further check and  confirm  these unusual characteristics of the linear peaked solitary waves using the exact  UWM  based on the symmetry and the fully nonlinear wave equations.            

\subsection{Peaked solitary waves by the nonlinear  UWM \label{NonlinearTheory}}

As shown in \S~\ref{LinearTheory}, the peaked solitary surface waves given by the linear  UWM with the free surface conditions (\ref{bc:phi:linear}) and (\ref{bc:zeta:linear}) have some unusual  characteristics that are quite different from the  smooth  periodic and solitary ones.  Does the  exact UWM  based on the symmetry (\ref{symmetry}) and the fully nonlinear wave equations (\ref{geq:phi}) -- (\ref{bc:bounded}) indeed admit such kind of peaked  solitary waves?    Do this kind of peaked solitary waves have the same unusual characteristics as those reported in \S~\ref{LinearTheory}?

To answer these questions, we consider here the solitary surface  waves  with a {\em finite} wave height in a {\em finite} water depth so that the nonlinear terms of the free surface conditions (\ref{bc:phi}) and (\ref{bc:zeta}) are not negligible, and besides $z=0$ is not a good approximation of the free surface $\zeta(x)$.  In other words, we had to accurately  solve the fully nonlinear wave equations (\ref{geq:phi}) - (\ref{bc:bounded}) in the interval $0<x<+\infty$  with the symmetry (\ref{symmetry}). 

In this section, an analytic approximation technique for highly nonlinear differential equations proposed by  Liao, namely the homotopy analysis method (HAM)  \cite{ Liao1997NLM, Liao1999JFM,   LiaoBook2003,  Liao2003JFM, Liao2006SAM,  LiaoBook2012, KVBook2012}, is applied to solve the exact UWM based on the symmetry (\ref{symmetry}) and the fully nonlinear wave equations (\ref{geq:phi}) -- (\ref{bc:bounded}).  Unlike perturbation techniques,  the HAM  does not need any assumptions of small/large physical parameters, since it is based on the homotopy, a basic concept  in topology.  Besides, the HAM provides us great freedom to choose base functions  and equation-type for high-order approximation equations.   Especially,  by means of the so-called ``convergence-control parameter'' that has no physical meanings,  the HAM provides us a convenient way to guarantee the convergence of approximation series.   In essence, it is the so-called ``convergence-control parameter'' that differs the HAM from all other analytic approximation techniques, as pointed out by Liao \cite{LiaoBook2012}.   Therefore, the HAM is valid for highly nonlinear problems, as shown by lots of successful applications in fluid mechanics, applied mathematics, physics and finance.  For example,  by means of the HAM,  Liao \cite{Liao1999JFM} gained, for the first time,  convergent series solution of Blasius and Falker-Skan boundary-layer flows, which are uniformly valid in the whole field of flow (Note that the traditional  power series given by Blasius \cite{Blasius1908}  is valid only in the near field, and thus had to be matched with another asymptotic approximation of flow in far field).   Besides,  using the HAM as a tool,  the exact Navier-Stokes equations were solved by Turkyilmazoglu \cite{Turkyilmazoglu2009PF} for  a compressible boundary layer flow due to a porous rotating disk, and by  Xu {\em et al.} \cite{Xu2010PF} for   viscous flows  in a porous channel with orthogonally moving walls.   Furthermore,   the limit cycle of  Duffing - van der Pol equation was solved by Turkyilmazoglu \cite{Turkyilmazoglu2011ASME}, and the two coupled Van der Pol equations were solved by Li  {\em et al.} \cite{Li2010JMP}.
Especially,  by means of the HAM,  some new boundary layer flows have  been  found by Liao \cite{Liao2005-IJHMT} and by  Liao \& Magyari \cite{Liao-Magyari2006}, which have been neglected by other analytic and even numerical techniques.   In addition, the HAM has been successfully applied to solve some nonlinear PDEs with moving boundary conditions,  such as those about the famous problem of the American put option in finance.   For example,  Zhu \cite{SPZhu2006B}  successfully applied the HAM to give a series approximation of the American put option and gained the optimal exercise boundary valid for a couple of years, while perturbative and/or asymptotic formulas are accurate only in a few days or weeks.  All of these illustrate the potential and validity of the HAM for highly nonlinear problems.

It should be emphasized that the HAM has been  successfully applied to solve the fully nonlinear wave equations, too.  Using the traditional base functions (\ref{base:traditional}),  Liao \& Cheung \cite{Liao2003JEM} applied the HAM to solve the smooth periodic progressive surface waves in deep water  and  obtained  convergent  solutions for waves with large amplitude even close to the limiting case.   Their  analytic results agree quite well  with  those given by  Schwartz \cite{Schwartz1974JFM} and  Longuet-Higgins \cite{Longuet-Higgins1975JFM}.     Besides,   using the same base functions (\ref{base:traditional}),
Tao {\em et al.} \cite{Tao2007CE} successfully applied the HAM  to  solve the fully nonlinear wave equations   for smooth periodic progressive waves in finite water depth, and their analytic results agree well  not only with the analytic ones given by  Cokelet \cite{Cokelet1977} and   Fenton \cite{Fenton1981JFM}  but  also  with the experimental ones reported by  Mehaute {\em et al.} \cite{Mehaute1968}.  Currently, Xu {\em et al.} \cite{Xudali-2012JFM} successfully applied the HAM to investigate the steady-state fully-resonant progressive waves in finite water depth, and found, for the first time,  the multiple steady-state  resonant waves whose wave spectrums are independent of time, i.e.  without exchange of wave energy.   All of these  demonstrate  the validity of the HAM  for  the  fully  nonlinear wave equations (\ref{geq:phi}) - (\ref{bc:bounded}).

\subsubsection{Analytic approach based on the HAM}

In short,  the fully nonlinear wave equations (\ref{geq:phi}) --  (\ref{bc:bounded}) in the domain $0 < x < +\infty$ can be solved by means of the HAM and the  evanescent  base functions (\ref{base:new})  in a rather similar way as  those  by Liao \& Cheung \cite{Liao2003JEM} and  Tao {\em et al.} \cite{Tao2007CE}, although   they   used  the traditional base functions (\ref{base:traditional}) with perfect smoothness.  Then, we expand these results to the whole domain $x\in(-\infty, +\infty)$  using  the symmetry (\ref{symmetry}).  In this way,  we can gain the peaked solitary waves in the whole domain $x\in(-\infty,+\infty)$ by means of the UWM.     

  Due to the symmetry (\ref{symmetry}),  we  need consider  the case  $x  >  0$ only.  As mentioned in \S~\ref{LinearTheory} for the linearized  UWM, given a phase speed $\alpha$, there exists an infinite number of peaked solitary waves $\zeta(x)= H_w \exp(-k_\nu x)$ with different decay-parameter $k_\nu$ satisfying the transcendental equation  (\ref{property:PhaseSpeed:B}), where $k_\nu \geq 0$ is an integer.  So, the general expression of the peaked solitary waves becomes rather complicated when the nonlinear boundary conditions are considered.  Thus, for the sake of simplicity, we consider here only the peaked solitary waves  with the primary decay-parameter $0 \leq k_0 \leq \pi/2$.   From now on, let $k$ denote the primary decay-parameter  $k_0$, if not mentioned.  Obviously, due to the nonlinearity of the two free surface conditions, the peaked solitary wave elevation should contain the terms $\exp(-n k x)$, and correspondingly the velocity potential function
 $\phi(x,z)$ should be expressed in the form
\begin{eqnarray}
\phi(x,z) &=& \sum_{n=1}^{+\infty} a_{n} \; \cos [n k ( z+1)] \; \exp(- n k  x),  \;\;\;  x > 0, \; k > 0,   \label{solution-expression:phi}
\end{eqnarray}
which automatically satisfies the Lapalce equation (\ref{geq:phi}) in the domain $0<x<+\infty$, the bottom condition (\ref{bc:bottom}) and the bounded condition (\ref{bc:bounded}), where $k = k_0 \in {\bf K}_\alpha$ defined by (\ref{def:set:K}) is a primary decay-parameter, and $a_{n}$ is a constant  to be determined.   We search for the solitary surface  waves  in the form
\begin{equation}
\zeta(x) =  \sum_{n=1}^{+\infty} b_{n} \exp\left(- n k  x \right),  \hspace{1.0cm}   x >  0, \; k=k_0  \in {\bf K}_\alpha, \label{solution-expression:zeta}
\end{equation}
where  $b_{n}$ is a constant coefficient to be determined.  The above expressions (\ref{solution-expression:phi}) and (\ref{solution-expression:zeta}) provide us the so-called solution-expression of $\phi(x,z)$ and $\zeta(x)$, respectively, which play important role in the frame of the HAM, as shown below.

Let $\phi_0(x,z), \zeta_0(x)$ denote the initial guess of the velocity potential  $\phi(x,z)$ and the wave elevation $\zeta(x)$ in the interval $x \in (0, +\infty)$,  respectively.   To  apply the HAM,  we should first of all construct  two continuous variations  from the initial guess $\phi_0(x,z),  \zeta_0(x)$  to the exact solution  $\phi(x,z), \zeta(x)$, respectively.   This can be easily done by means of the homotopy, a basic concept in topology, as shown below.

First,  according to the solution expression (\ref{solution-expression:phi}),   we choose
 \begin{equation}
 \phi_0(x,z) =  A_{0} \cos[k (z+1)]  \;  e^{-k  x}, \;\;\; x  >  0,  \; k  > 0,  \label{def:phi[0]}
 \end{equation}
 as the initial guess of the velocity potential $\phi(x,z)$,  where $A_{0}$ is a constant to be determined.   Note that, different from  Liao \& Cheung \cite{Liao2003JEM} and  Tao {\em et al.} \cite{Tao2007CE},  the evanescent-mode  base-function (\ref{base:new}) is used here.     Note also that  $\phi_0(x,z)$ automatically satisfies the Laplace equation (\ref{geq:phi}) in the interval $0<x<+\infty$, the bottom condition (\ref{bc:bottom}) and the bounded condition (\ref{bc:bounded}).    Besides,   following  Liao \& Cheung \cite{Liao2003JEM} and  Tao {\em et al.} \cite{Tao2007CE},  we choose
 \begin{equation}
 \zeta_0(x) = 0
 \end{equation}
as the initial guess of wave elevation  $\zeta(x)$.

Secondly,  according to (\ref{bc:phi}), we define a nonlinear operator
\begin{equation}
{\cal N}  \phi = \alpha^2 \frac{\partial^2 \phi}{\partial  x^2} +  \frac{\partial \phi}{\partial z} -\alpha\frac{\partial}{\partial x} \left(  \nabla \phi  \cdot \nabla \phi \right) +\nabla \phi \cdot \nabla \left( \frac{1}{2} \nabla \phi  \cdot \nabla \phi \right). \label{def:N}
\end{equation}
Let $q\in[0,1]$ denote an embedding parameter,  $c_\phi$ and $c_\eta$ be two non-zero auxiliary parameters without physical meanings,  called the convergence-control parameters,   and $\cal L$ denote an auxiliary linear operator, respectively.   Following  Liao \& Cheung  \cite{Liao2003JEM} and Tao {\em et al.} \cite{Tao2007CE},  we construct  the so-called zeroth-order deformation equation
 \begin{equation}
\nabla^2\Phi(x,z;q)=0,  \hspace{1.0cm} 0 < x <+\infty, \;\;  z \leq \eta(x;q),  \label{geq:phi:zero}
\end{equation}
subject to the  boundary conditions on the unknown free surface $z=\eta(x;q)$:
\begin{equation}
(1-q) {\cal L} \left[ \Phi(x,z;q) - \phi_0(x,z)\right] = c_\phi \; q \;   {\cal N}\left[ \Phi(x,z;q)\right], \;\;\; 0<x <+\infty,  \label{bc:zero:phi}
\end{equation}
\begin{equation}
(1-q) \eta(x;q)=c_\eta \; q \; \left[ \eta(x;q)  - \alpha \frac{\partial \Phi}{\partial x} + \frac{1}{2} \nabla \Phi  \cdot \nabla \Phi  \right], \;\;\; 0<x <+\infty,
\label{bc:zero:zeta}
\end{equation}
 and the boundary condition at the bottom
 \begin{equation}
 \frac{\partial \Phi}{\partial z} = 0, \hspace{1.0cm} z = -1, \;\; 0<x<+\infty. \label{bc:zero:bottom}
 \end{equation}
If  the wave height $H_w$ is given,  there exists  the additional  condition:
 \begin{equation}
\lim_{x\to 0}\eta(x;q) = H_w.  \label{bc:zero:Hw}
 \end{equation}
 Note that $\Phi(x,z;q)$ and $\eta(x;q)$ depend  not only on the original physical variables  $x,z$  but also  on  the embedding parameter  $q\in[0,1]$ and the two convergence-control parameter $c_\phi, c_\eta$,  though  $q, c_\phi$ and $c_\eta$ have no physical meanings at all.  It should be emphasized that  we have great freedom to choose the values of the convergence-control parameters $c_\phi$ and $c_\eta$.    Following  Liao \& Cheung  \cite{Liao2003JEM} and  Tao {\em et al.} \cite{Tao2007CE},  we  choose
the auxiliary linear operator
 \begin{equation}
 {\cal L}\phi =  \alpha^2 \frac{\partial^2 \phi}{\partial x^2} +   \frac{\partial \phi}{\partial z},   \label{def:L}
 \end{equation}
 which has the property  ${\cal L}[0] = 0$.  Note that ${\cal L} $  is exactly the linear part of the nonlinear operator $\cal N$ defined by (\ref{def:N}).    In this way,   the zeroth-order deformation equations (\ref{geq:phi:zero}) -- (\ref{bc:zero:Hw}) are well defined.

 When $q=0$,  we have from (\ref{bc:zero:zeta}) that
 \begin{equation}
 \eta(x;0) = 0= \zeta_0(x),  \label{eta:q=0}
 \end{equation}
 and then the corresponding  zeroth-order deformation equations become
  \begin{equation}
\nabla^2\Phi(x,z;0)=0,  \hspace{1.0cm} z \leq  0,  \;\;  0  <  x < +\infty,   \label{geq:phi:q=0}
\end{equation}
subject to the  boundary conditions on the known free surface
\begin{equation}
{\cal L} \left[ \Phi(x,z;0) - \phi_0(x,z)\right] = 0,  \hspace{1.0cm} \mbox{when $z=0$}, \;\; 0 <  x < +\infty,  \label{bc:zero:phi:q=0}
\end{equation}
 and the boundary condition at the bottom
 \begin{equation}
 \frac{\partial \Phi(x,z;0)}{\partial z} = 0, \hspace{1.0cm} z = -1, \;\; 0 <  x < +\infty. \label{bc:zero:bottom:q=0}
 \end{equation}
Since the auxiliary linear operator $\cal L$ has the property ${\cal L}[0]=0$ and besides the initial guess $\phi_0(x,z)$  defined by (\ref{def:phi[0]})  satisfies the Laplace equation (\ref{geq:phi}) and the bottom condition (\ref{bc:bottom}),  it is straightforward  that
\begin{equation}
\Phi(x,z;0) = \phi_0(x,z).   \label{Phi:q=0}
\end{equation}
 When $q=1$,   since $c_\phi \neq 0 $ and $c_\eta \neq 0$,   the zeroth-order  deformation equations (\ref{geq:phi:zero}) -- (\ref{bc:zero:Hw})  are equivalent to the original, fully nonlinear wave equations (\ref{geq:phi}) -- (\ref{bc:Hw}), respectively, so that we have  the relationship
 \begin{equation}
 \Phi(x,z;1) =\phi(x,z), \;\;\; \eta(x;1) =\zeta(x).
 \end{equation}
 Thus, as the embedding parameter  $q$  increases from 0 to 1,  $\Phi(x,z;q)$ and $\eta(x;q)$  indeed vary continuously from the initial guess $\phi_0(x,z), \zeta_0(x)$ to the exact solution $\phi(x,z), \zeta(x)$ of the fully nonlinear wave equations   (\ref{geq:phi}) -- (\ref{bc:Hw}), respectively.    Therefore, the zeroth-order deformation equations  (\ref{geq:phi:zero}) -- (\ref{bc:zero:Hw})  truly  construct such a kind of continuous variation that provides a base of our analytic approach, as shown below.

Since both of  $\Phi(x,z;q)$ and $\eta(x;q)$  are dependent upon the embedding parameter $q\in[0,1]$, we can expand them  in Maclaurin series with respect to $q$   to gain  the so-called  homotopy-Maclaurin series
 \begin{eqnarray}
 \Phi(x,z;q) &=&  \phi_0(x,z) +\sum_{m=1}^{+\infty} \phi_m(x,z) \; q^m,  \label{series:phi:q}\\
 \eta(x;q) &=& \sum_{m=1}^{+\infty} \zeta_m(x) \; q^m, \label{series:zeta:q}
 \end{eqnarray}
 where
 \[     \phi_m(x,z) = \frac{1}{m!} \left.  \frac{\partial^m \Phi(x,z;q)}{\partial q^m}\right|_{q=0},\;\;\;   \zeta_m(x) = \frac{1}{m!} \left.  \frac{\partial^m \eta(x;q)}{\partial q^m}\right|_{q=0}    \]
 and the relationship  (\ref{eta:q=0}) and (\ref{Phi:q=0})  are used.   However, it is well known that a Maclaurin series often has a finite radius of convergence.  Fortunately,  both of $\Phi(x,z;q)$ and $\eta(x;q)$ contain the two convergence-control parameters $c_\phi$ and $c_\eta$, which have great influence on the convergence of the Maclaurin series of $\Phi(x,z;q)$ and $\eta(x;q)$, as shown by  Liao \& Cheung \cite{Liao2003JEM} and  Tao {\em et al.} \cite{Tao2007CE}.   Here, it should be emphasized once again  that we have great freedom to choose the values of $c_\phi$ and $c_\eta$.  Thus,   if  the convergence-control parameters $c_\phi$ and $c_\eta$ are properly chosen so that the above homotopy-Maclaurin series are convergent at $q=1$, we have the homotopy-series solution
   \begin{eqnarray}
 \phi(x,z) &=&  \phi_0(x,z) +\sum_{m=1}^{+\infty} \phi_m(x,z) ,  \label{series:phi}  \\
 \zeta(x) &=& \sum_{m=1}^{+\infty} \zeta_m(x) .  \label{series:zeta}
 \end{eqnarray}

 The equations for the unknown $\phi_m(x,z)$ and $\zeta_m(x)$ can be derived directly from the zeroth-order deformation equations.  Like Liao \& Cheung \cite{Liao2003JEM} and Tao {\em et al.} \cite{Tao2007CE},  substituting the series  (\ref{series:phi:q}) and (\ref{series:zeta:q}) into the zeroth-order deformation equations (\ref{geq:phi:zero}) --  (\ref{bc:zero:Hw}), then equating the like-power of $q$,  we gain
 \begin{equation}
 \zeta_m( x) =   \left. \left\{ c_\eta  \; \Delta_{m-1}^\eta + \chi_m \; \zeta_{m-1} \right\}\right|_{z=0} ,  \;\;\;  m \geq 1, \;\;  0 < x < +\infty,  \label{eta:mth}
 \end{equation}
  where
 \begin{equation}
    \Delta_{m}^\eta = \zeta_{m}  - \alpha \; \bar\phi_{m,1}  + \Gamma_{m,0},    \label{def:Delta-eta}
\end{equation}
  and the $m$th-order deformation equation
  \begin{equation}
\nabla^2\phi_m(x,z)=0, \;\;\;  m\geq 1, \; z \leq  0,  \;\;  0 <  x < +\infty,   \label{geq:phi:mth}
\end{equation}
subject to the boundary condition on the known free surface $z=0$:
\begin{equation}
\bar{\cal L} \left( \phi_m \right) =  \left.  \left( \alpha^2 \;  \frac{\partial^2 \phi_m}{\partial x^2} +   \frac{\partial \phi_m}{\partial z} \right)\right|_{z=0}  =  R_m(x), \;\;  0 <  x < +\infty,   \label{bc:mth:phi}
\end{equation}
and the bottom condition
\begin{equation}
\frac{\partial \phi_m}{\partial z} = 0, \hspace{1.0cm} z = -1, \;\;  0  <  x < +\infty,   \label{bc:mth:bottom}
\end{equation}
where
\begin{eqnarray}
R_m(x) &=& \left. \left\{ c_\phi \; \Delta^\phi_{m-1} +\chi_m\; S_{m-1} -\bar{S}_m \right\}\right|_{z=0},  \;\;  0<  x < +\infty,\\
\chi_n  & = & \left\{
\begin{array}{cc}
0, & \mbox{when $n\leq 1$}, \\
1, & \mbox{when $n > 1$} .
\end{array} \right.  \label{def:chi}
\end{eqnarray}
The detailed derivations of  $\Delta_{m-1}^\eta,  \Delta^\phi_{m-1}, S_{m-1}$, $\bar{S}_m$ with all related formulas are  given  explicitly  in the Appendix.   These formulas are essentially  the same  as  those  for the smooth periodic  progressive  waves  used  by  Liao \& Cheung \cite{Liao2003JEM} and  Tao {\em et al.} \cite{Tao2007CE},  although we  explicitly  give  all  formulas  in  details  so  that  high-order approximations can be gained more  efficiently.  

Note that  the dimensionless  phase  speed $\alpha$  of  the peaked solitary  waves  is  unknown up to now.     According to the linear  UWM  described in \S~\ref{LinearTheory},  the peaked  solitary waves exist only when
 \begin{equation}
 \alpha^2 = \frac{\tan k}{k},  \hspace{1.0cm}  n\pi <  k < n \pi + \frac{\pi}{2},
 \end{equation}
 where $n\geq 0$ is an integer.  Note that here we consider only the primary decay-parameter $0 < k_0 <\pi/2$ with the definition  $k = k_0$.   If  the above expression also holds for the fully nonlinear wave equations of the  UWM,   the auxiliary linear operator defined by (\ref{def:L}) has the property
 \begin{equation}
 {\cal L}\left\{ \cos[k(z+1)] e^{-k x} \right\} = 0,\;\;\;  x\geq 0,  \; k = k_0 >0,   \label{L:property}
 \end{equation}
and the corresponding  inverse operator  of  $\bar{\cal L}$ defined by (\ref{bc:mth:phi})  has the  property
 \begin{equation}
   \bar{\cal L}^{-1}\left\{ \exp\left(-n k x \right) \right\} = \frac{\cos[n k (z+1)] \exp(-n k x)}{( n k )\left[ \alpha^2 (n k) \cos(n k)-\sin(n k)\right]} , \;\;\;   k>0, \;  n \neq 1, \; x >  0,
   \label{L:inverse:property}
 \end{equation}
 where $n\geq 2$ is an integer.     Note that the above expression does not hold when $n=1$.    Fortunately,  it is found that
   $R_m(x)$ indeed does not contain the term $\exp(-k x)$ as long as the  phase speed  is  given by   $\alpha^2=\tan (k)/k$, where $0< k <\pi/2$.    Mathematically, this is  because
   the nonlinear terms of  (\ref{def:N}) do not contain the term $\exp(-k x)$ at all,  since
   \[   \exp(-m k x) \times \exp(-n k x) = e^{-(m+n) k x} \]
with $m+n\geq 2$ for any integers $m\geq 1$ and $n \geq 1$.   So do the linear terms of the nonlinear operator (\ref{def:N}), since
\begin{eqnarray}
&& \left.\left\{ \left( \alpha^2 \frac{\partial^2 }{\partial  x^2} +  \frac{\partial }{\partial z} \right) \sum_{n=1}^{+\infty} b_{n} \; \cos [nk( z+1)] \; \exp(- n k x) \right\}\right|_{z=0}\nonumber\\
&=& \sum_{n=1}^{+\infty}  (n k) \left[ \alpha^2 (n k) \cos (nk) -\sin(n k)  \right] b_{n}  \exp(- n k x)  \nonumber\\
&=& \sum_{n=2}^{+\infty}  (n k) \left[ \alpha^2 (n k) \cos (nk) -\sin(n k)  \right] b_{n}  \exp(- n k x)  \nonumber
\end{eqnarray}
does not contain the term $\exp(-k x)$, too.   This is the essential reason why  the phase speed
\begin{equation}
\alpha = \sqrt{\frac{\tan (k)}{k}}   \label{def:alpha:nonlinear}
\end{equation}
given by the linear  UWM  {\em still} holds for the fully nonlinear wave equations (\ref{geq:phi}) -- (\ref{bc:Hw}) of the exact  UWM!     Thus,  unlike the traditional smooth periodic and solitary progressive waves that are dispersive with wave height,  the phase speed of the peaked solitary waves has nothing to do with the wave height:  it is dispersive with the so-called ``actual wavelength'' $\lambda_a$ when wave height is fixed.   More  discussions  about this point will be given in \S~5.

Keeping (\ref{def:alpha:nonlinear}) in mind and using the property (\ref{L:inverse:property}) of the inverse operator $\bar{\cal L}^{-1}$,  it is straightforward to gain the common solution of the high-order deformation equation (\ref{geq:phi:mth}) to (\ref{bc:mth:bottom}):
  \begin{equation}
  \phi_m(x,z) =  \phi_m^*(x,z)+ A_m  \cos\left[ k\left(1+z\right)  \right]  e^{-k x},  \;\;\;  \; x >  0,
  \end{equation}
  where $\phi_m^*(x,z)= \bar{\cal L}^{-1} [R_m(x)]$ is a special solution,  and the coefficient  $A_m$ is determined by
   the given  wave height
  \begin{equation}
  \sum_{n=1}^{m+1}\lim_{x\to 0}\zeta_{n}(x)  = H_w.
  \end{equation}
  This is mainly because,
according to (\ref{eta:mth}),  $\zeta_{m+1}(x)$  is dependent upon $\phi_{m}(x,z)$ that contains the unknown parameter $A_{m}$ for $m\geq 1$.  Note that, according to (\ref{L:inverse:property}), $\phi_m(x,z)$ is in the form of (\ref{solution-expression:phi}) and  thus  automatically  satisfies the Laplace equation
(\ref{geq:phi}) in the domain $0<x<+\infty$, the bottom condition (\ref{bc:bottom}) and the bounded condition (\ref{bc:bounded}).   Thus, using the explicit formulas given in the Appendix,  it is computationally efficient to gain high-order analytic approximations successively, especially by means of the computer algebra system such as Mathematica and Maple, since our analytic approach needs only algebra computations.

 For example,  using the initial guess (\ref{def:phi[0]}) and (\ref{eta:mth}), we directly have
 \begin{eqnarray}
\zeta_1(x) &=& -c_\eta \; \left. \left( \alpha \frac{\partial \phi_0}{\partial x} -\frac{1}{2} \nabla \phi_0  \cdot \nabla \phi_0 \right) \right|_{z=0}\nonumber\\
&=& c_\eta A_0 k \left[ \alpha\cos(k) e^{-k x} +\frac{A_0 k}{2} e^{-2 k x}\right], \;\;\;  x > 0.
\end{eqnarray}
Thus,  at the first-order of approximation,  we have an algebraic  equation for the given  wave height
\[   H_w = c_\eta k A_0 \left( \alpha \cos k +\frac{1}{2} k A_0   \right),      \]
which  gives  two  different solutions
\begin{equation}
 A_0 = k^{-1}\left[  -\alpha \cos k \pm \sqrt{\alpha^2 \cos^2 (k) + 2 H_w/c_\eta}\right].
\end{equation}
We simply choice
\begin{equation}
 A_0 =- k^{-1}\left[ \alpha \cos k - \sqrt{\alpha^2 \cos^2 (k) + 2 H_w/c_\eta}\right]
\end{equation}
to calculate $A_0$ for a given $H_w$,  since it has a smaller absolute value.

Furthermore,   using the initial guess (\ref{def:phi[0]}),  we have
 \[    \Delta_0^\phi = k A_0 \left(\alpha^2 k \cos k -\sin k \right) e^{-k x}+2\alpha  k^3  A_0^2 e^{-2kx} + k^4 A_0^3 \cos (k) e^{-3kx}.  \]
Using the phase speed  (\ref{def:alpha:nonlinear}),   the term $\exp(-k x)$ of the above expression disappears, say,
 \[    \Delta_0^\phi = 2\alpha  k^3  A_0^2 e^{-2kx} + k^4 A_0^3 \cos (k) e^{-3kx},\;\;\;   x  > 0.  \]
Thus,  the first-order deformation equation reads
   \begin{equation}
\nabla^2\phi_1(x,z)=0,  \hspace{1.0cm} z\leq 0, \;\; 0 < x < +\infty,  \label{geq:phi:1st}
\end{equation}
subject to the boundary condition on the known free surface $z=0$:
\begin{equation}
\bar{\cal L} \left( \phi_m \right) =  \left.  \left( \alpha^2 \;  \frac{\partial^2 \phi_m}{\partial x^2} +   \frac{\partial \phi_m}{\partial z} \right)\right|_{z=0}  =  c_\phi   \left[  2\alpha  k^3  A_0^2 e^{-2kx} + k^4 A_0^3 \cos (k) e^{-3kx}\right],  \label{bc:1st:phi}
\end{equation}
and the bottom condition
\begin{equation}
\frac{\partial \phi_1}{\partial z} = 0, \hspace{1.0cm} z = -1, \;\; 0  <  x < +\infty.  \label{bc:1st:bottom}
\end{equation}
 Using the property of the  inverse operator (\ref{L:inverse:property}),  it is easy to gain the common solution
\begin{eqnarray}
 \phi_1(x,z)  &=&  c_\phi \left\{  \frac{\alpha k^2 A_0^2 \cos[2 k(z+1)] e^{-2 k x}}{2\alpha^2 k \cos(2k) -\sin (2k) } +  \frac{ k^3 A_0^3 \cos k \cos[3 k(z+1)] e^{-3 k x}}{3[ 3\alpha^2 k \cos(3k) -\sin(3 k)]} \right\} \nonumber\\
 &+& A_1 \cos [k(z+1)] e^{-k x}, \;\;\; 0  < x < +\infty,  \label{phi:1st}
 \end{eqnarray}
where $A_1$ is an unknown constant to be determined.  Similarly,  using (\ref{eta:mth}),   we gain  $\zeta_2(x)$,   which contains the unknown constant $A_1$.   Then, for the  given wave height $H_w$,  we have a linear algebraic equation
\[    H_w = \lim_{x\to 0}\zeta_1(x) +  \lim_{x\to 0}\zeta_2(x), \]
which determines $A_1$.  Then,  $\phi_1(x,z)$ is completely determined.    Similarly, we further  gain  $\phi_2(x,z)$, $\zeta_3(x)$,  and so on.  Finally, using the symmetry (\ref{symmetry}), we  gain the  wave elevation $\zeta(x)$ and the velocities $u(x,z), v(x,z)$ in the interval $-\infty < x < 0$ and $0<x <+\infty$.   At $x=0$,  the  continuous  horizontal velocity $u(0,z) = U(z)$ is given by  (\ref{bc:left}), and the vertical velocity $v(0,z) = 0$ is  given directly by the symmetry condition (\ref{symmetry}).  The convergence of the solution series is guaranteed by properly choosing the two convergence-control parameters $c_\phi$ and $c_\eta$ in the frame of the HAM.   In this way, we gain the  convergent series solutions of the peaked solitary waves in the whole domain $-\infty < x < +\infty$ by means of the UWM.  

Our computations confirm  that,  for all $m \geq 0$,   $R_m(x)$ in (\ref{bc:mth:phi})  indeed  does {\em not} contain the term $\exp(-k x)$ at all.  Thus,  the fully nonlinear wave equations (\ref{geq:phi}) -- (\ref{bc:bounded})  indeed  give the {\em same} dimensionless phase speed $\alpha = \sqrt{\tan k/k}$ as that by the linear ones.   Therefore, the phase speed of  the peaked solitary wave  indeed has nothing to do with the wave height $H_w$: in fact,  it is dependent upon the decay-parameter $k$, which determines the decay-length, i.e. the so-called ``actual wavelength'' $\lambda_a$, when wave height is fixed.   In practice,  the peaked solitary waves can be regarded to be dispersive (for a fixed wave height) with the ``actual wavelength'' that is a characteristic length in the horizontal direction.  This is indeed completely  different from the  traditional periodic  and solitary  waves with smooth crest, which are dispersive with wave height that is a characteristic length in vertical direction.   This  unusual characteristic  clearly  demonstrates  the novelty of the peaked solitary surface waves.   We will  discuss this interesting characteristic of the  peaked solitary waves later.

Note that our HAM-based analytic approach mentioned above is rather similar to those by Liao \& Cheung \cite{Liao2003JEM} and  Tao {\em et al.} \cite{Tao2007CE} for the traditional smooth progressive waves in deep and finite water, except that we use here the symmetry condition (\ref{symmetry}),  the evanescent-mode base-function (\ref{base:new}),  and besides regard  the  dimensionless  phase  speed  $\alpha$ as a constant independent of wave height.

Finally, we should emphasize that, unlike perturbation methods,  our HAM-based  analytic approach does not need any assumptions about small/large physical parameters.  More importantly,  both of $\phi(x,z)$ and $\zeta(x)$ contain the two convergence-control parameters $c_\phi$ and $c_\eta$, which provide us a convenient way to guarantee the convergence of approximation series, as illustrated below.

\subsubsection{Convergence of series solution  \label{section:convergence}}

Note that, unlike perturbation results,  $\phi_m(x,z)$ and $\zeta_m(x)$ gained in the above-mentioned  HAM-based   approach  contain the two convergence-control parameters $c_\phi$ and $c_\eta$, which provide us a convenient way to  guarantee the convergence of the series (\ref{series:phi}) and (\ref{series:zeta}), as shown below.    

\begin{table}
\begin{center}
\def~{\hphantom{0}}
\begin{tabular}{c|ccccc}
 \hline \hline
  Order of approx.  & $U(-1)$ & $U(-0.5)$ & $U(-0.25)$ & $U(H_w)$ & $\zeta'(0_+)$ \\ [1pt]  \hline
  1 &   0.07222 &   0.06570    &   0.05762    &   0.04289    &   -0.04690 \\ [3pt]
  3 &   0.06833 &    0.06236 &   0.05466 &   0.04205 &   -0.04859    \\  [3pt]
  5 &   0.06796 &   0.06219 &   0.05489 &   \bf{0.04213} &   \bf{-0.04823}    \\  [3pt]
  10    &   \bf{0.06799} &   \bf{0.06221} &   \bf{0.05490} &   \bf{0.04213} &   \bf{-0.04823}    \\  [3pt]
 15    &   \bf{0.06799} &   \bf{0.06221} &   \bf{0.05490} &   \bf{0.04213} &   \bf{-0.04823}    \\  [3pt]
 20    &   \bf{0.06799} &   \bf{0.06221} &   \bf{0.05490} &   \bf{0.04213} &   \bf{-0.04823}    \\  [3pt]
 25    &   \bf{0.06799} &   \bf{0.06221} &   \bf{0.05490} &   \bf{0.04213} &   \bf{-0.04823}    \\  [3pt]
  \hline\hline
\end{tabular}
\caption{Analytic approximations of  $U(z) = u(0,z)$ and $\zeta'(0_+)$  in the case of   $H_w = 1/20$ and $k=1$  by means of $c_\phi=-1$ and $c_\eta=-1$. }
\label{Table:uK1Hw0d05}
\end{center}
\end{table}

\begin{table}
\def~{\hphantom{0}}
\begin{center}
\begin{tabular}{c|ccccc}
\hline  \hline
  Order of approx.  & $U(-1)$ & $U(-0.5)$ & $U(-0.25)$ & $U(H_w)$ & $\zeta'(0_+)$ \\   [1pt] \hline
  1 &   -0.07047    &   -0.06377    &   -0.05101    &   -0.03840    &   0.05248 \\    [3pt]
  3 &   -0.08133    &   -0.06737    &   -0.05218    &   -0.03779    &   0.05220 \\  [3pt]
  5 &   -0.08140    &   -0.06749    &   -0.05218    &   \bf{-0.03772}    &   0.05192 \\  [3pt]
  10&   \bf{-0.08145}    &   \bf{-0.06750}    &   \bf{-0.05218}    &   \bf{-0.03772}    &   \bf{0.05183} \\  [3pt]
  20&   \bf{-0.08145}    &   \bf{-0.06750}    &   \bf{-0.05218}    &   \bf{-0.03772}    &   \bf{0.05183} \\  [3pt]
  25&   \bf{-0.08145}    &   \bf{-0.06750}    &   \bf{-0.05218}    &   \bf{-0.03772}    &   \bf{0.05183} \\  [3pt]
  \hline\hline
\end{tabular}
\caption{Analytic approximations of $U(z) = u(0,z)$ and $\zeta'(0_+)$ in the case of   $H_w = -1/20$ and $k=1$  by means of $c_\phi=-1$ and $c_\eta=-1$. }
\label{Table:uK1Hw-0d05}
\end{center}
\end{table}

\begin{figure}
\centering
\includegraphics[scale=0.5]{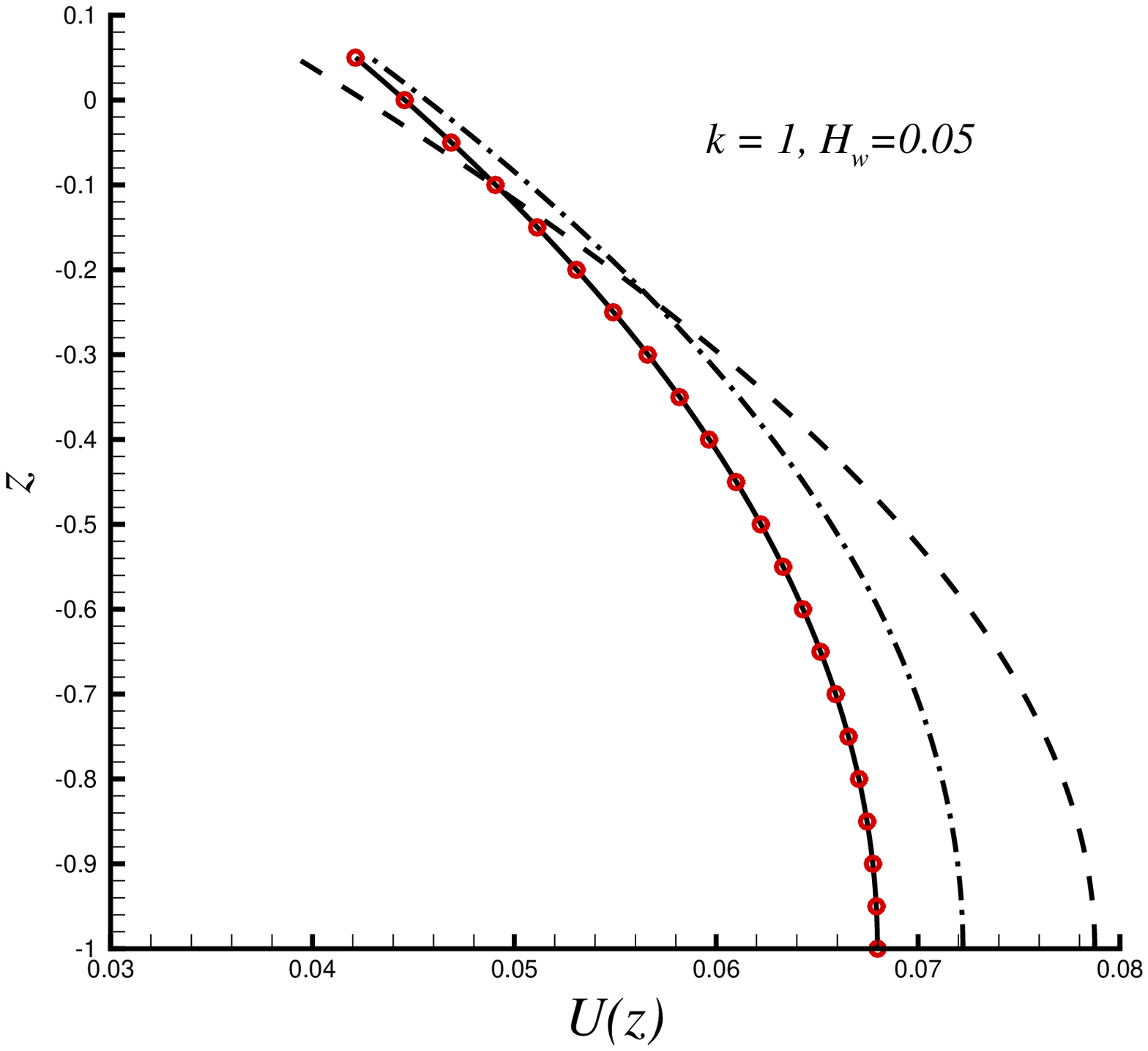}
\caption{Analytic approximations of the dimensionless horizontal velocity profile  $U(z) = u(0,z)$ beneath the crest  in the case of  $k=1$ and $H_w = 0.05$  given by $c_\phi = -1$ and $c_\eta=-1$.   Dashed-line: zeroth-order of approx.; Dash-dotted line: 1st-order of approx.; Solid line: 4th-order of approx.; Symbols: 25th-order of approximation.   }
\label{figure:uK1Hw0d05}
\end{figure}

First, let us consider the case of  $k=1$ and $H_w=1/20$, with the corresponding dimensionless  phase velocity $\alpha=\sqrt{\tan k/k}\approx 1.24796$.  Since the wave height is only 5\% of the water depth $D$, the nonlinearity is weak.  Thus, following  Liao \& Cheung \cite{Liao2003JEM} and  Tao {\em et al.} \cite{Tao2007CE},  we  choose $c_\phi=-1$ and $c_\eta=-1$ for such a kind of weakly nonlinear wave problem.  It is found that, the corresponding series of analytic approximations  indeed converge  quickly, as shown  in Table~\ref{Table:uK1Hw0d05} for $\zeta'(0_+)$ and the horizontal velocity $U(z) = u(0,z)$ at $x=0$ when $z=-1, z= -1/2, z = -1/4$ and $z = H_w$, respectively, where $0_{+}$  denotes $x\to 0$ from the right  along the $x$ axis.  It is found that the velocity potential $\phi(x,z)$ converges quickly in the domain $x\in(0,+\infty)$ and $z\leq \zeta(x)$, as shown  in Fig.~\ref{figure:uK1Hw0d05} for the corresponding horizontal velocity profile $U(z) = u(0,z)$ at crest ($x=0$).  This confirms that the  peaked solitary wave  is  indeed a solution of the  UWM  based on the symmetry and the fully nonlinear wave equations (\ref{geq:phi}) -- (\ref{bc:bounded}).  

\begin{figure}
\centering
\includegraphics[scale=0.5]{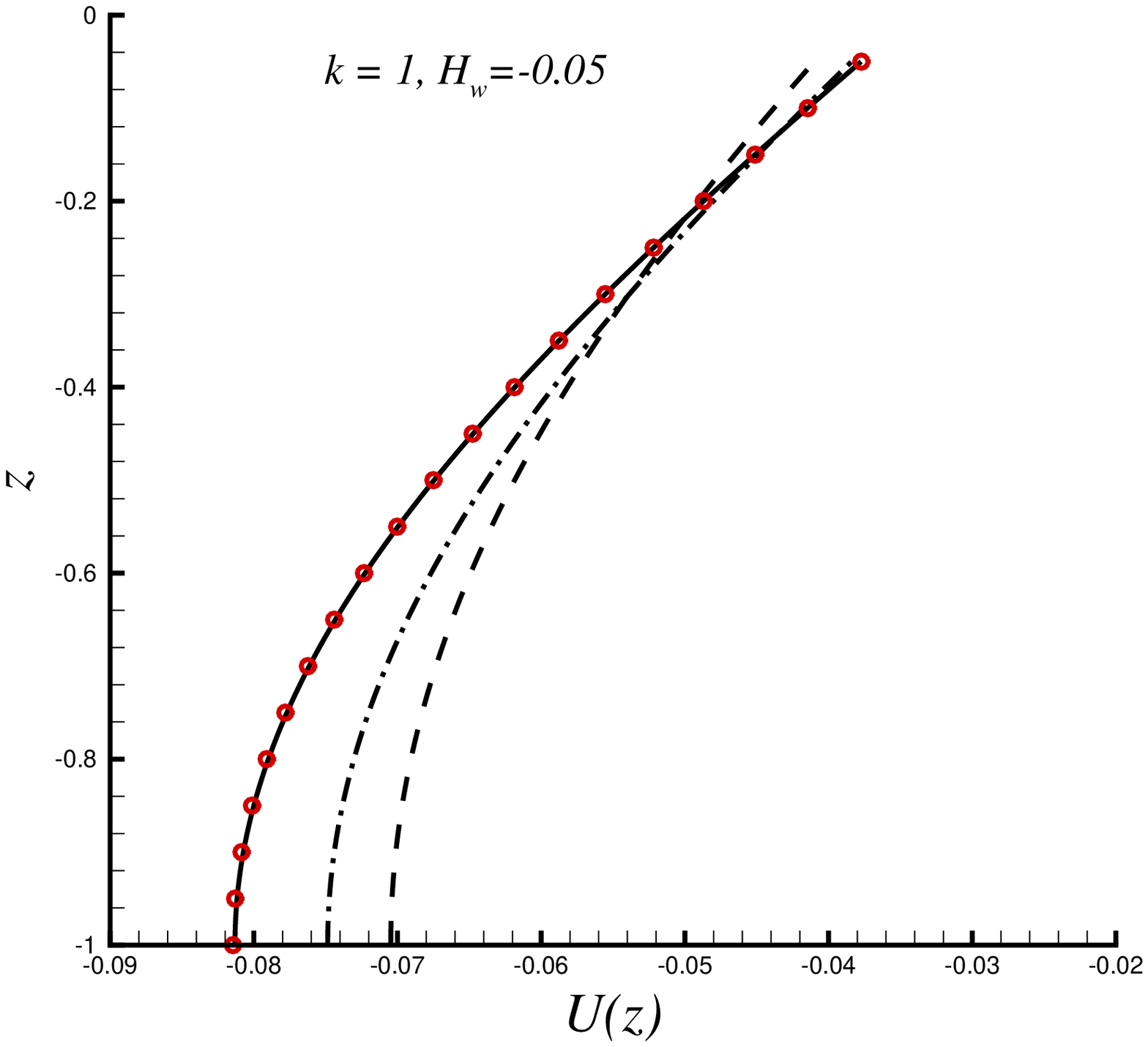}
\caption{Analytic approximations of the dimensionless horizontal velocity profile  $U(z) = u(0,z)$ beneath the crest  in the case of  $k=1$ and $H_w = -0.05$  given by $c_\phi = -1$ and $c_\eta=-1$.   Dashed-line: zeroth-order of approx.; Dash-dotted line: 1st-order of approx.; Solid line: 4th-order of approx.; Symbols: 25th-order of approximation.   }
\label{figure:uK1Hw-0d05}
\end{figure}

Secondly, let us consider the case with $k=1$ and $H_w=-1/20$, with the same dimensionless  phase velocity $\alpha=c/\sqrt{g D} \approx 1.24796$.  It is found that the corresponding series of analytic approximations given by $c_\phi=-1$ and $c_\eta=-1$ converge quickly in  the  domain $x\geq 0$,  as shown  in Table~\ref{Table:uK1Hw-0d05} for $\zeta'(0_+)$ and the horizontal velocity $U(z) = u(0,z)$ at crest ($x=0$) when $z=-1, -0.5, -0.25$ and $z=H_w$, respectively.  Besides, the corresponding  velocity potential $\phi(x,z)$ converges quickly in the  domain $x\in(0,+\infty)$ and $z\leq \zeta(x)$, as shown  in Fig.~\ref{figure:uK1Hw-0d05} for the horizontal velocity profile $U(z) = u(0,z)$ at  crest.  This confirms that the  peaked solitary wave in the form of depression  is also a solution of the  UWM, too.  

\begin{table}
\begin{center}
\def~{\hphantom{0}}
\begin{tabular}{c|ccccc}
\hline  \hline
  Order of approx.  & $U(-1)$ & $U(-0.5)$ & $U(-0.25)$ & $U(H_w)$ & $\zeta'(0_+)$ \\ [1pt] \hline
  1 &   0.1696 &   0.1543    &   0.1347    &   0.09362    &   -0.08561 \\  [3pt]
  3 &   0.1246  &   0.1180  &   0.1090  &   0.08837 &   -0.09332    \\ [3pt]
  5 &   0.1270  &   0.1196  &   0.1098  &   0.08813 &   -0.09142    \\ [3pt]
  10    &   0.1254  &   0.1183  &   0.1090  &   0.08788 &   -0.09285    \\ [3pt]
  15    &   \bf{0.1254}  &   \bf{0.1183}  &   \bf{0.1090}  &   \bf{0.08789} &   \bf{-0.09299}    \\ [3pt]
  20    &   \bf{0.1254}  &   \bf{0.1183}  &   \bf{0.1090}  &   \bf{0.08789} &   \bf{-0.09299}    \\ [3pt]
 25    &   \bf{0.1254}  &   \bf{0.1183}  &   \bf{0.1090}  &   \bf{0.08789} &   \bf{-0.09299}    \\ [3pt]
  \hline\hline
\end{tabular}
\caption{Analytic approximations of  $U(z) = u(0,z)$ and $\zeta'(0_+)$  in the case of   $H_w = 1/10$ and $k=1$  by means of $c_\phi=-1/2$ and $c_\eta=-1$. }
\label{Table:uK1Hw0d1}
\end{center}
\end{table}

\begin{table}
\begin{center}
\def~{\hphantom{0}}
\begin{tabular}{c|ccccc}
  \hline\hline
  Order of approx.  & $U(-1)$ & $U(-0.5)$ & $U(-0.25)$ & $U(H_w)$ & $\zeta'(0_+)$ \\ [1pt] \hline
  1 &   -0.1418 &   -0.1191    &   -0.09328    &   -0.07474    &   0.1091 \\ [3pt]
  3 &   -0.1704 &   -0.1343 &   -0.09684    &   -0.07211    &   0.1130  \\ [3pt]
  5 &   -0.1768 &   -0.1368 &   -0.09664    &   -0.07070    &   0.1107  \\ [3pt]
  10    &   -0.1801 &   -0.1379 &   -0.09650    &   -0.07016    &   0.1079  \\ [3pt]
  15    &   -0.1805 &   -0.1380 &   -0.09649    &   -0.07013    &   0.1075  \\ [3pt]
  20    &   \bf{-0.1806} &   \bf{-0.1380} &   \bf{-0.09648}    &   \bf{-0.07012}    &   \bf{0.1075}  \\ [3pt]
  25    &   \bf{-0.1806} &   \bf{-0.1380} &   \bf{-0.09648}    &   \bf{-0.07012}    &   \bf{0.1075}  \\ [3pt]
  \hline\hline
\end{tabular}
\caption{Analytic approximations of  $U(z) = u(0,z)$ and $\zeta'(0_+)$  in the case of   $H_w = -1/10$ and $k=1$  by means of $c_\phi=-3/4$ and $c_\eta=-1$. }
\label{Table:uK1Hw-0d1}
\end{center}
\end{table}

 \begin{figure}
\centering
\includegraphics[scale=0.5]{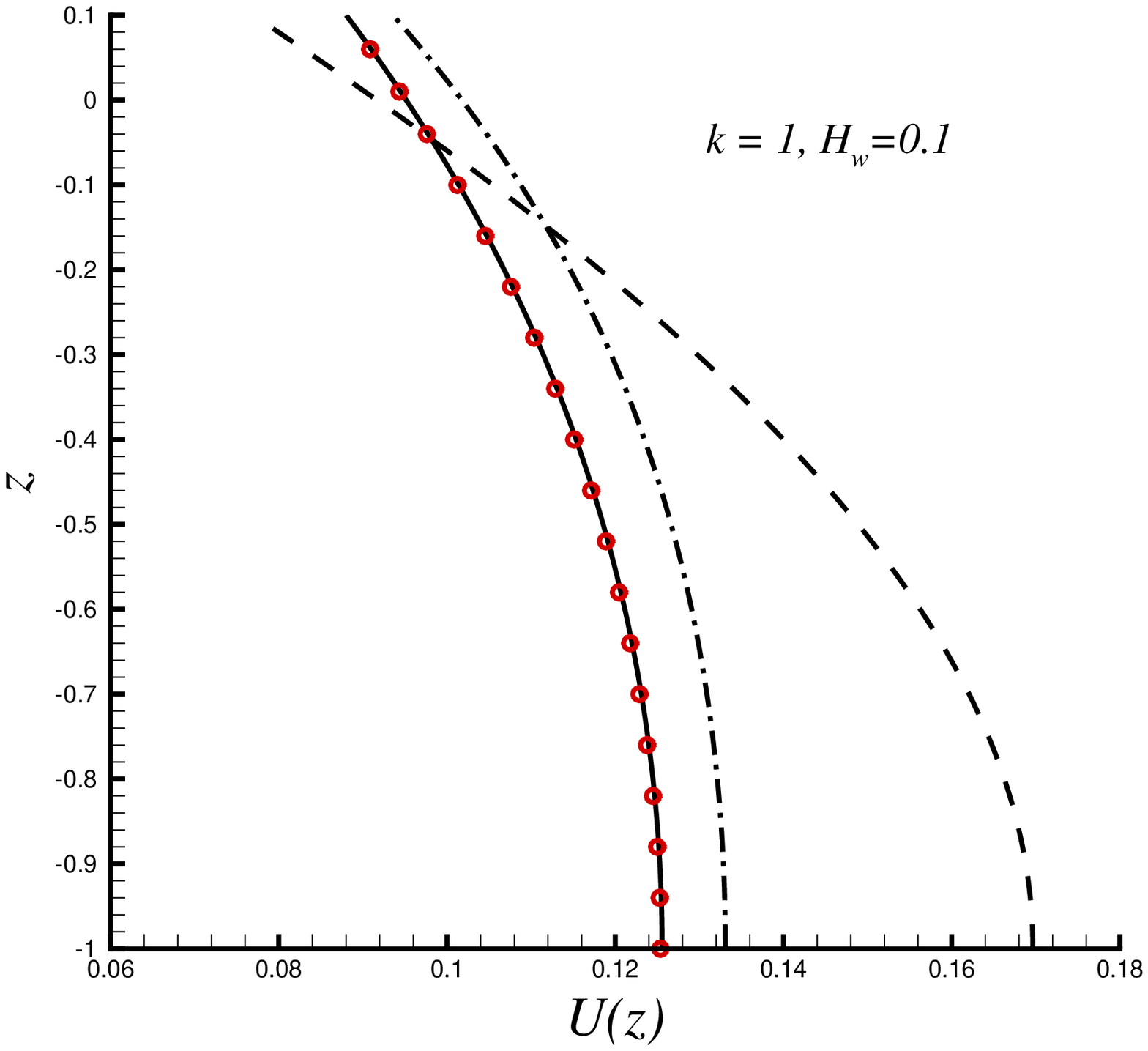}
\caption{Analytic approximations of the dimensionless horizontal velocity profile  $U(z) = u(0,z)$ beneath the crest  in the case of  $k=1$ and $H_w = 0.1$  given by $c_\phi = -1/2$ and $c_\eta=-1$.   Dashed-line: zeroth-order of approx.; Dash-dotted line: 2nd-order of approx.; Solid line: 6th-order of approx.; Symbols: 25th-order of approximation.   }
\label{figure:uK1Hw0d1}

\centering
\includegraphics[scale=0.5]{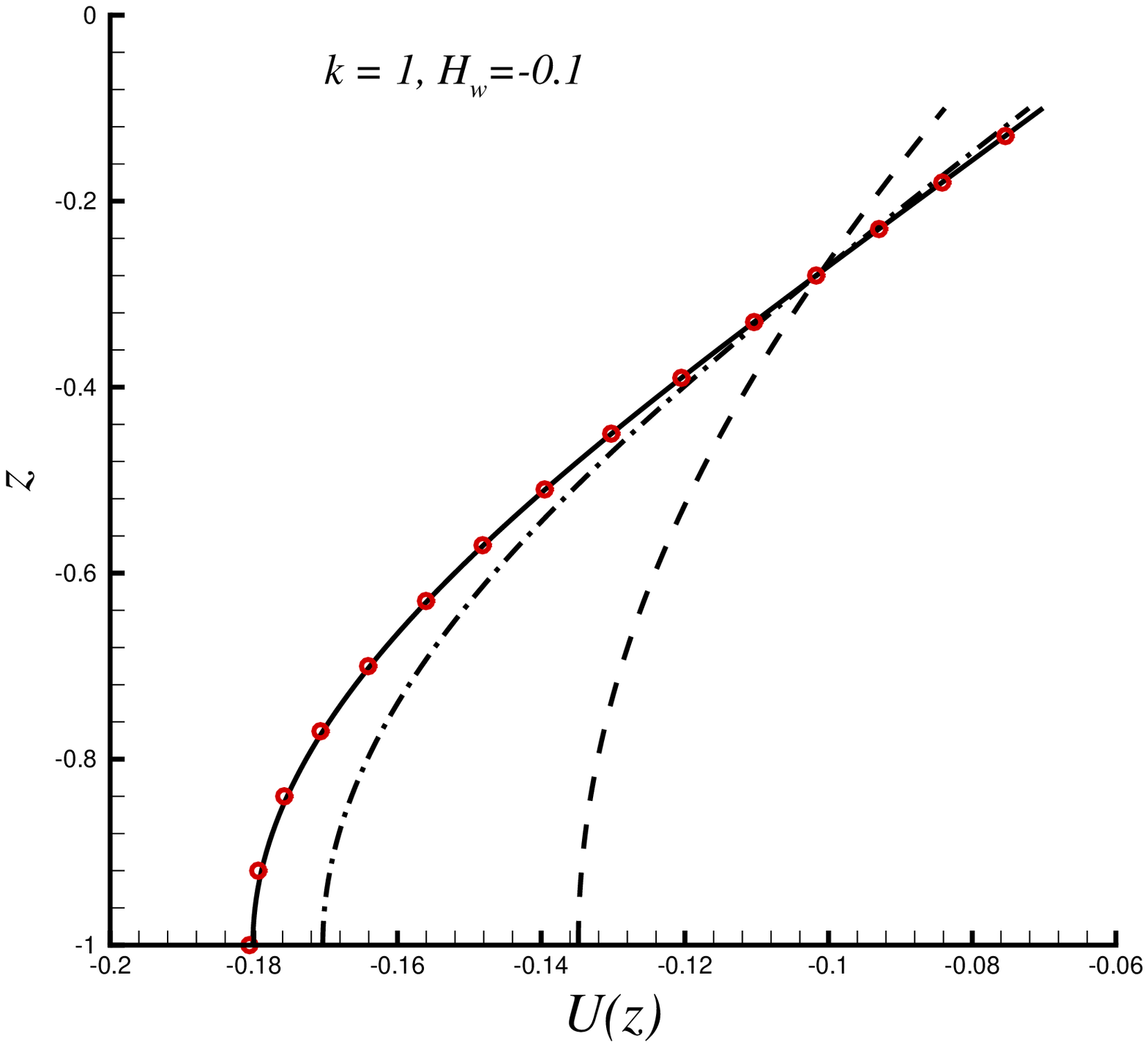}
\caption{Analytic approximations of the dimensionless horizontal velocity profile  $U(z) = u(0,z)$ beneath the crest  in the case of  $k=1$ and $H_w = -0.1$  given by $c_\phi = -3/4$ and $c_\eta=-1$.   Dashed-line: zeroth-order of approx.; Dash-dotted line: 1st-order of approx.; Solid line: 10th-order of approx.; Symbols: 25th-order of approximation.   }
\label{figure:uK1Hw-0d1}
\end{figure}

 \begin{figure}
\centering
\includegraphics[scale=0.5]{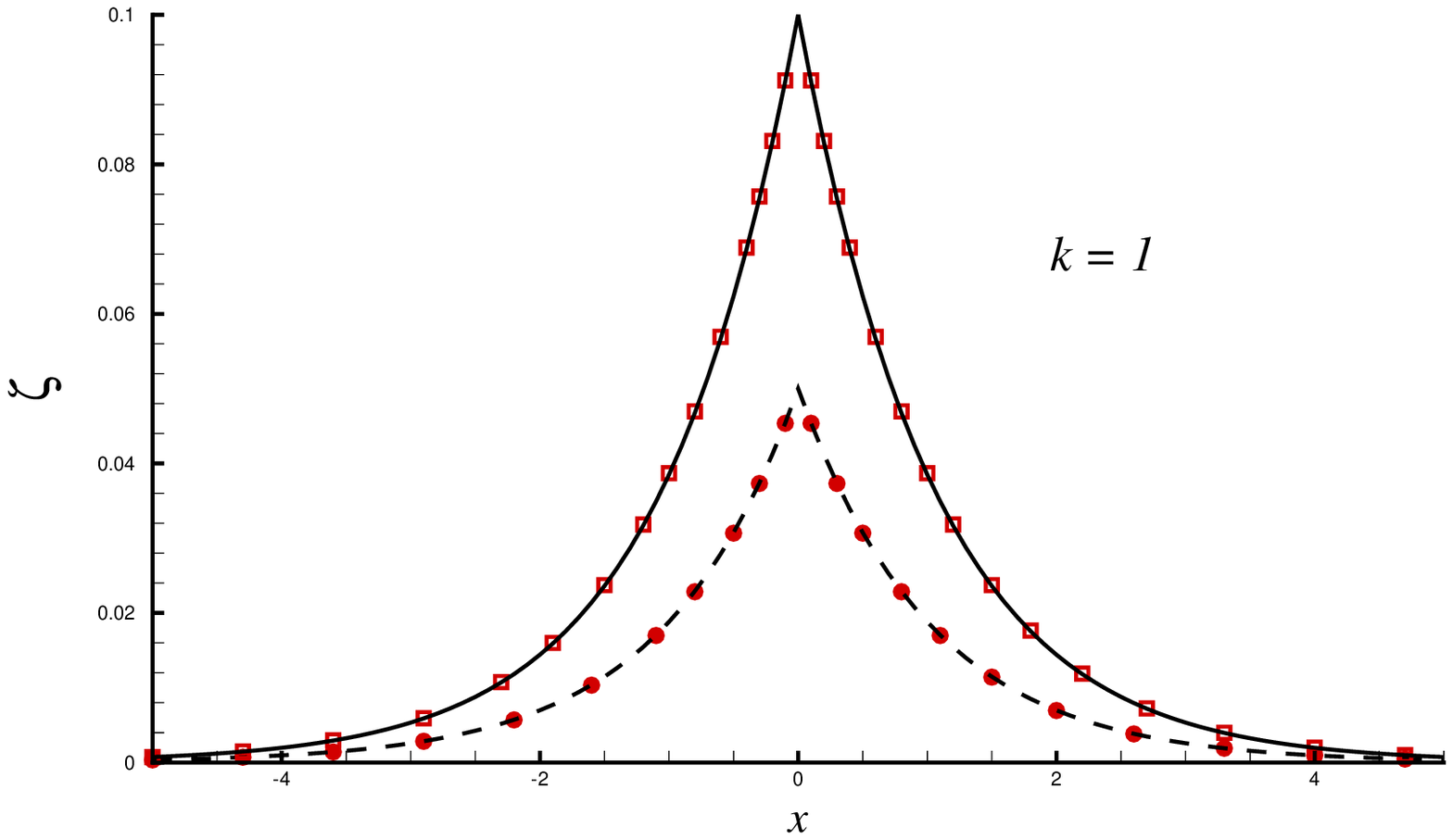}
\caption{Analytic approximations of  elevation of the peaked solitary waves when $k=1$ (corresponding to $c/\sqrt{g D} = 1.24796$).  Solid line:  5th-order approximation when $H_w = 0.1$ given by $c_\phi = -0.5$ and $c_\eta=-1$;  Filled circles: 25th-order approximation when $H_w = 0.1$ given by $c_\phi = -0.5$ and $c_\eta=-1$;    Dashed line:  5th-order approximation when $H_w = 0.05$ given by $c_\phi = -1$ and $c_\eta=-1$;  Open circles: 25th-order approximation when $H_w = 0.05$ given by $c_\phi = -1$ and $c_\eta=-1$.     }
\label{figure:K1Hw}
\end{figure}

  \begin{figure}
\centering
\includegraphics[scale=0.5]{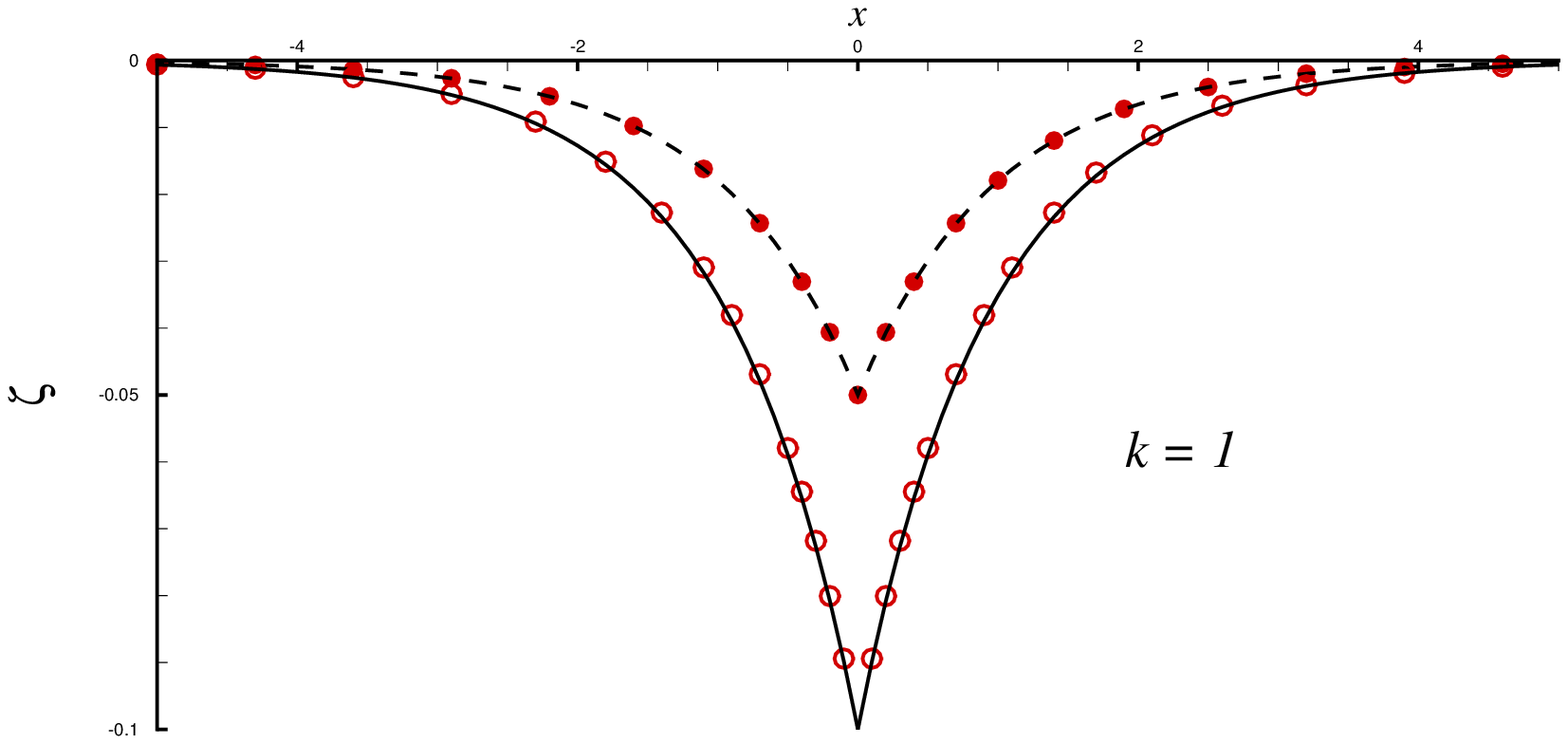}
\caption{Analytic approximations of  elevation of the peaked solitary  waves when $k=1$ (corresponding to $c/\sqrt{g D} = 1.24796$).  Solid line:  5th-order approximation when $H_w = -0.1$ given by $c_\phi = -0.75$ and $c_\eta=-1$;  Filled circles: 25th-order approximation when $H_w = -0.1$ given by $c_\phi = -0.75$ and $c_\eta=-1$;    Dashed line:  5th-order approximation when $H_w =- 0.05$ given by $c_\phi = -1$ and $c_\eta=-1$;  Open circles: 25th-order approximation when $H_w = -0.05$ given by $c_\phi = -1$ and $c_\eta=-1$.     }
\label{figure:K1-Hw}
\end{figure}

  \begin{figure}
\centering
\includegraphics[scale=0.5]{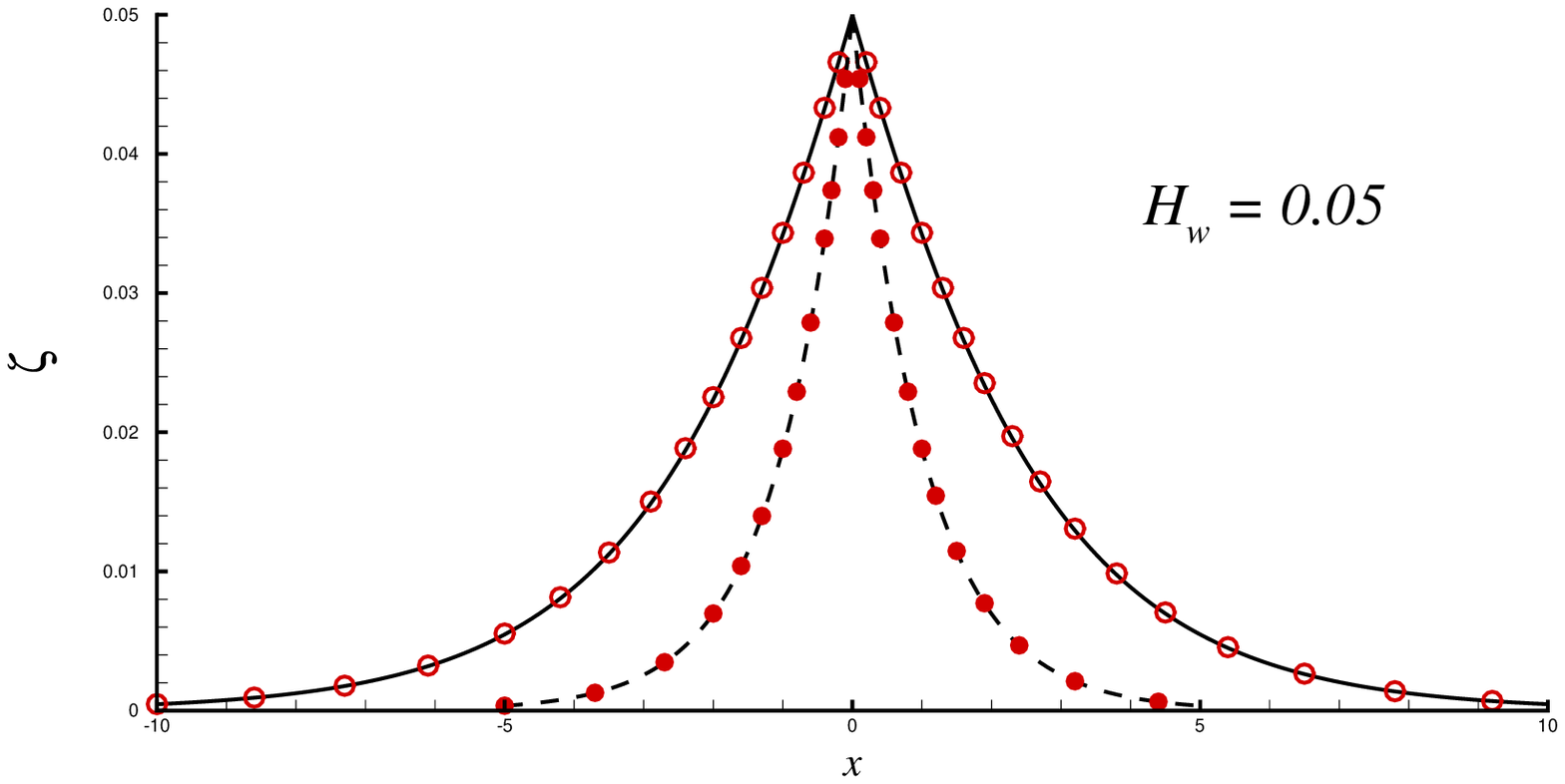}
\caption{Analytic approximations of  $\zeta(x)$ of the peaked solitary waves when $H_w=0.05$ by means of $c_\phi = -1$ and $c_\eta=-1$.  Solid line:  5th-order approximation when $k = 1/2$ (corresponding  to $c/\sqrt{g D} = 1.04528$);  Filled circles: 25th-order approximation when $k=1/2$;    Dashed line:  5th-order approximation when $k=1$ (corresponding  to $c/\sqrt{g D} = 1.24796$);  Open circles: 25th-order approximation when $k=1$.     }
\label{figure:Hw0d05}
\end{figure}

  \begin{figure}
\centering
\includegraphics[scale=0.5]{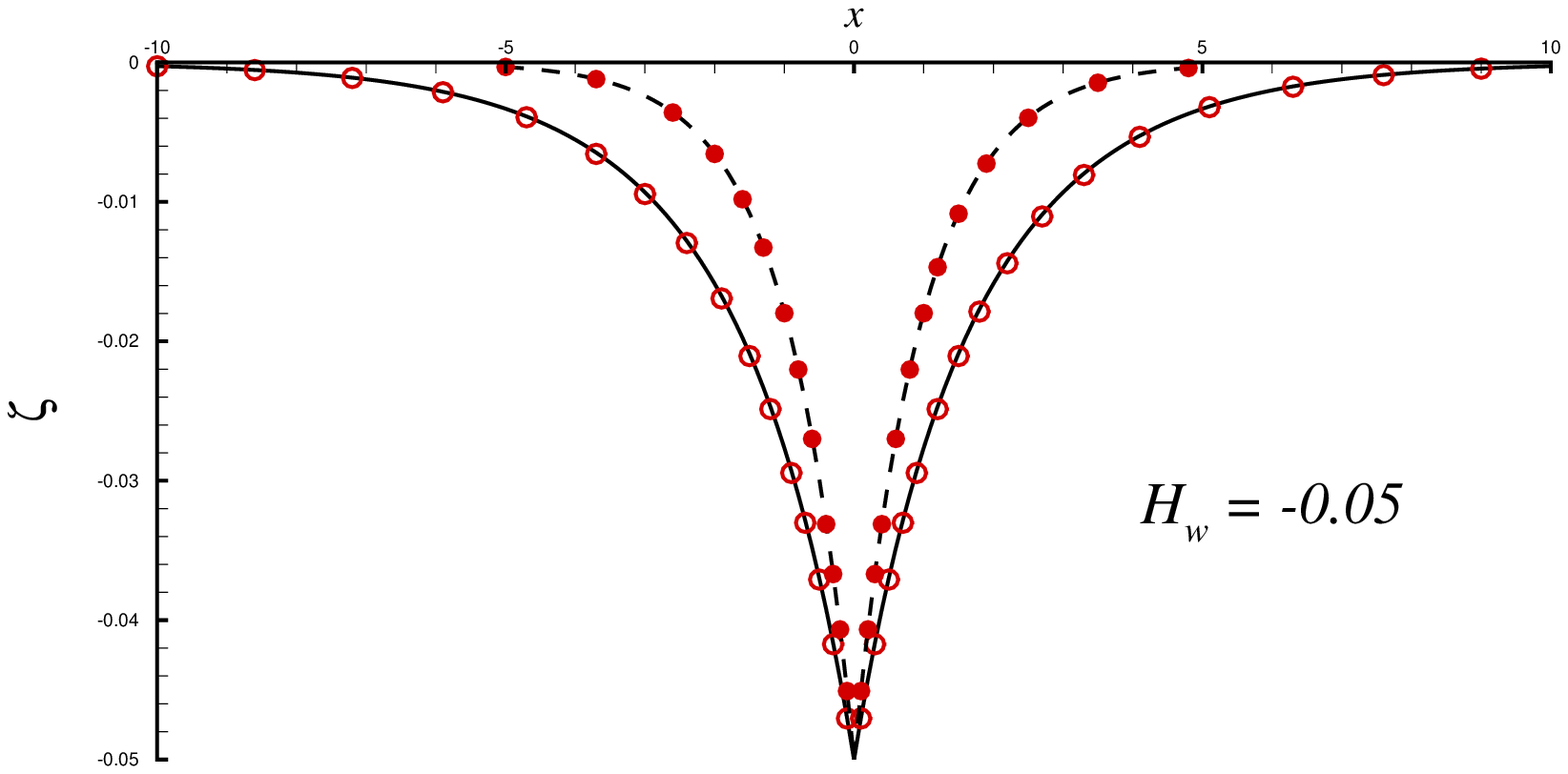}
\caption{Analytic approximations of  $\zeta(x)$ of the peaked solitary waves when $H_w=-0.05$ by means of $c_\phi = -1$ and $c_\eta=-1$.  Solid line:  5th-order approximation when $k = 1/2$ (corresponding  to $c/\sqrt{g D} = 1.04528$);  Filled circles: 25th-order approximation when $k=1/2$;    Dashed line:  5th-order approximation when $k=1$ (corresponding  to $c/\sqrt{g D} = 1.24796$;  Open circles: 25th-order approximation when $k=1$.     }
\label{figure:Hw-0d05}
\end{figure}

Similarly, in the case of  $k = 1/2$  and $H_w = \pm 1/20$, with the corresponding dimensionless phase speed $\alpha  \approx 1.04528$,  we gain convergent  series solutions of $\phi(x,z)$ and $\zeta(x)$ in the  domain $x\in(0,+\infty)$ and $z\leq \zeta(x)$ by means of $c_\phi=-1$ and $c_\eta=-1$, respectively.  

Note that, using the symmetry conditions  (\ref{symmetry}) and (\ref{v:x=0}), we can expand all of these convergent series solutions  into the {\em whole} domain $-\infty < x < +\infty$.  Thus, the UWM indeed admits the peaked solitary waves!   In other words, the UWM {\em unifies} both of the smooth and peaked waves, for the first time, to the best of the author's knowledge.   Mathematically, it reveals that the peaked solitary waves are {\em consistent}  with the traditional smooth waves and thus are as acceptable as them in the frame of inviscid fluid.   

Furthermore, let us consider the case of $k=1$ and $H_w = 0.1$, with the corresponding  dimensionless  phase velocity $\alpha \approx 1.24796$.  Since the wave weight increases to 10\% of water depth, the nonlinearity becomes stronger.  As suggested by Liao \& Cheung \cite{Liao2003JEM} and  Tao {\em et al.} \cite{Tao2007CE}, we should choose  the convergence-control parameters $c_\phi$ and $c_\eta$ with smaller absolute values for high nonlinearity.  It is found that the series of the analytic approximations given by $c_\phi=-1/2$ and $c_\eta=-1$ converges quickly, as shown in Table~\ref{Table:uK1Hw0d1} and Fig.~\ref{figure:uK1Hw0d1}  for the horizontal velocity profile $U(z) = u(0,z)$ at crest.  Similarly, in the case of  $k=1$ and $H_w=-0.1$, we gain convergent series solution by means of $c_\phi=-3/4$ and $c_\eta=-1$,  as shown in Table~\ref{Table:uK1Hw-0d1} and Fig.~\ref{figure:uK1Hw-0d1}.  This illustrates that the two convergence-control parameters $c_\phi$ and $c_\eta$ indeed provide us a convenient way to guarantee the convergence of approximation series.  Note that the absolute value of the horizontal velocity on bottom in the case of  $H_w=-0.1$ is  44\%  larger than that in the case of  $H_w=0.1$.    So, due to the nonlinearity,  there does not exist a symmetry between these two wave elevations for $H_w =  +0.1$ and $H_w=-0.1$, respectively.

It should be emphasized that, in the case of  $k = 1$, we gain convergent series solutions of the peaked solitary waves with the {\em same} phase speed but  {\em different} positive and negative values of $H_w$, such as $H_w = \pm 0.05$ and $H_w=\pm 0.1$, respectively.   This confirms that the phase speed of the peaked solitary waves indeed has nothing to do with the wave height.  As mentioned in \S~4.1,  for a given wave height, the value of  $k$ determines the  decay length, which is the characteristic length of the peaked solitary waves in the horizontal direction, corresponding to  the so-called ``actual wavelength'' $\lambda_a$ defined by (\ref{def:L:a}).   Thus,  unlike the traditional smooth waves that are dispersive with wave height,  the peaked solitary waves  are dispersive (for a fixed wave height) with the ``actual wavelength'', although their phase speed has nothing to do with the wave height.    This is an unusual characteristic of the peaked solitary waves.

Finally, using the symmetry (\ref{symmetry}), it is straightforward  to gain the wave elevation in the whole interval $-\infty < x < +\infty$.   As shown in Figs.~\ref{figure:K1Hw}  to \ref{figure:Hw-0d05},  the wave elevations $\zeta(x)$ also converge  quickly in all of above-mentioned cases.
In case of $k=1$,  the wave elevations  for $ H_w = \pm 0.1$ are  compared with  those  for $H_w=\pm 0.05$, as shown in Figs.~\ref{figure:K1Hw} and \ref{figure:K1-Hw}.   It is found that,  for the same value of $k$, the larger the value of $|H_w|$, the faster $\zeta(x)$ decays to zero.
Note also that the wave elevation $\zeta(x)$ with larger $k$ decays to 0 more quickly, as shown in Figs.~\ref{figure:Hw0d05} and \ref{figure:Hw-0d05}.   In other words, the larger the value of $k$, the faster $\zeta(x)\to 0$. This provides us a physical meaning of the parameter $k$.  For this reason, we call $k$ the decay-parameter,  which determines the so-called decay-length $\lambda_a$, defined by (\ref{def:L:a}).

Note that, according to the convergence theorem generally proved by Liao \cite{LiaoBook2003, LiaoBook2012} in the frame of the HAM, each series solution of the peaked solitary waves given by the HAM satisfies its original equations,  as long as it is convergent.  So, all of these convergent series of $\phi(x,z)$ and $\zeta(x)$ are solutions of the  unified  wave model  (UWM)  based on the symmetry and the fully nonlinear wave equations (\ref{geq:phi}) - (\ref{bc:bounded}), as further confirmed below.

\subsubsection{Validation check of analytic approximations  \label{section:check}}%

Note that the velocity potential $\phi(x,z)$ is expressed in the form (\ref{solution-expression:phi}), which {\em automatically} satisfies the Laplace equation (\ref{geq:phi}) in the interval $0<x<+\infty$, the bottom  condition (\ref{bc:bottom}), and the bounded condition (\ref{bc:bounded}).  Thus, it is only necessary for us to check the two nonlinear boundary conditions (\ref{bc:phi}) and (\ref{bc:zeta})  defined  on the unknown wave elevation $\zeta(x)$.

 To check the validation of our analytic approximations, we define the averaged residual squares of the two free surface boundary conditions
\begin{eqnarray}
{\cal E}^\phi_m(c_\phi,c_\eta) &=& \frac{1}{M} \sum_{n=1}^{M} \left. \left( {\cal N}\left[\check{\phi}(x,z) \right]\right)^2\right|_{x=x_n, z=\zeta(x_n)} , \\
{\cal E}^\zeta_m(c_\phi,c_\eta) &=& \frac{1}{M} \sum_{n=1}^{M} \left. \left[ \check\zeta(x)  - \alpha \frac{\partial \check\phi}{\partial x} + \frac{1}{2} \nabla \check\phi  \cdot \nabla \check\phi  \right]^2 \right|_{x=x_n, z=\zeta(x_n)},
\end{eqnarray}
where
\[    \check\phi(x,z)  = \sum^{m}_{n=0}\phi_n(x,z),\;\;  \check\zeta = \sum_{n=1}^{m }\zeta_n(x) \]
are the $m$th-order approximation of $\phi(x,z)$ and $\zeta(x)$, respectively, and
\[  x_n = n \left(\frac{x_R}{M}\right), \;\;\; 1 \leq n \leq M,    \]
with large enough $x_R$ and $M$.  For all results given below, we choose $x_R=10$ and $M=100$, if not mentioned.  Since the potential velocity $\phi(x,z)$ and the wave elevation $\zeta(x)$ decay exponentially in the horizontal direction, $x_R=10$ is large enough.

In case of $k = 1$ and $H_w=\pm 0.05$, the averaged residual squares ${\cal E}^\phi_m$ and ${\cal E}^\zeta_m$ of the corresponding analytic approximations obtained by $c_\phi=-1$ and $c_\eta=-1$ decay quickly to the level $10^{-25}$ as the order of approximation increases to 25, as shown in Table~\ref{Table:Err-K1Hw0.05}.  In other words, our 25th-order approximation of $\phi(x,z)$ and $\zeta(x)$ satisfies the Laplace equation (\ref{geq:phi}) in the interval $0<x<+\infty$, the bottom condition (\ref{bc:bottom}) and the bounded condition (\ref{bc:bounded}) {\em exactly}, and besides the two nonlinear free surface boundary conditions (\ref{bc:phi}) and (\ref{bc:zeta}) {\em very accurately} (to the level $10^{-25}$).  In addition,   $u(0,z)$  determined by (\ref{bc:left}) converges  quickly  so  that   $U(z)$  is  uniquely determined.    Therefore,  our convergent analytic approximation is a very accurate solution of the  UWM  based on the symmetry and the fully  nonlinear wave equations (\ref{geq:phi}) -- (\ref{bc:bounded}).  Similarly,   ${\cal E}^\phi_m$ and ${\cal E}^\zeta_m$ decays to the level $10^{-13}$ in the case of  $k=1$ and $H_w=\pm 0.1$, and to the level $10^{-18}$ in the case of  $k=1/2$ and $H_w = \pm 0.05$, respectively, as shown in Tables~\ref{Table:Err-K1Hw0.1} and \ref{Table:Err-K0.5}.   Theses guarantee that the corresponding analytic approximations of $\phi(x,z)$ and $\zeta(x)$ are indeed quite accurate solutions of the  UWM, respectively.  All of these confirm once again the  convergence theorem  generally  proved by  Liao \cite{LiaoBook2003, LiaoBook2012} in the frame of the HAM:  each convergent series solution given by the HAM satisfies its original equations.

\begin{table}
\begin{center}
\def~{\hphantom{0}}
\begin{tabular}{c|cc|cc}
\hline\hline
  Order of approx.  &   \multicolumn{2}{c|}{$H_w=0.05$} & \multicolumn{2}{c}{$H_w=-0.05$}\\ 
  $m$ &  ${\cal E}^\phi_m$ & ${\cal E}^\zeta_m$  &   ${\cal E}^\phi_m$ & ${\cal E}^\zeta_m$   \\ \hline
1	&	6.59   $\times 10^{-7}$	&	9.21  $\times 10^{-9}$	&	 2.25   $\times 10^{-6}$	&	 9.39  $\times 10^{-9}$	 	\\
3	&	1.40 $\times 10^{-8}$		&	1.57 $\times 10^{-9}$		&	8.09  $\times 10^{-9}$	&	 5.90  $\times 10^{-10}$	 	\\ [3pt]
5	&	3.32 $\times 10^{-11}$	&	2.90 $\times 10^{-12}$	&	7.80 $\times 10^{-11}$	&	1.53  $\times 10^{-11}$		\\ [3pt]
10	&	9.71 $\times 10^{-15}$	&	7.00 $\times 10^{-16}$	&	8.96 $\times 10^{-16}$	&	2.42  $\times 10^{-16}$	\\ [3pt]
15	&	7.33 $\times 10^{-19}$	&	1.68 $\times 10^{-19}$	& 6.97 $\times 10^{-20}$	&	1.05  $\times 10^{-20}$  	\\ [3pt]
20	&	4.43 $\times 10^{-22}$	&	4.83 $\times 10^{-23}$ 	& 2.40 $\times 10^{-24}$	&	6.45  $\times 10^{-25}$   	\\ [3pt]
25	&	2.23 $\times 10^{-25}$	&	2.09 $\times 10^{-27}$ 	& 4.56 $\times 10^{-28}$	&	3.96  $\times 10^{-29}$   	\\ [3pt]
 \hline\hline
\end{tabular}
\caption{Averaged residual squares of the two nonlinear free boundary conditions  (\ref{bc:phi}) and (\ref{bc:zeta})  in the case of  $k=1$ and $H_w =  \pm 0.05$ by means of $c_\phi=-1$ and $c_\eta=-1$,  with the corresponding dimensionless phase speed $c/\sqrt{g D} = 1.24796$. }
\label{Table:Err-K1Hw0.05}
\end{center}
\end{table}%

\begin{table}
\begin{center}
\def~{\hphantom{0}}
\begin{tabular}{c|cc|cc}
\hline\hline
  Order of approx.  &   \multicolumn{2}{c|}{$H_w=0.1$} & \multicolumn{2}{c}{$H_w=-0.1$}\\
 $m$ &  \multicolumn{2}{c|}{$(c_\phi=-0.5, c_\eta=-1)$} & \multicolumn{2}{c}{$(c_\phi=-0.75, c_\eta=-1)$}   \\  
   &  ${\cal E}^\phi_m$ & ${\cal E}^\zeta_m$  &  ${\cal E}^\phi_m$ & ${\cal E}^\zeta_m$   \\ \hline
1	&	1.48   $\times 10^{-4}$	&	7.21  $\times 10^{-7}$	&  5.89   $\times 10^{-5}$	&	3.67  $\times 10^{-7}$ 	\\   [3pt]
3	&	1.63  $\times 10^{-7}$	&	8.63 $\times 10^{-8}$		&	5.84   $\times 10^{-7}$	&	 1.21  $\times 10^{-7}$		\\	 [3pt]
5	&	1.39 $\times 10^{-7}$		&	3.96 $\times 10^{-10}$	&	6.83   $\times 10^{-7}$	&	 1.11  $\times 10^{-8}$		\\  [3pt]
10	&	5.96 $\times 10^{-10}$	&	3.09 $\times 10^{-11}$	&	8.63   $\times 10^{-9}$	&	2.31  $\times 10^{-10}$	\\  [3pt]
15	&	1.30 $\times 10^{-12}$	&	3.70 $\times 10^{-14}$	&	3.29   $\times 10^{-11}$	&	 1.95  $\times 10^{-12}$	\\  [3pt]
20	&	4.34 $\times 10^{-13}$	&	1.88 $\times 10^{-15}$	&    1.17   $\times 10^{-12}$	&	 3.69  $\times 10^{-14}$		\\  [3pt]
25	&	2.25 $\times 10^{-13}$	&	4.52 $\times 10^{-16}$	&    2.51   $\times 10^{-14}$  & 	 8.11   $\times 10^{-16}$ \\ [3pt]
\hline  \hline
\end{tabular}
\caption{Averaged residual squares of the two nonlinear free boundary conditions  (\ref{bc:phi}) and (\ref{bc:zeta})  in the case of  $k=1$ and $H_w = \pm 0.1$, with the corresponding dimensionless phase speed $c/\sqrt{g D} = 1.24796$. }
\label{Table:Err-K1Hw0.1}
\end{center}
\end{table}%

\begin{table}
\begin{center}
\def~{\hphantom{0}}
\begin{tabular}{c|cc|cc}
\hline\hline
  Order of approx.  &   \multicolumn{2}{c|}{$H_w=0.05$} & \multicolumn{2}{c}{$H_w=-0.05$}\\ 
  $m$ &  ${\cal E}^\phi_m$ & ${\cal E}^\zeta_m$  &  ${\cal E}^\phi_m$ & ${\cal E}^\zeta_m$   \\ \hline
1	&	2.75   $\times 10^{-8}$	&	1.04  $\times 10^{-6}$	&	3.74   $\times 10^{-7}$	&	9.09  $\times 10^{-7}$ 	\\   [3pt]
3	&	3.76   $\times 10^{-10}$	&	4.85  $\times 10^{-8}$	&	4.48   $\times 10^{-10}$	&	 9.61  $\times 10^{-9}$	 	\\    [3pt]
5	&	4.13   $\times 10^{-12}$	&	2.00  $\times 10^{-9}$	&	4.85   $\times 10^{-13}$	&	 4.18  $\times 10^{-12}$	 	\\     [3pt]
10   &	1.80   $\times 10^{-14}$	&	2.21  $\times 10^{-12}$	&	1.08   $\times 10^{-16}$	&	 3.16  $\times 10^{-17}$ 	\\     [3pt]
15   &	5.23   $\times 10^{-15}$	&	1.55  $\times 10^{-15}$	&	7.93   $\times 10^{-20}$	&	 1.06  $\times 10^{-19}$	\\     [3pt]
20   &	1.86   $\times 10^{-16}$	&	1.50  $\times 10^{-16}$	&	1.67   $\times 10^{-23}$	&	 1.52  $\times 10^{-24}$ 	\\     [3pt]
25   &	2.56   $\times 10^{-18}$	&	9.46  $\times 10^{-18}$	&	3.87   $\times 10^{-29}$	&	 2.57  $\times 10^{-28}$ 	\\     [3pt]
\hline		\hline
\end{tabular}
\caption{Averaged residual squares of the two nonlinear free boundary conditions  (\ref{bc:phi}) and (\ref{bc:zeta})  in the case of  $k=1/2$ and $H_w = \pm 0.05$ by means of $c_\phi=-1$ and $c_\eta=-1$,  with the corresponding dimensionless phase speed $c/\sqrt{g D} = 1.04528$. }
\label{Table:Err-K0.5}
\end{center}
\end{table}%

In fact,  one can choose the optimal values of $c_\phi$ and $c_\eta$ by the minimum of  ${\cal E}^\phi_m(c_\phi,c_\eta)$ and ${\cal E}^\zeta_m(c_\phi,c_\eta)$, say,
\begin{equation}
\frac{\partial {\cal E}^\phi_m(c_\phi,c_\eta)}{\partial c_\phi} = 0, \;\;  \frac{\partial {\cal E}^\zeta_m(c_\phi,c_\eta)}{\partial c_\zeta} = 0.
\end{equation}
It is found that, by means of the optimal values of $c_\phi$ and $c_\eta$, the corresponding series of analytic approximations often converge more quickly.

All of these demonstrate that  the convergent series of the peaked solitary waves obtained by our HAM-based approach are indeed the solutions of the  unified  wave model  (UWM)  based on the symmetry and the  fully nonlinear wave equations (\ref{geq:phi}) -- (\ref{bc:bounded}).

 \subsection{Characteristics of peaked solitary surface waves\label{Characteristic}}

Based on  the so-called evanescent base-functions (\ref{base:new}),  the  peaked  solitary waves governed by the exact  UWM  have  some unusual characteristics that are quite different from those of the traditional  smooth waves.  

  First,  the  peaked  solitary waves have a peaked wave crest, since $\zeta'(x)$ is discontinuous at $x=0$, i.e. $\zeta'(0_+) =- \zeta'(0_{-}) \neq 0$, where $0_{+}$  and $0_{-}$ denote $x\to 0$ from the right and left along the $x$ axis, respectively.   For example,     in the case of  $H_w= + 0.1$ and $k=1$,  we have  $\zeta'(0_{+}) = -0.09299$, but  $\zeta'(0_{-}) = 0.09299$, respectively.  This is quite different from traditional smooth periodic and solitary waves which are infinitely differentiable everywhere.  

  Secondly,  the  peaked  solitary waves may be in the form of depression, which has been reported for internal  waves but  never for surface  ones, to the best of the author's knowledge.   Mathematically, it is  straightforward to  gain such kind of  solitary  waves  in a depression form  even by means of  the linear  UWM, as shown in \S~\ref{LinearTheory}.

Third, unlike traditional smooth periodic and solitary waves which are dispersive with wave height,  the dimensionless phase speed of the  peaked  solitary waves has nothing to do with the wave height,  but  depends  only  upon the so-called decay-parameter $k$.   For a fixed wave height,  the decay-parameter $k$ determines the decay-length  $\lambda_a$, which is a characteristic length of the peaked solitary waves in the horizontal direction, called the ``actual wavelength''.      Thus, unlike the traditional smooth waves, the peaked solitary waves are dispersive (for a fixed wave height) with ``actual wavelength'' $\lambda_a$, defined by (\ref{def:L:a}).    So, in the same water depth $D$, the peaked solitary waves with the same $k$ but different wave height  $H_w$  may propagate with the same phase speed,  where $H_w$ may be either positive or negative.    For example,  it is found that,  in the case of  $k=1$,  all of the peaked solitary waves with $H_w  = \pm 0.1$  or  $H_w=\pm 0.5$  propagate with the {\em same} phase speed $c\approx 1.24796\sqrt{g D}$:  in these cases,  we gain {\em different} series solutions with the {\em same} phase speed, as shown in \S~\ref{section:convergence}.    On the other side,   the peaked  solitary waves with the same wave height $H_w$  but different decay-parameter $k$ (corresponding to different ``actual wavelengths'')  may propagate with different phase speed!    These are completely different from the traditional periodic and solitary waves with smooth crest.  In summary,  unlike the traditional smooth periodic and solitary waves  which are dispersive with wave height that is a characteristic length in the vertical direction, the phase speed of peaked solitary waves has nothing to do with wave height:  in fact, they are dispersive (for a fixed wave height) with ``actual wavelength'', a characteristic length in the horizontal direction.

\begin{figure}
\centering
\includegraphics[scale=0.5]{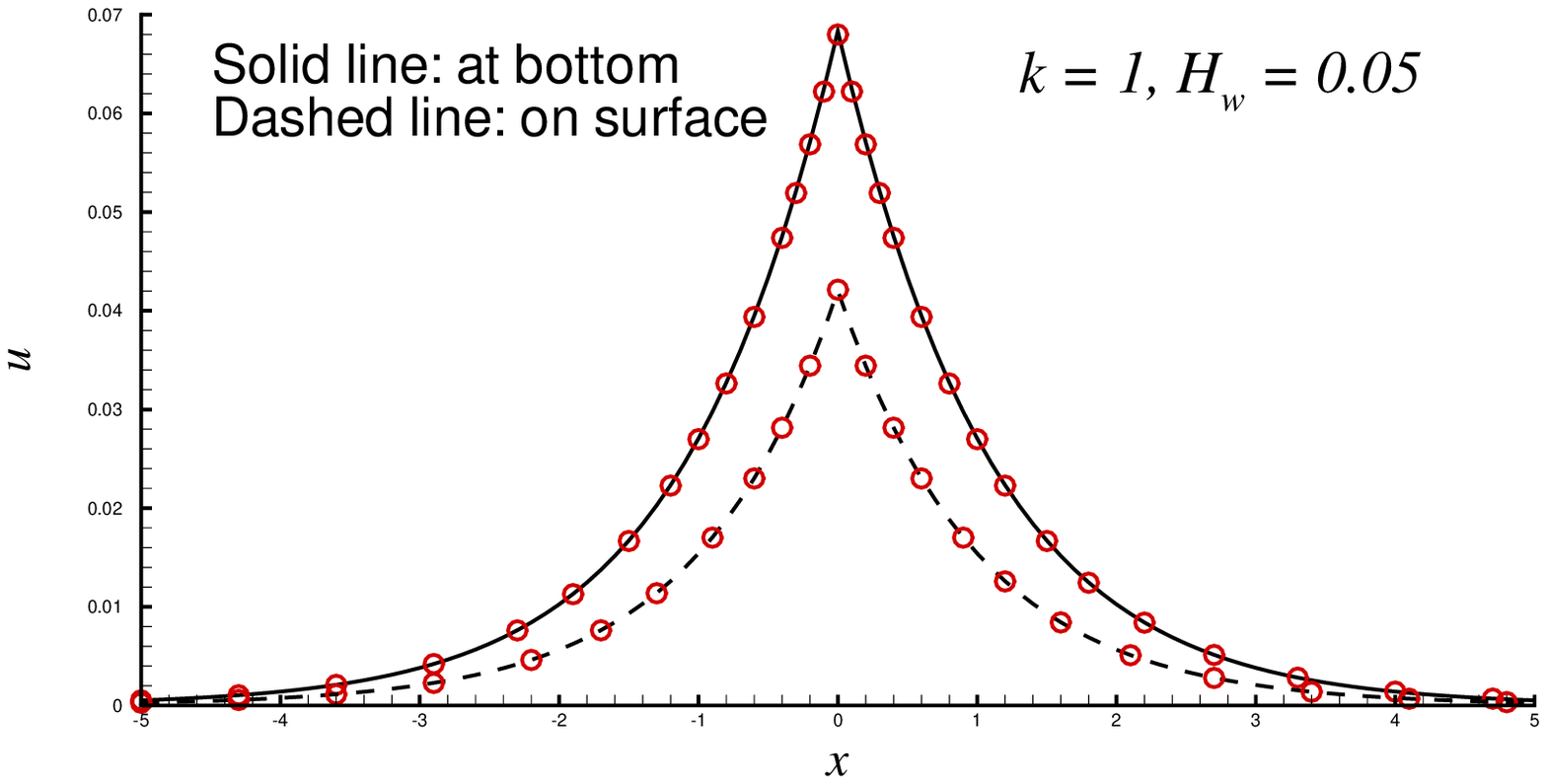}
\caption{  Horizontal velocity at bottom and on free surface when $k=1$ and $H_w=0.05$ by means of $c_\phi=-1$ and $c_\eta=-1$.  Solid line:   3rd-order approx. of $u(x,-1)$ (at bottom);  Dashed line: 3rd-order approx. of $u(x,\zeta(x))$ (on free surface);  Symbols: the corresponding 25th-order approximations.  }
\label{figure:ubHw0d05}
\end{figure}

\begin{figure}
\centering
\includegraphics[scale=0.5]{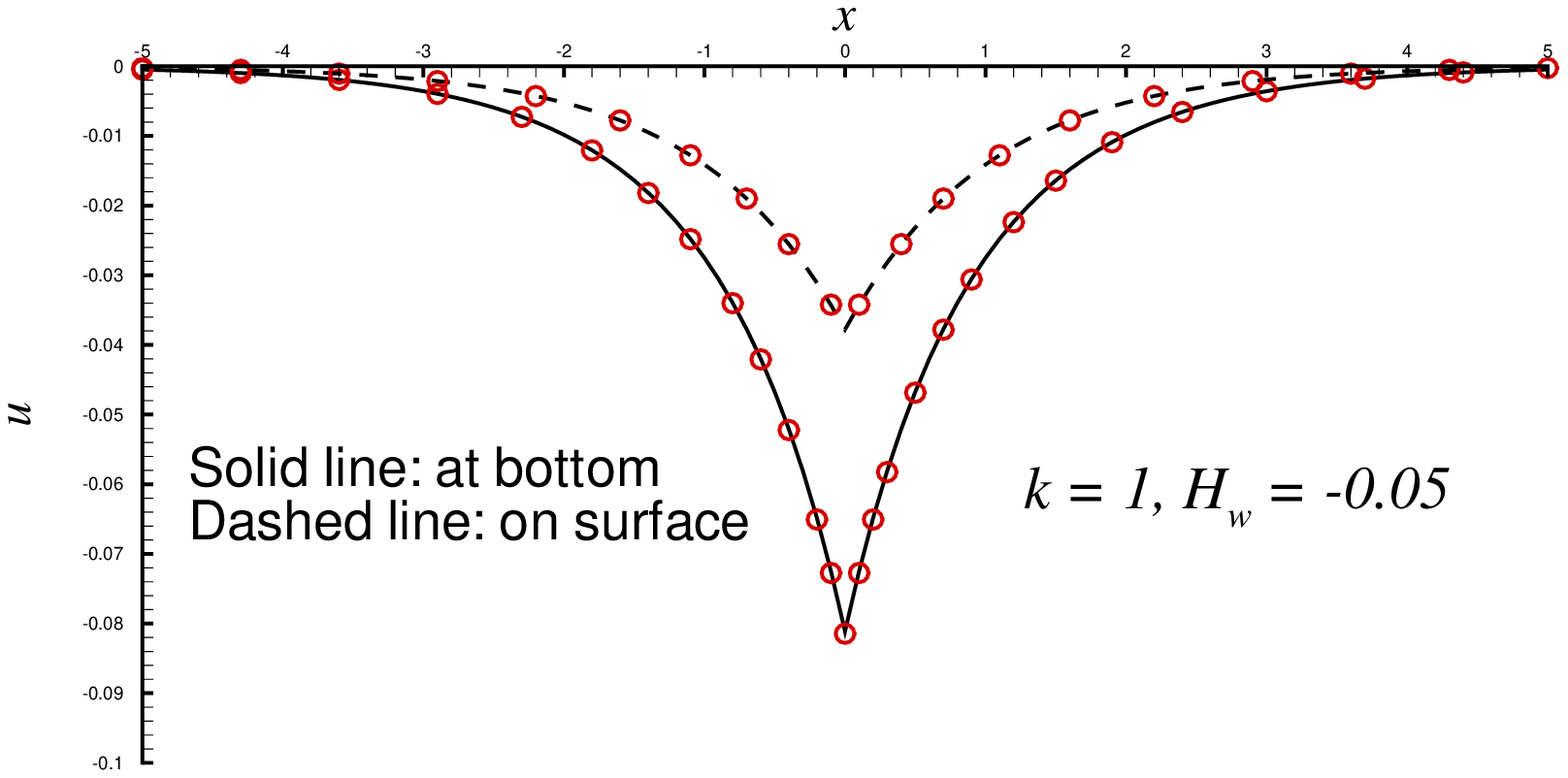}
\caption{  Horizontal velocity at bottom and on free surface when $k=1$ and $H_w=-0.05$ by means of $c_\phi=-1$ and $c_\eta=-1$.  Solid line:   3rd-order approx. of $u(x,-1)$ (at bottom);  Dashed line: 3rd-order approx. of $u(x,\zeta(x))$ (on free surface);  Symbols: the corresponding 25th-order approximations.    }
\label{figure:ubHw-0d05}
\end{figure}

 Furthermore,  as shown in Tables \ref{Table:uK1Hw0d05} to \ref{Table:uK1Hw-0d1} and Figs. 2 to \ref{figure:uK1Hw-0d1},  the horizontal bottom velocity of the peaked solitary waves is  always  larger than that on free surface.  For example, in the case of  $k=1$ and $H_w =  0.1$, the bottom horizontal velocity at crest is 43\%  larger than that on free surface, as shown in Table~\ref{Table:uK1Hw0d1}.  Especially, as shown in Table~\ref{Table:uK1Hw-0d1},  in the case of  $k=1$ and $H_w=-0.1$, the horizontal bottom velocity at crest  is even 158\% larger than that on free surface!   In general,  for the same $x$,   the  horizontal bottom velocity $u(x,-1)$  has always a larger absolute value than $u(x,\zeta)$ on the free surface, as shown  in Figs.~\ref{figure:ubHw0d05} and \ref{figure:ubHw-0d05}.   This is quite different from the traditional smooth  periodic and solitary waves,  whose horizontal bottom velocity  always decays exponentially from free surface to bottom.  The similar characteristic was reported for the peaked solitary waves given by the linear UWM in \S~4.1.   
 
In addition, as mentioned in \S~4.1,  the kinetic energy of peaked  solitary  waves given by the linear  UWM  is  {\em independent} of $z$.     Using the exact  UWM,  we obtain the similar conclusions.  For example,  in case of $k=1$, it is found that the kinetic energy of peaked solitary waves on bottom may be 13.1\% larger than that on free surface when $H_w=-1/20$, and is at most  only 10.4\% smaller when $H_w=1/20$.  Therefore,  from  free surface to bottom,  the kinetic energy of peaked solitary waves either increases or decays a little.   This is quite different from the traditional smooth waves whose kinetic energy decreases {\em exponentially} from free surface to bottom.   

All of these  unusual characteristics   clearly  indicate  the novelty of the peaked solitary waves.  Note  that the  peaked solitary waves  given  by the linear   UWM  in \S~\ref{LinearTheory}  have the rather  similar characteristics as mentioned above.

\subsection{Kelvin's theorem for peaked solitary waves }

As mentioned above, at crest,  the peaked solitary waves given by the  unified  wave model  (UWM) have a discontinuous vertical velocity  and the discontinuous 1st-derivative of wave elevation.  Does Kelvin's theorem still hold for them?    

Without loss of generality,  let us consider the interface of  two fluids in general.   Let ${\bf u}_i$ and $\rho_i$  denote the velocity vector and density of the two fluids,  where $i=1,2$.    Taking a closed curve $\mathcal{C}$ moving with the fluid, we have the circulation 
\begin{equation}
\Gamma = \oint_{\mathcal{C}}  {\bf u} \cdot  {\bf dr}. 
\end{equation} 
 Obviously,  it holds the Kelvin's theorem $d \Gamma/d t = 0$ in each fluid.   Then,  consider a closed curve $\mathcal{C}$  cross the interface of the two fluids.   Let $Q_1$ and  $Q_2$ denote the two points of intersection of the curve $\mathcal{C}$ and the interface.    The circulation reads     
\begin{equation}
\Gamma =  \int_{\mathcal{C}_1}  {\bf u}_1 \cdot  {\bf dr} + \int_{\mathcal{C}_2}  {\bf u}_2 \cdot  {\bf dr}, 
\end{equation} 
where $\mathcal{C}_i$ denotes the part of the curve in the $i$th fluid.   Then,  using the Euler equation in each fluid,  it  holds  
 \begin{eqnarray}
\frac{d\Gamma }{dt} &=& \int_{\mathcal{C}_1} \{\frac{d{\bf u}_1}{dt} \cdot d{\bf r} + {\bf u}_1. d{\bf u}_1\} + 
\int_{\mathcal{C}_2} \{\frac{d{\bf u}_2}{dt} \cdot d{\bf r} + {\bf u}_2. d{\bf u}_2 \} \nonumber \\ 
& =&\left. \left( \frac{p}{\rho_{1}} + gz + \frac{|{\bf u}_1|^2 }{2}\right)\right|^{Q_{2}}_{Q_{1}} +  \left. \left( 
\frac{p}{\rho_{2}} + gz + \frac{|{\bf u}_2|^2 }{2}\right)\right|^{Q_{1}}_{Q_{2}} \nonumber \\
& =&\left[ p(Q_2 ) -p(Q_1) \right] \left(\frac{1}{\rho_2 }- \frac{1}{\rho_1}\right) \nonumber\\
&+&\frac{1}{2} \left\{  \left(\left|{\bf u}_1(Q_2) \right|^2   -  \left|{\bf u}_2(Q_2)\right|^2\right)  -  \left(\left|{\bf u}_1(Q_1) \right|^2   -  \left|{\bf u}_2(Q_1)\right|^2\right)  \right\}, \label{Kelvin-2}
\end{eqnarray}
where $p(Q_i)$ denotes the pressure at $Q_i$, and ${\bf u}_j(Q_i)$  denotes the velocity of the $j$th fluid at the point $Q_i$, respectively,  where $i,j = 1,2$.  

Let us first consider the traditional smooth interfacial waves of two layer flows with $\rho_2 > \rho_1$.     It is well-known \cite{Grue-JFM1997, Choi1999,  Kataoka-JFM2008, Alam-JFM2012} that, on the  free  surface of  the  interfacial waves,  although the velocity normal to the interface is continuous,  the tangential velocity is discontinuous, so that 
\begin{equation}  
\left| {\bf u}_1(Q_i) \right| \neq \left| {\bf u}_2(Q_i) \right|, \;\;\; i = 1,2.   
\end{equation} 
Besides, it holds  $p(Q_1) \neq p(Q_2) $ in general.    Thus, for the traditional  interfacial waves with smooth crest,  it holds  $d \Gamma/dt \neq 0$ when the closed curve $\mathcal{C}$ crosses the interface.  Therefore,  the Kelvin theorem does {\em not} hold for the traditional  interfacial waves with smooth crest.     
  
However, for the peaked solitary waves governed by the  UWM,  we have $\rho_1 =\rho_2$  so that the pressure term in (\ref{Kelvin-2}) vanishes.  Besides,  due to the symmetry (\ref{symmetry}),    it holds
\begin{equation} 
 \left| {\bf u}_1(Q_i) \right| = \left| {\bf u}_2(Q_i) \right|, \;\;\; i = 1,2,   
 \end{equation}
so that the last term in (\ref{Kelvin-2}) vanishes, too!   Therefore, the Kelvin's theorem is still valid {\em everywhere} for the peaked solitary waves given by the UWM, although there exists the discontinuity at crest!    This is indeed a little surprise:  from the viewpoint of Kelvin's theorem,  the peaked solitary waves  are  even  more acceptable than the traditional interfacial waves of two-layer fluids with smooth crest!    

 \section{Concluding remarks and discussions\label{Discussion}}
 
 Many  wave  models  such as  the Camassa-Holm (CH) equation  \cite{Camassa1993PRL}  admit peaked solitary waves, and thousands of articles about peaked solitary waves have been published.   However,  it was an open question whether the fully nonlinear wave equations admit peaked solitary waves or not.   This is rather strange,  since most of wave models such as the CH equation  \cite{Camassa1993PRL}  were derived from the fully nonlinear wave equations under some assumptions.     
 
In this paper, we propose a  unified  wave model  (UWM) for progressive gravity waves with symmetric and permanent form in a finite water depth.  The UWM is based on the symmetry and the fully nonlinear wave equations.   Especially, unlike other wave models,  its flow is {\em unnecessary} to be irrotational at crest ($x=0$).   Thus, the  UWM  is more general:  it  admits  not only all traditional periodic and solitary waves with infinitely differential surface (see \S~3), but also  the peaked solitary waves (see \S~4) which  include the famous  peaked solitary wave (\ref{res:u:CH}) of the CH equation (\ref{geq:CH}) and process many unusual characteristics.    In addition,  it is proved (see \S~4.4) that  Kelvin's theorem still  holds  {\em everywhere}  for  the the peaked solitary waves,  although there exists a vortex sheet  at  crest ($x=0$).    Therefore, the  UWM {\em unifies} the traditional smooth waves and the peaked solitary ones, for the first time, to the best of the author's knowledge.  In other words, the peaked solitary waves are  {\em consistent} with the traditional smooth waves: they are as acceptable as the traditional smooth waves!   It is  fantastic that the two completely {\em different} types of waves  can  be  derived from the {\em same} wave model, i.e. the UWM.      

These peaked solitary waves expressed by the evanescent  base-functions (\ref{base:new}) have many unusual characteristics quite different from the traditional  periodic and solitary waves expressed by the smooth base-functions (\ref{base:traditional}).   First,  it has a peaked crest with a discontinuous vertical velocity at crest ($x=0$).   Secondly,   it  may be in the form of depression,  corresponding to a negative wave height $H_w$,  which  has been reported for interfacial solitary waves  but  never for free-surface solitary waves,  to the best of author's knowledge.    Third,  from free surface to bottom, its horizontal velocity always increases, and besides its kinetic energy either increases or decays a little.    Especially,  unlike the traditional smooth waves, which are dispersive with wave height that is a characteristic length of smooth wave in the vertical direction, the phase speed of peaked solitary waves  has nothing to do with wave weight:  in fact, they are dispersive (for a fixed wave height) with ``the actual wavelength'' $\lambda_a$, defined by (\ref{def:L:a}), which is a characteristic length of peaked solitary wave in the horizontal direction.   All of these are so  different from the smooth periodic and solitary waves that they clearly indicate the novelty  of the  peaked  solitary waves  reported in this article.  
 
  All of these unusual characteristics come from the so-called evanescent  base-functions (\ref{base:new}) of the peaked solitary waves,  which are essentially different from
 the smooth base-functions (\ref{base:traditional}) for the traditional periodic and solitary waves,  although both of them  automatically  satisfy  the Laplace equation (\ref{geq:phi}) in $0<x<+\infty$, the bottom condition (\ref{bc:bottom}) and the bounded condition (\ref{bc:bounded}).  It should be emphasized that both of them are widely used and can be found in the textbook of water waves \cite{MeiBook2005}, although the smooth base-functions (\ref{base:traditional}) are currently  more familiar and thus  regarded as the mainstream.     Note that  the traditional base-functions (\ref{base:traditional}) are infinitely differentiable {\em everywhere}, and {\em automatically} satisfy  the symmetry conditions (\ref{symmetry}) and (\ref{v:x=0}).   However,  the  evanescent  base-functions    (\ref{base:new})   have  a  {\em discontinuous} 1st-derivative (with  respect  to  $x$)  at crest $x=0$, and besides  do {\em not}  automatically  satisfy  the symmetry conditions (\ref{symmetry}) and (\ref{v:x=0}).   Mathematically, these  essential  differences  of  the  two  kinds  of  base-functions  (\ref{base:traditional}) and (\ref{base:new}) are the origin of the completely different characteristics between the smooth and peaked  waves.  
 
Unlike traditional smooth waves,  the peaked solitary waves admit not only the discontinuous 1st-derivative of elevation at crest but also the discontinuous velocity there: 
the vertical velocity $v$ changes sign as we cross  the  plane  $x=0$.    Thus, there exists a vorticity sheet at  crest $x=0$.   However,  such kind of discontinuity of velocity is acceptable even in the frame of the traditional wave theories for smooth waves, since  it  also exists  for the traditional  interfacial  waves expressed by the smooth base-functions (\ref{base:traditional}): ``the tangential velocity changes sign as we cross the surface'' of interfacial waves,  and ``in reality the discontinuity, if it could ever be originated, would be immediately abolished by viscosity'',  as mentioned by Lamb \cite{Lamb1932} (\S 231, page 371).   Similarly,   in reality,  the discontinuity of the vertical velocity (at $x=0$) of the peaked solitary waves, if it could ever be originated, would be immediately abolished by viscosity.   So,  in the frame of the inviscid fluid,  the  peaked  solitary  waves  should be as  reasonable  as the widely accepted interfacial  waves  with smooth crest.   

Besides,  as  shown in \S~4.4,  for the traditional interfacial waves with smooth crest,  a  circulation (expressed by $\Gamma$) along a closed curve crossing the interface of two-layer fluids  {\em disobeys} the Kelvin's theorem, i.e. $d\Gamma/d t \neq 0$.   However,  it is proved that  the Kelvin's theorem still holds {\em everywhere} for the peaked solitary waves in general cases.  Therefore,  from the viewpoint of Kelvin's theorem, the peaked solitary waves are even more reasonable and thus more acceptable than the traditional interfacial waves with smooth crest.  

In a sense,  such kind of  discontinuity (or singularity)  of the peaked solitary waves can be removed.  Let $U(z)$ denote the  horizontal velocity   of a progressive wave (with symmetry and permanent form) at crest ($x=0$),  governed by the  UWM based on the symmetry and the fully nonlinear wave equations  (\ref{geq:phi}) -- (\ref{bc:bounded}).   Assume that, in the frame moving with the solitary wave,   one could  instantaneously  replace  the  boundary $x=0$  by  a  porous vertical plate, and at the same time  could  enforce a horizontal velocity $ U(z)= u(0,z)$ through the porous plate.  Then,  the  corresponding  velocity potential $\phi$  and wave elevation $\zeta(x)$  are governed by the {\em  same}  fully nonlinear wave equations (\ref{geq:phi}) -- (\ref{bc:bounded}), but only in the domain $x\in(0,+\infty)$.    Therefore, for  a properly given   $U(z)$,  one gains either the traditional periodic/solitary waves with smooth crest,  if the smooth base-functions (\ref{base:traditional}) are used, or the  peaked solitary waves, if the evanescent base-functions (\ref{base:new})  are  employed, respectively\footnote{Note that the corresponding horizontal velocities $U(z)$ at $x=0$ for the traditional smooth progressive waves with permanent form are different from that for the peaked solitary ones.}.  In this case,  the Laplace equation (\ref{geq:phi}) is satisfied  in the {\em total} physical domain $x\in (0,+\infty)$  so that {\em no} discontinuity (or singularity) exists at all, since $x=0$ becomes a physical boundary.   For this wave propagation problem,  the peaked solitary waves (although only in the region $0 \leq  x < +\infty$) have very clear physical meanings and are  physically as reasonable as the traditional smooth waves.  
 
The peaked solitary surface waves found in this paper  may provide us not only  new explanations of some natural phenomenon but also  a few theoretical predictions.   First, the peaked  solitary waves have an unusual and interesting characteristic: its phase speed has nothing to do with the wave height $H_w$.  It should be emphasized that,  according to the transcendental  equation (\ref{property:PhaseSpeed:B}),  the peaked solitary waves with a small wave height $H_w$ may propagate very quickly, since $\tan(k)/k\to +\infty$ as $ k \to n \pi + \pi/2$ for an integer $n \geq 0$.   Thus,  all of these peaked solitary waves with small wave height but different phase speed may create a huge  solitary surface wave somewhere: this gives a new theoretical explanation about the so-called  ``rogue wave'' that can {\em suddenly} appear on ocean,  even  when ``the weather was good, with clear skies and glassy swells'', as reported by Graham \cite{Graham2000} and mentioned by  Kharif \cite{Kharif2003}.   
Secondly,   our peaked solitary waves  predict  free surface  in  the form  of depression, corresponding to a negative wave height $H_w$.   To the best of the author's knowledge,  solitary waves of depression have been reported only for interfacial waves with smooth crest,  but never for surface gravity waves.  Such  kind  of  peaked solitary surface waves of depression might be more difficult to create in practice and/or by experiment than the traditional ones with smooth crest.  However, if this theoretical prediction would be physically correct, we should observe it in laboratory  and/or in  practice, sooner or later.    This is an interesting  but  challenging  work:  it could enrich and deepen our understanding about solitary waves, no matter whether the conclusions are positive or not.  
Thirdly, according to the traditional wave theories,  the  kinetic energy  of  traditional waves decreases {\em exponentially} from free surface to bottom,  thus  a submarine far  beneath  surface  is  safe even if  there are huge waves on  ocean.    However,  different from traditional smooth periodic and solitary waves,   the horizontal velocity  of the peaked  solitary waves  always increases from free surface to bottom.  Besides,   its kinetic energy  either increases or decays a little from free surface to bottom.   However,  due to the viscosity of fluid in practice,  the horizontal velocity of water waves must be zero at bottom  so that such kind of the peaked solitary waves might not exist exactly in its theoretical form as reported in this article, since there always exists a thin viscous boundary layer near  bottom.  However,  if such kind of peaked solitary  waves with large horizontal velocity and kinetic energy  near  bottom  would  indeed  exist  in practice,  they should be quite dangerous to  submarines,   platforms  and  equipments  in underwater  engineering, even though they are not exactly in the  same form as found  in this paper.

Possibly,  the new kind of peaked  solitary  waves might change some  traditional view-points. For example,  solitary waves are often regarded as a nonlinear phenomenon.  However,   we illustrate in \S~4.1 that solitary surface waves may exist even  in a system  of  {\em linear} differential equations.     Besides,  it  is  widely  believed  that solitary water waves exist only in {\em shallow} water.   But,  we indicate here  that  solitary waves can exist  even   in a {\em finite} water depth, say, $D$ is {\em unnecessary} to be small.  For instance,  in the case of  $D = 100$ (meter), $k=1$ and the dimensionless wave height $H_w=0.05$,  the corresponding  peaked solitary  wave propagates  with  the 5 meter  wave-height in the phase  speed $c = 1.24796 \sqrt{g D} \approx 39.1$ meter per second, which is not very dangerous.    However,  in the case of  $D=1000$ (meter) with the same  dimensionless parameters (i.e. $k=1$  and  $H_w=0.05$),  the corresponding peaked  solitary  wave propagates with the 50 meter wave-height and the phase speed $c = 123.5 $ meter per second,  which  is  destructive  if it  could  indeed  occur on the earth!   Here, it may be worth mentioning that, the peaked solitary waves contain large kinetic energy only in a small domain near crest, since the kinetic energy either increases or decays a little from free surface to bottom, but decays exponentially in the horizontal  direction.    Thus, the peaked solitary waves should be more dangerous than the traditional smooth waves, if they  could  indeed  exist in practice.              

However, it should be emphasized  that  the peaked  solitary waves found in this article are obtained under the assumptions  that the fluid is {\em inviscid} and incompressible,  the flow is irrotational in the domain $x > 0$ and $x<0$ (it implies that the flow is {\em not} necessarily irrotational at $x=0$), the surface tension is neglected,  and the wave elevation has a symmetry.  Although part of  these assumptions  are used for the traditional periodic and solitary waves with smooth crest, the {\em physical}  reasonableness of the peaked solitary waves should  be studied deeply in future.  Note that surface tension might have an important influence on the crest of the peaked solitary waves.   Especially,  the viscosity of fluid might have an important influence  on  the discontinuous vertical velocity  and the vortex sheet  at crest ($x=0$):  it is unknown  whether or not    viscosity  of  fluid and surface tension  might  essentially change the properties of the peaked solitary waves.  The author  personally believes  that  these open questions can not be answered in the frame of inviscid fluid, and thus must be studied in the frame of viscous fluid.   In the past, peaked solitary waves were  investigated mainly from mathematical viewpoints.   Now, it is the time  to  consider them more from physical viewpoints.     

Note that the peaked solitary waves found in this article by the  unified  wave model  logically include the famous peaked solitary wave (\ref{res:u:CH}) of the CH equation (\ref{geq:CH}).   It should be emphasized that the newly found peaked solitary waves have a discontinuous velocity and a vortex sheet at crest, and especially their phase speed has nothing to do with wave height.   These unique characteristics  are however {\em not} predicted by the traditional simplified wave models such as the CH equation.   So, if  the newly found  peaked  solitary  waves  are  indeed {\em physically} correct and reasonable, they could deepen our understandings about the traditional peaked solitary waves.  However, if these newly found peaked waves are physically {\em not} reasonable, the traditional  peaked/cusped solitary waves given by these  simplified wave models must be checked carefully from the {\em physical}  viewpoints, although thousands of related articles have been published mainly from the {\em mathematical} viewpoints.    

Note also that the symmetry plays an important role in the  unified  wave model (UWM).   Such kind of symmetry has been successfully applied to give a theoretical explanation \cite{Liao2013-WM} for the newly found standing solitary waves by physical experiment \cite{PRL2011}.   Note that both of the odd and even symmetry were used in \cite{Liao2013-WM},  but  only even symmetry (\ref{symmetry})  is  considered  in  this  article.   In fact, it is straightforward to include both of the odd and even symmetry in the UWM.   However, it is a pity that such kind of symmetry does not exist for two or more progressive waves.   This  restricts  the  applications  of  the  UWM  based  on  the  symmetry and fully nonlinear wave equations.   To further  extend the  UWM,  some new mathematical methods for discontinuous functions  must be proposed.    

Finally,  this  paper  is  a clear demonstration of the power of the HAM, for the discovery of the new kind of peaked solitary waves with many unusual characteristics, which have never been reported, to the best of the author's knowledge.   Indeed, a truly new method  always  gives  something new or different.  So, this paper also illustrates the novelty of the HAM, too.   Independent of small/large physical parameters, the HAM is a useful tool  for  lots of nonlinear problems.         

Without doubts, further theoretical, numerical and experimental studies, and especially practical/experimental observations about the solitary surface waves with peaked crest,  are needed  in future.  Indeed, discontinuity and/or singularity are  more  difficult to handle.   But, they  should  not  be  evaded easily,  since they might open some new fields of research, and  greatly enrich  and  deepen  our understandings about the real world.
 
\section*{Acknowledgement}
 
Thanks to  professor C.C. Mei (M.I.T., USA) for the helpful discussions, and  professor Roger Grimshaw (Loughborough University, UK) for the discussion on the Kelvin's theorem for a closed curve crossing the interface of two layer fluids (about the traditional interfacial waves).   This work is partly supported by the State Key Laboratory of Ocean Engineering (Approval No. GKZD010061) and the National Natural Science Foundation of China (Approval No. 11272209).

\bibliography{NewSoliton}

\begin{thebibliography}{10}

\bibitem{Russell1845}
J.~S. Russell.
\newblock On waves.
\newblock {\em Report of the fourteenth meeting of the British Association for
  the Advancement of Science}, pages 311--390, 1845.

\bibitem{Boussinesq1872}
J.~Boussinesq.
\newblock Th\'{e}orie des ondes et des remous qui se propagent le long d'un
  canal rectangulaire horizontal, en communiquant au liquide contenu dans ce
  canal des vitesses sensiblement pareilles de la surface au fond.
\newblock {\em Journal de Math'{e}matiques Pures et Appliqu\'{e}es.
  Deuxi\'{e}me S\'{e}rie}, 17:55 -- 108, 1872.

\bibitem{KdV1895}
D.~J. Korteweg and G.~de~Vries.
\newblock On the change of form of long waves advancing in a rectangular canal,
  and on a new type of long stationary waves.
\newblock {\em Philosophical Magazine}, 39:422 -- 443, 1895.

\bibitem{Benjamin1972}
B.~Benjamin, J.L. Bona, and J.J. Mahony.
\newblock Model equations for long waves in nonlinear dispersive systems.
\newblock {\em Philos. Trans. Roy. Soc. London}, 272:47 -- 78, 1972.

\bibitem{Stokes1894}
G.G. Stokes.
\newblock On the theory of oscillation waves.
\newblock {\em Trans. Cambridge Phil. Phys.}, 8:441--455, 1894.

\bibitem{Camassa1993PRL}
R.~Camassa and D.~D. Holm.
\newblock An integrable shallow water equation with peaked solitons.
\newblock {\em Phys. Rev. Lett.}, 71:1661 -- 1664, 1993.

\bibitem{Constantin2000}
A.~Constantin.
\newblock Existence of permanent and breaking waves for a shallow water
  equation: a geometric approach.
\newblock {\em Ann. Inst. Fourier, Grenoble}, 50(2):321 -- 362, 2000.

\bibitem{Fushssteiner1996}
B.~Fuchssteiner.
\newblock Some tricks from the symmetry-toolbox for nonlinear equations:
  generalizations of the {Camassa-Holm} equation.
\newblock {\em Physica D}, 95:296 -- 343, 1996.

\bibitem{Wu-2008IJNMF}
Y.Y. Wu and K.F. Cheung.
\newblock Explicit solution to the exact riemann problem and application in
  nonlinear shallow-water equations.
\newblock {\em Int. J. Numer. Meth. Fluids}, 57:1649 -- 1668, 2008.

\bibitem{Rosatti-2010JCP}
R.~Rosatti and L.~Begnudelli.
\newblock The riemann problem for the one-dimensional, free-surface shallow
  water equations with a bed step: Theoretical analysis and numerical
  simulations.
\newblock {\em J. Computational Phys.}, 229:760 -- 787, 2010.

\bibitem{Liao2012-SciChinaG}
S.J. Liao.
\newblock Two kinds of peaked solitary waves of the {KdV, BBM and Boussinesq}
  equations.
\newblock {\em Science China- Physics, Mechanics \& Astronomy}, 55:2469 --
  2475, 2012.

\bibitem{Liao2013-SciChinaG}
S.J. Liao.
\newblock On peaked solitary waves of the {Degasperis-Procesi} equation.
\newblock {\em Science China- Physics, Mechanics \& Astronomy}, 56:418 -- 422,
  2013.

\bibitem{Kraenkel1999}
R.~A. Kraenkel and A.~Zenchuk.
\newblock {Camassa - Holm} equation: transformation to deformed {sinhÐGordon}
  equations, cuspon and soliton solutions.
\newblock {\em J. Phys. A: Math. Gen.}, 32:4733 -- 4747, 1999.

\bibitem{Liao1997NLM}
S.J. Liao.
\newblock An approximate solution technique which does not depend upon small
  parameters ({Part} 2): an application in fluid mechanics.
\newblock {\em Int. J. Non-Linear Mechanics}, 32:815--822, 1997.

\bibitem{Liao1999JFM}
S.J. Liao.
\newblock A uniformly valid analytic solution of {2D} viscous flow past a
  semi-infinite flat plate.
\newblock {\em J. Fluid Mechanics}, 385:101--128, 1999.

\bibitem{LiaoBook2003}
S.J. Liao.
\newblock {\em {Beyond Perturbation: Introduction to the Homotopy Analysis
  Method}}.
\newblock CRC Press, Boca Raton, 2003.

\bibitem{Liao2003JFM}
S.J. Liao.
\newblock On the analytic solution of magnetohydrodynamic flows of
  non-newtonian fluids over a stretching sheet.
\newblock {\em J. Fluid Mechanics}, 488:189--212, 2003.

\bibitem{Liao2006SAM}
S.J. Liao.
\newblock Series solutions of unsteady boundary-layer flows over a stretching
  flat plate.
\newblock {\em Studies in Applied Mathematics}, 117:239--263, 2006.

\bibitem{LiaoBook2012}
S.J. Liao.
\newblock {\em {Homotopy Analysis Method in Nonlinear Differential Equations}}.
\newblock Springer $\&$ Higher Education Press, Heidelberg, 2012.

\bibitem{KVBook2012}
K.~Vajravelu and R.~Van~Gorder.
\newblock {\em {Nonlinear Flow Phenomena and Homotopy Analysis: Fluid Flow and
  Heat Transfer}}.
\newblock Springer \& Higher Education Press, Heidelberg, 2013.

\bibitem{Cokelet1977}
E.D. Cokelet.
\newblock Steep gravity waves in water of arbitrary uniform depth.
\newblock {\em Philosophical Transaction of the Royal Society of London - A},
  286:286, 1977.

\bibitem{MeiBook2005}
C.C Mei, M.~Stiassnie, and D.K.P. Yue.
\newblock {\em Theory and Applications of Ocean Surface Waves}.
\newblock World Scientific, New Jersey, 2005.

\bibitem{Rayleigh1876}
J.W. Rayleigh.
\newblock On waves.
\newblock {\em Phil. Mag.}, 1:257 -- 271, 1876.

\bibitem{Fenton1972JFM}
J.D. Fenton.
\newblock A ninth-order solution for the solitary wave.
\newblock {\em J. Fluid Mech.}, 53:257--271, 1972.

\bibitem{Fenton1979JFM}
J.D. Fenton.
\newblock A high-order cnoidal wave theory.
\newblock {\em J. Fluid Mech.}, 94:129--161, 1979.

\bibitem{McKee1988}
W.D. McKee.
\newblock Calculation of evenescent wave modes.
\newblock {\em Proc. ASCE, J. Waterway Port Coastal Ocean Eng.}, 114:373 --
  378, 1988.

\bibitem{Massek-CE1983}
S.R. Massel.
\newblock Harmonic generation by waves propagating over a submerged step.
\newblock {\em Coastal Engineering}, 7:357 -- 380, 1983.

\bibitem{Kirby-JFM1983}
J.T. Kirby and R.A. Dalrymple.
\newblock Propagation of obliquely incident water waves over a trench.
\newblock {\em J. Fluid Mech.}, 133:47 -- 63, 1983.

\bibitem{Kirby-JFM1986}
J.T. Kirby.
\newblock A general wave equation for waves over rippled beds.
\newblock {\em J. Fluid Mech.}, 162:171 -- 186, 1986.

\bibitem{Porter-JFM1995}
D.~Porter and D.J. Staziker.
\newblock Extensions of the mild-slope equation.
\newblock {\em J. Fluid Mech.}, 300:367 -- 382, 1995.

\bibitem{Mattioli-1990}
F.~Mattioli.
\newblock Resonant reflection of a series of submerged breakwaters.
\newblock {\em Nuovo Cimento}, 13:823--833, 1990.

\bibitem{Mattioli-1991}
F.~Mattioli.
\newblock Resonant refection of surface waves by non-sinusoidal bottom
  undulations.
\newblock {\em Appl. Ocean Res.}, 13:49 -- 52, 1991.

\bibitem{Massel-CE1993}
S.R. Massel.
\newblock Extended refraction-diffration equation for surface waves.
\newblock {\em Coastal Engineering}, 19:97 -- 126, 1993.

\bibitem{Grue-JFM1997}
J.~Grue, H.A. Friis, E.~Palm, and P.O. Rusas.
\newblock A method for computing unsteady fully nonlinear interfacial wave.
\newblock {\em J. Fluid Mech.}, 351:223 -- 252, 1997.

\bibitem{Choi1999}
W.~Choi and R.~Camassa.
\newblock Fully nonlinear internal waves in a two-fluid system.
\newblock {\em J. Fluid Mech.}, 396:1 -- 36, 1999.

\bibitem{Kataoka-JFM2008}
T.~Kataoka.
\newblock Transverse instability of interfacial solitary waves.
\newblock {\em J. Fluid Mech.}, 611:255 -- 282, 2008.

\bibitem{Alam-JFM2012}
M.R. Alam.
\newblock A new triad resonance between co-propagating surface and interfacial
  waves.
\newblock {\em J. Fluid Mech.}, 691:267 -- 278, 2012.

\bibitem{Lamb1932}
H.~Lamb.
\newblock {\em Hydrodynamics}.
\newblock Cambridge University Press, Cambridge, 1932.

\bibitem{Blasius1908}
H.~Blasius.
\newblock Grenzschichten in fl\"{u}ssigkeiten mit kleiner reibung.
\newblock {\em Z. Math. u. Phys.}, 56:1--37, 1908.

\bibitem{Turkyilmazoglu2009PF}
M.~Turkyilmazoglu.
\newblock Purely analytic solutions of the compressible boundary layer flow due
  to a porous rotating disk with heat transfer.
\newblock {\em Physics of Fluids}, 21:106104, 2009.

\bibitem{Xu2010PF}
H.~Xu, Z.L. Lin, S.J. Liao, J.Z. Wu, and J.~Majdalani.
\newblock Homotopy-based solutions of the {NavierÐStokes} equations for a
  porous channel with orthogonally moving walls.
\newblock {\em Physics of Fluids}, 22(5):053601, 2010.

\bibitem{Turkyilmazoglu2011ASME}
M.~Turkyilmazoglu.
\newblock An optimal analytic approximate solution for the limit cycle of
  {Duffing - van der Pol} equation.
\newblock {\em ASME J. Appl. Mech.}, 78:021005, 2011.

\bibitem{Li2010JMP}
Y.J. Li, B.T. Nohara, and S.J. Liao.
\newblock Series solutions of coupled {Van der Pol} equation by means of
  homotopy analysis method.
\newblock {\em J. Mathematical Physics}, 51:063517, 2010.

\bibitem{Liao2005-IJHMT}
S.J. Liao.
\newblock A new branch of solutions of boundary-layer flows over an impermeable
  stretched plate.
\newblock {\em Int. J. Heat Mass Tran.}, 48:2529 -- 2539, 2005.

\bibitem{Liao-Magyari2006}
S.J. Liao and E.~Magyari.
\newblock Exponentially decaying boundary layers as limiting cases of families
  of algebraically decaying ones.
\newblock {\em Z. angew. Math. Phys.}, 57:777 -- 792, 2006.

\bibitem{SPZhu2006B}
S.P. Zhu.
\newblock An exact and explicit solution for the valuation of american put
  options.
\newblock {\em Quant. Financ.}, 6:229 -- 242, 2006.

\bibitem{Liao2003JEM}
S.~J. Liao and K.~F. Cheung.
\newblock Homotopy analysis of nonlinear progressive waves in deep water.
\newblock {\em Journal of Engineering Mathematics}, 45(2):105--116, 2003.

\bibitem{Schwartz1974JFM}
L.W. Schwartz.
\newblock Computer extension and analytic continuation of stokes' expansion for
  gravity waves.
\newblock {\em J. Fluid Mech.}, 62:553--578, 1974.

\bibitem{Longuet-Higgins1975JFM}
M.S. Longuet-Higgins.
\newblock Integral properties of periodic gravity waves of finite amplitudes.
\newblock {\em Proc. R. Soc. London - A}, 342:157 -- 174, 1975.

\bibitem{Tao2007CE}
L.~B. Tao, H.~Song, and S.~Chakrabarti.
\newblock Nonlinear progressive waves in water of finite depth - an analytic
  approximation.
\newblock {\em Coastal Engineering}, 54(11):825--834, 2007.

\bibitem{Fenton1981JFM}
M.M. Rienecker and J.D. Fenton.
\newblock A fourier approximation method for steady water waves.
\newblock {\em J. Fluid Mech.}, 104:119?37, 1981.

\bibitem{Mehaute1968}
B.~Le~M\'{e}haut\'{e}, D.~Divoky, and A.~Lin.
\newblock Shallow water waves: a comparison of theories and experiments.
\newblock {\em Proceedings of 11th Conference on Coastal Engineering}, pages 86
  -- 107, 1968.

\bibitem{Xudali-2012JFM}
D.L. Xu, Z.L. Lin, S.J. Liao, and M.~Stiassnie.
\newblock On the steady-state fully resonant progressive waves in water of
  finite depth.
\newblock {\em J. Fluid Mechanics}, 710:379 -- 418, 2012.

\bibitem{Graham2000}
D.M. Graham.
\newblock {NOAA} vessel swamped by rogue wave.
\newblock {\em Oceanspace}, 2000.
\newblock No. 284.

\bibitem{Kharif2003}
C.~Kharif and E.~Pelinovsky.
\newblock Physical mechanisms of the rogue wave phenomenon.
\newblock {\em Euro. J. Mech. B/Fluids}, 22:603--634, 2003.

\bibitem{Liao2013-WM}
S.J. Liao.
\newblock Two new standing solitary waves in shallow water.
\newblock {\em Wave Motion}, 50:785 -- 792, 2013.

\bibitem{PRL2011}
J.~Rajchenbach, A.~Leroux, and D.~Clamond.
\newblock New standing solitary waves in water.
\newblock {\em Phys. Rev. Lett.}, 107:24502, 2011.

\end{thebibliography}

\bibliographystyle{unsrt}

\newpage

\begin{center}
{\bf\large Appendix }
{\bf\large Detailed derivation of formulas  (\ref{def:Delta-eta}) - (\ref{def:chi}) }\label{appA}
\end{center}

Write
\begin{equation}
\left(\sum_{i=1}^{+\infty}  \zeta_i \; q^i \right)^m =\sum_{n=m}^{+\infty} \mu_{m,n} \; q^n, \label{def:mu:0}
\end{equation}
with the definition
\begin{equation}
\mu_{1,n}( x) = \zeta_{n}( x), \;\; n\geq 1.   \label{def:mu:1}
\end{equation}
Then,
\begin{eqnarray}
&&\left(\sum_{i=1}^{+\infty}  \zeta_i \; q^i \right)^{m+1}  =  \left( \sum_{n=m}^{+\infty} \mu_{m,n} \; q^n \right) \left(\sum_{i=1}^{+\infty}  \zeta_i \; q^i \right) = \sum_{n=m+1}^{+\infty} \mu_{m+1,n} \; q^n,
\end{eqnarray}
which gives
\begin{equation}
\mu_{m,n}( x) = \sum_{i=m-1}^{n-1} \mu_{m-1,i}( x) \; \zeta_{n-i}( x), \;\;\; m\geq 2, \; n\geq m. \label{def:mu:2}
\end{equation}

Define
\[      \psi_{i}^{n,m}( x)  =  \frac{\partial ^i}{\partial  x^i } \left(\frac{1}{m!} \; \left. \frac{\partial^m \phi_n}{\partial z^m} \right|_{z=0}\right).  \]
By Taylor series,  we have  for {\em any} $z$ that
\begin{equation}
    \phi_n( x, z) = \sum_{m=0}^{+\infty} \left(\left.\frac{1}{m!} \; \frac{\partial^m \phi_n}{\partial z^m}  \right|_{z=0}\right) z^m  =
     \sum_{m=0}^{+\infty} \psi_{0}^{n,m} \; z^m \label{def:phi[n]:Taylor:0}
\end{equation}
and
\begin{equation}
    \frac{\partial^{i} \phi_n}{\partial  x^i } = \sum_{m=0}^{+\infty} \frac{\partial^{i} }{\partial  x^i } \left(\left.\frac{1}{m!} \; \frac{\partial^m \phi_n}{\partial z^m}  \right|_{z=0}\right) z^m  =
     \sum_{m=0}^{+\infty} \psi_{i}^{n,m} \; z^m. \label{def:phi[n]:Taylor:1}
\end{equation}
Then, on the {\em unknown} free surface $z = {\eta}( x;q) $,  we have using (\ref{def:mu:0}) that
\begin{eqnarray}
\frac{\partial^{i}\phi_n}{\partial  x^i }
&=& \sum_{m=0}^{+\infty} \psi_{i}^{n,m}  \; \left( \sum_{s=1}^{+\infty} \zeta_{s} \; q^s\right)^m = \psi_{i}^{n,0}+ \sum_{m=1}^{+\infty} \psi_{i}^{n,m} \; \left( \sum_{s=m}^{+\infty} \mu_{m,s} \; q^s \right)  \nonumber\\
&=&  \sum_{m=0}^{+\infty}\beta_{i}^{n,m}( x) \; q^m, \label{series:phi:n}
\end{eqnarray}
where
\begin{eqnarray}
\beta_{i}^{n,0} &=& \psi_{i}^{n,0},\label{def:beta:1}\\
\beta_{i}^{n,m} &=& \sum_{s=1}^{m} \psi_{i}^{n,s} \; \mu_{s,m}, \;\; m\geq 1. \label{def:beta:2}
\end{eqnarray}

Similarly,  on  the unknown free surface $z = {\eta}( x;q) $, it holds
\begin{eqnarray}
\frac{\partial^{i}}{\partial x^i }\left(\frac{\partial \phi_n}{\partial z}\right)
&=& \sum_{m=0}^{+\infty} \gamma_ {i}^{n,m}( x)\; q^m,  \label{series:phi:n:z} \\
\frac{\partial^{i}}{\partial x^i }\left(\frac{\partial^2 \phi_n}{\partial z^2}\right)
&=& \sum_{m=0}^{+\infty} \delta_ {i}^{n,m}( x)\; q^m,  \label{series:phi:n:zz}
\end{eqnarray}
where
\begin{eqnarray}
\gamma_ {i}^{n,0}  &=& \psi_ {i}^{n,1}, \label{def:gamma:1}\\
\gamma_ {i}^{n,m} &=& \sum_{s=1}^{m} (s+1) \psi_ {i}^{n,s+1} \; \mu_{s,m}, \;\; m \geq 1,     \label{def:gamma:2}\\
\delta_ {i}^{n,0} &=& 2\psi_ {i}^{n,2},  \label{def:delta:1}\\
\delta_ {i}^{n,m} &=& \sum_{s=1}^{m} (s+1)(s+2) \psi_ {i}^{n,s+2} \; \mu_{s,m}, \;\; m \geq 1.    \label{def:delta:2}
\end{eqnarray}

Then, on the unknown free surface $z = {\eta}( x;q) $, it holds  using (\ref{series:phi:n}) that
\begin{eqnarray}
\Phi( x, \zeta; q) &=& \sum_{n=0}^{+\infty} \phi_n( x,  \zeta) \; q^n
=\sum_{n=0}^{+\infty} q^n \left[\sum_{m=0}^{+\infty} \beta_{0}^{n,m}( x) \; q^m \right]\nonumber\\
&=&\sum_{n=0}^{+\infty}\sum_{m=0}^{+\infty} \beta_{0}^{n,m}( x)\; q^{m+n} =\sum_{s=0}^{+\infty}q^s \left[ \sum_{m=0}^{s} \beta_{0}^{s-m,m}( x) \right] \nonumber\\
&=&\sum_{n=0}^{+\infty} \bar{\phi}_{n,0} (x) \; q^n,\label{def:phi:surface}
\end{eqnarray}
where
\begin{eqnarray}
 \bar{\phi}_{n,0} ( x)  = \sum_{m=0}^{n} \beta_{0}^{n-m,m}.  \label{def:phi:bar}
 \end{eqnarray}
Similarly, we have
\begin{eqnarray}
\frac{\partial ^{i}\Phi}{\partial x^i } &=&  \sum_{n=0}^{+\infty} \bar{\phi}_{n,i}( x) \; q^n, \label{def:phi:bar:2}\\
\frac{\partial ^{i}}{\partial x^i } \left(\frac{\partial \Phi}{\partial z}\right) &=& \sum_{n=0}^{+\infty}
\bar{\phi}^z_{n,i}( x) \; q^n,\label{def:phi:z}\\
\frac{\partial ^{i}}{\partial x^i } \left(\frac{\partial^2 \Phi}{\partial z^2} \right)&=& \sum_{n=0}^{+\infty} \bar{\phi}^{zz}_{n,i}( x) \; q^n, \label{def:phi:zz}
\end{eqnarray}
where
\begin{eqnarray}
 \bar{\phi}_{n,i} ( x) & = & \sum_{m=0}^{n} \beta_{i}^{n-m,m}, \label{def:phi:bar:3} \\
  \bar{\phi}^{z}_{n,i} (x)  &=& \sum_{m=0}^{n} \gamma_{i}^{n-m,m} , \label{def:phi:z:bar}\\
 \bar{\phi}^{zz}_{n,i} (x)  &=& \sum_{m=0}^{n} \delta_{i}^{n-m,m}. \label{def:phi:zz:bar}
 \end{eqnarray}

 Then, on the unknown free surface $z =\eta( x;q)$,   it holds using (\ref{def:phi:bar:2}) and (\ref{def:phi:z}) that
 \begin{eqnarray}
f &=& \frac{1}{2} \nabla \Phi \cdot   \nabla \Phi\nonumber\\
 &=&\frac{ 1}{2}\left[ \left(\frac{\partial \Phi}{\partial  x}\right)^2 +  \left(\frac{\partial \Phi}{\partial z}\right)^2 \right]\nonumber\\
 &=& \sum_{m=0}^{+\infty} \Gamma_{m,0}( x) \; q^m,  \label{def:f:bar}
 \end{eqnarray}
 where
 \begin{eqnarray}
 \Gamma_{m,0}( x) &=& \frac{1}{2} \sum_{n=0}^{m} \left( \bar\phi_{n,1} \; \bar\phi_{m-n,1}+\bar\phi^{z}_{n,0} \; \bar\phi^z_{m-n,0}\right) . \label{def:Gamma}
 \end{eqnarray}

Similarly,  it holds on $z =\eta( x;q)$ that
\begin{eqnarray}
\frac{\partial  f}{\partial  x} &=& \nabla \Phi \cdot   \nabla \left(\frac{\partial \Phi}{\partial  x}\right)\nonumber\\
&=& \frac{\partial \Phi}{\partial  x} \frac{\partial^2 \Phi}{\partial  x^2} + \frac{\partial \Phi}{\partial z} \frac{\partial }{\partial  x}\left( \frac{\partial \Phi}{\partial z}\right) \nonumber\\
&=&\sum_{m=0}^{+\infty} \Gamma_{m,1}( x) \; q^m, \label{def:Gamma:x1}
\end{eqnarray}
where
\begin{eqnarray}
\Gamma_{m,1}( x) &=&  \sum_{n=0}^{m} \left( \bar\phi_{n,1}  \; \bar\phi_{m-n,2}
 +  \bar\phi^z_{n,0} \; \bar\phi^z_{m-n,1}\right).
\label{def:GAMMA:x1:coefficient}
\end{eqnarray}
Besides,  on $z  =\eta( x;q)$, we have by means of (\ref{def:phi:bar:2}), (\ref{def:phi:z}) and (\ref{def:phi:zz}) that
 \begin{eqnarray}
 \frac{\partial  f}{\partial z} &=&  \nabla \Phi \cdot \nabla \left(\frac{\partial \Phi}{\partial z}\right) \nonumber\\
 &=&  \frac{\partial \Phi}{\partial  x} \frac{\partial}{\partial  x} \left( \frac{\partial \Phi}{\partial z}\right) +\frac{\partial \Phi}{\partial z}  \frac{\partial^2 \Phi}{\partial z^2}  \nonumber\\
 &=& \sum_{m=0}^{+\infty} \Gamma_{m,3}( x) \; q^m, \label{def:Gamma:z}
 \end{eqnarray}
 where
\begin{eqnarray}
\Gamma_{m,3}( x) &=&  \sum_{n=0}^{m} \left( \bar\phi_{n,1} \; \bar\phi^z_{m-n,1}
+  \bar\phi^z_{n,0} \; \bar\phi^{zz}_{m-n,0} \right).
\label{def:GAMMA:z:coefficient}
\end{eqnarray}
Furthermore, using (\ref{def:phi:bar:2}), (\ref{def:Gamma:x1}) and (\ref{def:Gamma:z}),  we have on $z  =\eta( x;q)$ that
\begin{eqnarray}
\nabla \Phi \cdot \nabla  f &=&  \frac{\partial \Phi}{\partial x} \frac{\partial  f}{\partial  x} + \frac{\partial \Phi}{\partial z} \frac{\partial  f}{\partial z}
=\sum_{m=0}^{+\infty} \Lambda_m( x) \; q^m,  \label{def:Lambda}
\end{eqnarray}
where
\begin{eqnarray}
\Lambda_{m}( x ) &=&  \sum_{n=0}^{m} \left(  \bar\phi_{n,1} \; \Gamma_{m-n,1}
+  \bar\phi^z_{n,0} \; \Gamma_{m-n,3}\right)
\label{def:Lambda:coefficient}
\end{eqnarray}

 Then, using  (\ref{def:phi:bar:2}), (\ref{def:phi:z}), (\ref{def:Gamma:x1})  and (\ref{def:Lambda}),  we have on $z =\eta( x;q)$ that
 \begin{eqnarray}
&& {\cal N}  \left[ \Phi(x,z;q) \right] \nonumber\\
 &=& \alpha^2 \frac{\partial^2 \Phi}{\partial x^2} +\frac{\partial \Phi}{\partial z} - 2 \alpha \frac{\partial f}{\partial x}  +\nabla \Phi \cdot \nabla f
 \nonumber\\
 &=&  \sum_{m=0}^{+\infty} \Delta^\phi_m( x) \; q^m, \label{series:N:q}
 \end{eqnarray}
  where
  \begin{eqnarray}
  \Delta^\phi_m( x) &=& \alpha^2 \; \bar\phi_{m,2} + \bar\phi^z_{m,0} - 2  \alpha \; \Gamma_{m,1}+\Lambda_m  \label{def:Delta:phi}
  \end{eqnarray}
for $ m\geq 0 $.

Using (\ref{series:phi:q}) and (\ref{series:phi:n}), we have  on $z=\eta( x;q)$ that
\begin{eqnarray}
\frac{\partial^2 }{\partial x^2}\left( \Phi-\phi_0\right) &=& \sum_{n=1}^{+\infty} \frac{\partial^2 \phi_n( x,\eta)}{\partial x^2} \; q^n  = \sum_{n=1}^{+\infty} q^n
\left(\sum_{m=0}^{+\infty} \beta_{2}^{n,m} \; q^m \right)\nonumber\\
&=&\sum_{n=1}^{+\infty} q^n \left( \sum_{m=0}^{n-1}\beta_{2}^{n-m,m}\right),
\end{eqnarray}
and similarly
\begin{eqnarray}
\frac{\partial }{\partial z}\left( \Phi-\phi_0 \right)&=& \sum_{n=1}^{+\infty}    \frac{\partial \phi_n(x,\eta)}{\partial z} \; q^n  = \sum_{n=1}^{+\infty} q^n \left(\sum_{m=0}^{+\infty} \gamma_{0}^{n,m} \; q^m \right)\nonumber\\
&=&\sum_{n=1}^{+\infty} q^n \left( \sum_{m=0}^{n-1}\gamma_{0}^{n-m,m}\right),
\end{eqnarray}
respectively.    Then, on $z=\eta( x;q)$, it holds due to the linear property of the operator (\ref{def:L}) that
\begin{equation}
{\cal L}\left(\Phi -\phi_0 \right) =  \sum_{n=1}^{+\infty} S_n( x)\; q^n,
\end{equation}
where
\begin{eqnarray}
S_n( x)= \sum_{m=0}^{n-1}\left( \alpha^2 \; \beta_{2}^{n-m,m} +  \; \gamma_{0}^{n-m,m}\right) . \label{def:S}
\end{eqnarray}
Then, on $z=\eta( x;q)$, it holds
\begin{eqnarray}
(1-q) {\cal L}\left(\Phi -\phi_0 \right) = (1-q) \;\sum_{n=1}^{+\infty} S_n\; q^n = \sum_{n=1}^{+\infty} \left( S_n-\chi_n \; S_{n-1}\right) q^n,\label{geq:zero:left}
\end{eqnarray}
where
\begin{equation}
\chi_n = \left\{
\begin{array}{cc}
0, & \mbox{when $n\leq 1$}, \\
1, & \mbox{when $n > 1$} .
\end{array} \right.
\end{equation}
Substituting (\ref{geq:zero:left}), (\ref{series:N:q}) into (\ref{bc:zero:phi}) and equating the like-power of $q$, we have the boundary condition:
\begin{equation}
S_m( x)-\chi_m \; S_{m-1}( x) = c_\phi \; \Delta^\phi_{m-1}( x),\;\;\;\;   m\geq 1. \label{bc:phi:mth:0}
\end{equation}

Define
\begin{eqnarray}
\bar{S}_n( x) = \sum_{m=1}^{n-1}\left(  \alpha^2 \; \beta_{2}^{n-m,m} + \gamma_{0}^{n-m,m}\right) . \hspace{1.0cm} \label{def:S:bar}
\end{eqnarray}
Then,
\begin{eqnarray}
S_n &=& \left( \alpha^2 \; \beta_{2}^{n,0}  +  \gamma_{0}^{n,0}  \right) +\bar{S}_n = \left.  \left( \alpha^2 \;  \frac{\partial^2 \phi_n}{\partial x^2} +\frac{\partial  \phi_n}{\partial z} \right)\right|_{z=0} + \bar{S}_n. \hspace{1.0cm}
\end{eqnarray}
Substituting the above expression into (\ref{bc:phi:mth:0}) gives the boundary condition  on $z=0$:
\begin{equation}
\left.  \left( \alpha^2\; \frac{\partial^2 \phi_m}{\partial x^2} + \frac{\partial  \phi_m}{\partial z} \right)\right|_{z=0}  =  \left. \left\{ c_\phi \; \Delta^\phi_{m-1} +\chi_m\; S_{m-1} -\bar{S}_m \right\}\right|_{z=0} ,  \;\;\;  m \geq 1. \label{bc:phi:m:0}
\end{equation}

Substituting the series (\ref{series:zeta:q}), (\ref{def:phi:bar:2}) and (\ref{def:f:bar}) into (\ref{bc:zero:zeta}), equating the like-power of $q$, we have on $z=0$ that
 \begin{equation}
 \zeta_m( x) =   \left. \left\{ c_\eta  \; \Delta_{m-1}^\eta + \chi_m \; \zeta_{m-1} \right\}\right|_{z=0} , \;\;\;  m\geq 1,
 \end{equation}
where
\[   \Delta_{m}^\eta = \zeta_{m}  - \alpha \; \bar\phi_{m,1}
 + \Gamma_{m,0}.    \]

\end{document}